\documentclass[nohyper]{JHEP3}

\usepackage{amsmath}
\usepackage{amsfonts}
\usepackage{slashed}
\usepackage{graphicx}
\usepackage{verbatim}
\usepackage{subfigure}
\usepackage{epstopdf}
\newcommand \E {\text{E}}
\newcommand \NE {\text{NE}}

\newcommand \slsh [1] {\not\!{#1}}

\newcommand{\pb}{\ensuremath{\slsh{\bar{p}}}}

\def\beq{\begin{equation}}
\def\eeq{\end{equation}}
\def\beqa{\begin{eqnarray}}
\def\eeqa{\end{eqnarray}}

\def\sect#1{Sec.~{\ref{#1}}}
\def\fig#1{fig.~{\ref{#1}}}
\def\Fig#1{Fig.~{\ref{#1}}}

\def\eqn#1{eq.~(\ref{#1})}
\def\Eqn#1{Equation~(\ref{#1})}
\def\eqns#1#2{eqs.~(\ref{#1}) and~(\ref{#2})}

\title{Next-to-eikonal corrections to soft gluon radiation: a
diagrammatic approach}

\author{Eric Laenen\\
 ITFA, University of Amsterdam,
Science Park 904, 1090 GL Amsterdam, \\
 ITF, Utrecht University, Leuvenlaan 4, 3584 CE Utrecht\\
Nikhef, Science Park, Amsterdam, the Netherlands\\
E-mail: \email{Eric.Laenen@nikhef.nl}}
  
\author{Lorenzo Magnea\\
Dipartimento di Fisica Teorica, Universit\`{a} di Torino, and\\
INFN, Sezione di Torino, Via P. Giuria 1, I-10125 Torino, Italy\\
E-mail: \email{magnea@to.infn.it}}
 
\author{Gerben Stavenga\\
Fermi National Accelerator Laboratory, MS106, P.O. Box 500, 
IL 60510, U.S.A.\\
E-mail: \email{stavenga@fnal.gov}}

\author{Chris D. White\\
Department of Physics and Astronomy, University of Glasgow, 
Glasgow G12 8QQ, Scotland, UK;\\
Institute for Particle Physics Phenomenology, Department of Physics, 
Durham University, Durham DH1 3LE, United Kingdom\\
E-mail: \email{c.white@physics.gla.ac.uk}}

\abstract{
We consider the problem of soft gluon resummation for gauge theory 
amplitudes and cross sections, at next-to-eikonal order, using a 
Feynman diagram approach. At the amplitude level, we prove 
exponentiation for the set of factorizable contributions, and construct
effective Feynman rules which can be used to compute next-to-eikonal emissions directly in the logarithm of the amplitude, finding agreement 
with earlier results obtained using path-integral methods.  For cross
sections, we also consider sub-eikonal corrections to the phase space 
for multiple soft-gluon emissions, which contribute to next-to-eikonal 
logarithms. To clarify the discussion, we examine a class of $\log(1 - x)$ 
terms in the Drell-Yan cross-section up to two loops. Our results are
the first steps towards a systematic generalization of threshold 
resummations to next-to-leading power in the threshold expansion.
}


\preprint{NIKHEF/2010-032, ITP-UU-10/32\\
ITFA-2010-20, DFTT 13/2010\\
FERMILAB-PUB-10-353-T\\
IPPP/10/75, DCPT/10/150}

\begin{document}

\section{Introduction}
\label{introduction}

The eikonal approximation has played an important role throughout
the history of gauge theory scattering amplitude calculations. It 
embodies the universal behavior of such amplitudes in the limit 
where massless gauge boson momenta become soft, and it encodes 
the semiclassical nature of soft, long-wavelength radiation. Early
applications to potential scattering and Regge theory (see for 
example ~\cite{Levy:1969cr,Abarbanel:1969ek,Wallace:1973iu}) 
predate the standard model of particle physics. In the context 
of modern applications to QED and QCD, the eikonal approximation 
is typically phrased in terms of Wilson lines: when radiated 
gauge bosons are soft, one may neglect the recoil of energetic 
particles; these can then be replaced by straight Wilson 
lines along the classical trajectories of the emitters (often called
eikonal lines for this reason); this expresses the fact that soft gauge fields can affect energetic particles only by dressing them with a 
gauge phase. The fact that the eikonal approximation can be recast
in terms of expectation values of (non-local) operators such as 
Wilson lines points to a deep connection between the perturbative 
and non-perturbative aspects of gauge theory scattering, a fact
that was recently highlighted by remarkable results in maximally
superymmetric ${\cal N} = 4$ Yang-Mills  theory (for a review, 
see~\cite{Alday:2008zza}). Recent significant developments in 
the understanding of the all-order structure of infrared divergences
in non-abelian gauge theories~\cite{Gardi:2009qi,Becher:2009qa,Becher:2009kw,Gardi:2009zv} were also derived with 
the help of the special properties of the eikonal approximation.

A crucial feature of the eikonal approximation is the fact that 
scattering amplitudes exponentiate, which means that it is possible 
to establish a set of simplified rules to compute the logarithm 
of the amplitude. As a consequence, low-order perturbative 
calculations can be employed to gain access to all-order 
information, which is of great interest both for theory and for
phenomenology. In QED, for example, a scattering amplitude 
${\cal M}$ dressed with multiple soft photon emission
may be written~\cite{Yennie:1961ad} in the schematic form
\beq
  {\cal M} \, = \, {\cal M}_0 \, \exp \left[ \sum G_c \right] \, ,
  \label{Aexp}
\eeq
where ${\cal M}_0$ is the amplitude without soft photon 
radiation, and the sum in the exponent is over connected 
Feynman diagrams for soft photon emission. Exponentiation was 
later shown to hold also for non-abelian gauge theory 
amplitudes~\cite{Sterman:1981jc,Gatheral:1983cz,Frenkel:1984pz}, 
where however the structure of the exponent is more complicated, due 
to the non-commutativity of  color matrices. The non-abelian analogue 
of \eqn{Aexp} for soft gluon emissions from two energetic (hard) 
partons connected by a color singlet vertex is
\beq 
  {\cal M} \, = \, {\cal M}_0 \, \exp \left[\sum \bar{c}_{W}
  \, W \right] \, ,
  \label{Aexp2}
\eeq
where the sum in the exponent involves Feynman diagrams $W$ 
which are two-particle irreducible with respect to the hard emitting
partons, and are called {\it webs} in the literature. The factors 
$\bar{c}_W$ are modified color factors, which differ from the 
conventional color factors $c_W$ associated with the same 
diagrams through the standard Feynman rules. Recently, \eqn{Aexp2}
was extended~\cite{Gardi:2010rn,Mitov:2010rp} to more general
multiparton amplitudes. In that case webs do not allow for a simple
topological characterization, but it remains true that the logarithm 
of the amplitude can be computed in terms of a subset of Feynman
diagrams with modified color factors.

The exponential structure of the amplitudes is a necessary ingredient
for the resummation of soft gluon effects in many different high-energy
cross sections.  In general, in order to perform a resummation for a
given observable, one must also require that the phase space for
multiple soft gauge boson emission should factorize. If that happens,
potentially large logarithms of ratios of physical scales, which often
jeopardize the applicability of perturbation theory, can be resummed
in exponential form. Soft gluon resummation is well understood, and
has been widely investigated using a variety of techniques~\cite{Sterman:1986aj,Catani:1989ne,Korchemsky:1993uz,Contopanagos:1997nh,Forte:2002ni,Hill:2004if,Becher:2006nr,Gardi:2007ma,Schwartz:2007ib,Bauer:2008dt,Chiu:2009mg}, 
leading to a vast array of phenomenological applications to many 
scattering processes in perturbative QCD. In this connection, it is 
important to recall that some of the large logarithms that arise in 
high-energy cross sections, for example when scattering takes 
place near a partonic threshold, originate from the emission of 
hard collinear gluons, and thus cannot be captured by the soft 
approximation. In what follows we will try to clearly distinguish 
these two effects.

The aim of this paper is to clarify the structure of {\it next-to-eikonal}
(NE) contributions to scattering amplitudes (and eventually cross 
sections). For an amplitude involving $n$ soft gauge bosons with
momenta $k_i$ ($i = 1, \ldots, n$), we imagine rescaling all soft 
momenta by a common factor $\eta$. Then the eikonal approximation 
to the amplitude keeps the leading power in $\eta$ as $\eta \to 0$, 
which is typically $\eta^{- n}$ for an amplitude involving $n$ soft 
gauge bosons. The NE approximation organizes terms proportional 
to the next-to-leading power in this expansion, $\eta^{- (n - 1)}$.
Essentially, one takes $k_i \rightarrow 0$ for all but one gauge
boson, whose momentum is kept to one power beyond the eikonal
approximation. This eikonal expansion is related, but not identical,
to the expansion of partonic cross sections near kinematic thresholds,
which is used to organize Sudakov logarithms. Near partonic thresholds,
one can typically parametrize the distance from threshold in terms
of a single dimensionless variable $\xi$, such that $\xi \to 0$ is the 
threshold limit. Examples include $\xi = W^2/Q^2 \sim (1 - 
x_{\rm Bj}) $ in DIS, $\xi = 1 - Q^2/s$ in Drell-Yan, and $\xi =
1- T$ for, say, the thrust distribution in electron-positron annihilation.
In all these cases, the partonic differential cross section receives contributions of the form
\beq
  \frac{d \sigma}{d \xi} \, = \, \sum_{n = 0}^\infty 
  \sum_{m = 0}^{2 n - 1} \alpha_s^n \left[a_{n m}
  \frac{\log^m(\xi)}{\xi} + b_{n m} \log^m (\xi) +
  {\cal O} (\xi) \right] \, .
\label{soft}
\eeq
We call the expansion in powers of $\xi$ in \eqn{soft} a ``threshold'' 
expansion. In this terminology, NE corrections to scattering amplitudes
start contributing at the level of $b_{n m}$ terms, that is at 
next-to leading order in the threshold expansion (NLT). On the other 
hand, in the presence of final-state jets of hard particles, there are 
hard collinear emissions in the matrix elements which contribute at 
NLT but are not captured by the eikonal expansion: they must be dealt 
with using collinear evolution equations. NLT contributions can be 
phenomenologically important (see for example~\cite{Kramer:1996iq}) 
and are conceptually interesting since they probe the reach of the 
universal  properties of soft and collinear radiation. As a consequence, 
these contributions have been recently studied in some detail: 
Ref.~\cite{Eynck:2003fn} showed how $\xi$-independent terms 
completely exponentiate for simple color-singlet QCD cross-sections
(see also~\cite{Friot:2007fd,Ahrens:2008qu}); subsequently, 
Ref.~\cite{Dokshitzer:2005bf} proposed a modification of collinear 
evolution equations that allows the exponentiation of a subset of NLT 
terms; this was exploited in Ref.~\cite{Laenen:2008ux} to propose 
an ansatz for NLT Sudakov resummation for color-singlet QCD cross 
sections, that was shown to reproduce the bulk of these contributions 
through NNLO. Other 
studies~\cite{Moch:2009hr,Soar:2009yh,Vogt:2010pe,Vogt:2010ik,Grunberg:2009yi,Grunberg:2009vs,Grunberg:2010sw} 
have uncovered intriguing relations between the coefficients 
of NLT contributions in finite order calculations, and have proposed 
all-order generalizations. The ultimate goal of our work, as well as of 
Refs.~\cite{Laenen:2008gt,Laenen:2010kp}, is to reach a complete
understanding of NLT terms and organize them in resummed form
to the extent to which this is possible. For cross sections, this will
also require an understanding of multi-gluon phase space beyond
the eikonal approximation, and some progress in this direction is
described already in this paper.

As a first step in this program, we intend to establish to what 
extent the exponentiation properties of the eikonal approximation 
extend to NE level, and eventually to derive effective Feynman 
rules and iterative methods to compute directly the perturbative 
exponent to the required accuracy. Our results will reproduce 
and clarify from a diagrammatic viewpoint those obtained 
in~\cite{Laenen:2008gt}, where the problem of NE contributions 
was first considered using path integral methods. 
In that paper, a factorised form was assumed for Green functions 
with a fixed number of hard outgoing particles, which may emit any 
number of soft gluons. By recasting the propagators for the external
particles (in the background of a soft gauge field) in terms of
first-quantised path integrals, an effective field theory for the soft
gauge field arises, with source vertices localised along the external
lines. Within that formalism, exponentiation of soft photon 
contributions (in an abelian field theory) was shown to be
straightforwardly related to the exponentiation of connected 
diagrams in quantum field theory. In the non-abelian case, the 
field theory for the soft gauge field was more complicated, due to 
the fact that the source terms are matrix-valued in color space and, 
thus, non-commuting. It was, however, possible to ascertain that a
subset of diagrams (indeed the webs of~\cite{Frenkel:1984pz}) 
does exponentiate. The exponentiation derived in~\cite{Laenen:2008gt}
leaves out those contributions to the matrix elements that violate
the assumed factorization properties of the correlators. We similarly observe that our analysis applies to both real and virtual gluons, 
provided all emissions can be considered soft with respect to hard 
scale of the problem; in the case of hard virtual gluons, however,
the known factorization properties of soft radiation do not fully 
generalize to NE level, and there are further contributions arising 
from soft gluons emitted by lines internal to the hard subamplitude. 
Here we will concentrate on the structure of factorizable contributions,
which do exponentiate.

To be specific, the diagrammatic techniques developed in 
the present paper, in agreement with the final result 
of~\cite{Laenen:2008gt}, organize next-to-eikonal contributions 
to matrix elements in the schematic form 
\beq 
  {\cal M} \, = \, {\cal M}_0 \, \exp \left[{\cal M}_{\rm eik} 
  + {\cal M}_{\NE} \right] \, (1 + {\cal M}_r) + 
  {\cal O} \left( {\rm NNE} \right) \, .
  \label{Mstruc}
\eeq
Here ${\cal M}_0$ is the Born contribution, and ${\cal M}_{{\rm eik},
\, \NE}$ collect factorizable contributions due to emissions of soft gluons external to the hard interaction. Non-factorizable contributions,
arising from internal emission graphs, are collected in the function 
${\cal M}_r$. This contribution does not formally exponentiate, but 
has an iterative structure to all orders in perturbation theory. At NE 
level, as suggested in~\cite{Laenen:2008gt}, non-factorizable
contributions can be organized by means of the Low-Burnett-Kroll 
theorem~\cite{Low:1958sn,Burnett:1967km}, as extended 
in~\cite{DelDuca:1990gz} to encompass collinear divergences. 
In the present paper, we neglect non-factorizable contrbutions
contained in ${\cal M}_r$ and concentrate on the structure of
the exponent in \eqn{Mstruc} at NE level. Note that 
Ref.~\cite{Laenen:2008gt} addressed explicitly the case of two 
external hard partons connected by a color-singlet interaction.
The structure of the exponent in the case of multi-parton amplitudes 
was recently examined in~\cite{Gardi:2010rn,Mitov:2010rp}, including 
a first analysis  of next-to-eikonal corrections using path integral 
methods in~\cite{Gardi:2010rn}.

We emphasize that the diagrammatic analysis presented in this 
paper is a necessary step, not only in order to test the results
derived with path-integral methods in~\cite{Laenen:2008gt}, but
also in order to place these results in the context of general proofs
of factorization theorems, and to clarify the precise limits of
applicability of the formalism. As an example, the factorization 
arguments organizing Low's theorem contributions at NE level in
Ref.~\cite{DelDuca:1990gz} are formulated in diagrammatic
language, and a proper matching with factorizable contributions
in order to avoid double counting must also employ diagrammatic
techniques. Similarly, possible extensions of this formalism to
collinear contributions at next-to-leading order in the threshold 
expansion will need to be mapped to existing diagrammatic 
analyses. Our results thus strengthen the validity of the conclusions
reached by the path integral method, and place them in the more
general context of factorization studies with diagrammatic methods.

In order to test our results and to illustrate and clarify our technique
in a concrete application, in the last part of this paper we consider
a subset of the real-emission contributions to the Drell-Yan cross
section up to NNLO. Processes involving parton annihilation into an 
electroweak final state, such as Drell-Yan or weak boson production,
or Higgs production in the gluon fusion channel, are ideal testing 
grounds for our formalism: indeed, near threshold all radiation in
these processes is soft, so that one does not need to worry about 
hard collinear emissions. Furthermore, by considering real emission
diagrams only, we don't need to include non-factorizable contributions
related to Low's theorem, and therefore we should be able to reproduce
exactly the fixed-order results upon applying our effective Feynman 
rules at the level of the exponent. We perform this test on abelian-like
contributions (proportional to $C_F^n$ in QCD) up to two loops
(the first non-trivial order where exponentiation has an impact), finding
the expected result. Applying our formalism to a hadronic cross 
section forces us to tackle the issue of the factorizability of soft 
gluon phase space, and we present some preliminary results suggesting
that the factorization properties of the eikonal phase space may
admit a simple generalization at NE level.

The structure of the paper is as follows. In \sect{eikonal} we review 
the derivation of the exponentiation of soft corrections in the abelian 
and non-abelian cases, using an iterative Feynman diagram method. 
In \sect{NErules} we derive effective Feynman rules for NE emissions. 
Crucial to this result is the fact that the sum over all NE diagrams has a
factorisable form. In \sect{sec:exp} we then demonstrate the
exponentiation of NE corrections, extending the methods used in 
\sect{eikonal}. In \sect{NErulespath} we perform a detailed comparison 
of the rules thus obtained with the results found using the path integral 
method of~\cite{Laenen:2008gt}. Finally, in \sect{sec:DY} we present
the application of the ideas derived in this paper to the case of Drell-Yan 
production.  In \sect{sec:conclusion} we briefly summarize our results,
while some technical aspects are discussed in the Appendices: notably,
Appendix B discusses the structure of phase space for multiple soft 
gluon emission at arbitrary order, providing arguments for a form of
factorization that could be applied at NE level.

\section{Introduction to eikonal exponentiation}
\label{eikonal}

Exponentiation of soft gauge boson corrections at the eikonal 
level was first demonstrated for abelian gauge theories 
in~\cite{Yennie:1961ad}, and later generalised to non-abelian 
theories in~\cite{Sterman:1981jc,Gatheral:1983cz,Frenkel:1984pz}. 
Here we review the derivation of this result using Feynman diagram 
methods (see also~\cite{Berger:2003zh} for a pedagogical 
exposition), both in order to make our paper reasonably 
self-contained and to introduce methods and notations that 
will prove useful when generalising the results to NE order.

The proof of exponentiation is two-fold. First, one establishes 
a set of effective Feynman rules in the eikonal approximation,
showing that they capture the leading-power behavior of the 
amplitude as soft momenta become vanishingly small, to all 
orders in perturbation theory; then one uses these effective 
Feynman rules to classify a subset of Feynman diagrams that
generate the full eikonal amplitude upon exponentiation. 
We will do this first for the simple case of an abelian theory,
and then discuss the non-trivial modifications that are necessary
in the nonabelian case.

\subsection{Abelian eikonal exponentiation}
\label{sec:abelianexp}

\begin{figure}
  \begin{center}
  \includegraphics[scale=0.6]{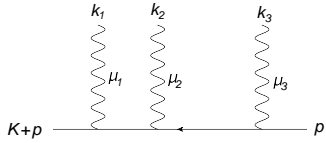}
  \caption{Soft photon emission from an energetic line.} 
  \label{fig_basic_line}
  \end{center}
\end{figure}

In order to derive the eikonal Feynman rules it is sufficient to 
consider a single hard massless external line of final on-shell 
momentum $p$, originating from some unspecified hard interaction
described by a matrix element ${\cal M}_0(p)$. The hard line 
may emit a number $n$ of soft photons with momenta $k_i$, as 
depicted in \fig{fig_basic_line}, where we choose the ordering so 
that momentum $k_1$ is emitted closest to the hard interaction.
If the emitting particle is a Lorentz scalar, such an external line
dresses the hard interaction according to
\beq
\label{eq_spin0_string}
  {\cal M}^{\mu_1 \ldots \mu_n} (p, k_i)  = {\cal M}_0 (p) \,
  \frac1{(p + K_1)^2}(2 p + K_1 + K_2)^{\mu_1} \ldots
  \frac1{(p + K_n)^2}(2 p + K_n)^{\mu_n} \, ,
\eeq
where we have introduced the partial momentum sums 
$K_i = \sum_{m = i}^n k_m$.

The eikonal approximation in this case can simply be defined as
the leading-power contribution to the amplitude when the photon
momenta $k_i^{\mu_i} \to 0$, $\forall i$. In this limit,
\eqn{eq_spin0_string} becomes
\beq
  {\cal M}^{\mu_1 \ldots \mu_n} (p, k_i)  = {\cal M}_0 (p) \,
  \frac{p^{\mu_1} \ldots p^{\mu_n}}{(p \cdot K_1) 
  \ldots  (p \cdot K_n)} \, .
\label{eq_spin0_most}
\eeq
If the emitter is a Dirac fermion, \eqn{eq_spin0_string} becomes
\beq
  {\cal M}^{\mu_1 \ldots \mu_n} (p, k_i)  = {\cal M}_0 (p) \,
  \frac{\slashed p + \slashed K_1}{(p + K_1)^2} 
  \gamma^{\mu_1} \ldots \frac{\slashed p + 
  \slashed K_n}{(p + K_n)^2} \gamma^{\mu_n} u(p) \, ,
\label{spin1/2mat}
\eeq
where $u(p)$ is the spinor associated with the final state on-shell 
particle. At leading power in the soft momenta, this reduces to
\beq
  {\cal M}^{\mu_1 \ldots \mu_n} (p, k_i)  = {\cal M}_0 (p) \,
  \frac{\slashed p}{2 p \cdot K_1} \gamma^{\mu_1} \ldots 
  \frac{\slashed p}{2 p \cdot K_n} \gamma^{\mu_n} u(p).
\label{spin1/2eik}
\end{equation}
This appears to be more complicated than the scalar case due 
to the non-trivial spinor structure. One may however use the 
anticommutation properties of Dirac matrices and the Dirac 
equation to reduce \eqn{spin1/2eik} precisely to the form of 
\eqn{eq_spin0_most}, with the spinor $u(p)$ reabsorbed
into the radiationless matrix element ${\cal M}_0(p)$.
The eikonal factor is the same in both cases, displaying the 
well-known result that the eikonal approximation is insensitive 
to the spin of the emitting particles.  One may also notice that 
the eikonal factor does not depend on the energy of the
emitter, since it is invariant under rescalings of the hard 
momentum $p^\mu$: at leading power in the soft momenta,
one is effectively neglecting the recoil of the hard particle 
against soft radiation.

The eikonal factor can be further simplified by employing Bose
symmetry. Indeed, in constructing any physical quantity 
depending on the amplitude ${\cal M}^{\mu_1 \ldots \mu_n} 
(p, k_i)$, one will need to sum over all diagrams corresponding
to permutations of the emitted photons along the hard line. 
Having done this, the eikonal factor multiplying ${\cal M}_0 
(p)$ on the {\it r.h.s.} of \eqn{eq_spin0_most} will be replaced 
by the symmetrized expression
\beq
  E^{\mu_1 \ldots \mu_n} (p, k_i) \, = \, \frac{1}{n!} \, 
  p^{\mu_1} \ldots p^{\mu_n} \, \sum_\pi
  \frac{1}{p \cdot k_{\pi_1}} \frac{1}{p \cdot (k_{\pi_1} + 
  k_{\pi_2})} \, \ldots \, \frac{1}{p \cdot(k_{\pi_1} + 
  \ldots + k_{\pi_n})} \, ,
\label{eikfact2}
\eeq
where the sum is over all permutations of the photon momenta, 
and $k_{\pi_i}$ is the $i^\text{th}$ momentum in a given 
permutation. There are $n!$ permutations, and each gives the 
same contribution to any physical observable. This becomes 
manifest using the {\it eikonal identity}
\beq
  \sum_\pi \frac{1}{p \cdot k_{\pi_1}} \frac{1}{p \cdot 
  (k_{\pi_1} + k_{\pi_2})} \ldots \frac{1}{p \cdot (k_{\pi_1}
  + \ldots k_{\pi_n})} = \prod_i \frac{1}{p \cdot k_i} \, .
\label{eikonalid}
\eeq
Using \eqn{eikonalid}, the eikonal factor $E^{\mu_1 \ldots 
\mu_n} (p, k_i)$ arising from $n$ soft emissions on an external 
hard line becomes simply
\beq
  E^{\mu_1 \ldots \mu_n} (p, k_i) \, = \, \prod_i \, 
  \frac{p^{\mu_i}}{p\cdot k_i} \, ,
\label{eikfactfin}
\eeq
which is manifestly Bose symmetric and invariant under rescalings
of the momenta $\{p_i\}$.
In practice, each eikonal emission can be expressed by the 
effective Feynman rule
\beq
\label{eikrule}
  \includegraphics[width=2cm]{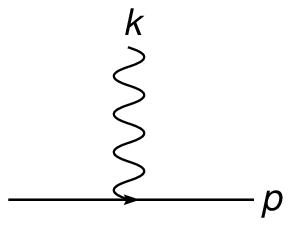}   \qquad
  {\raisebox{3ex}{\ensuremath{{\, = \, \, \,
  \Large\frac{p^\mu}{p \cdot k}}}}}
\eeq
As is well known, these Feynman rules can be obtained by 
replacing the hard external line with a Wilson line along the 
classical trajectory of the charged particle. In abelian quantum 
field theories this is given by the operator
\beq
  \Phi_\beta (0, \infty) =
  \exp \left[\, {\rm i} e \int_{0}^{\infty} d \lambda \,
  \beta \cdot A (\lambda \beta) \, \right]~,
\eeq
where $\beta$ is the dimensionless four-velocity corresponding 
to the momentum $p$. This expresses the fact that soft emissions 
affect the hard particle only by dressing it with a gauge phase.

\begin{figure}
  \begin{center}
  \includegraphics[scale=0.8]{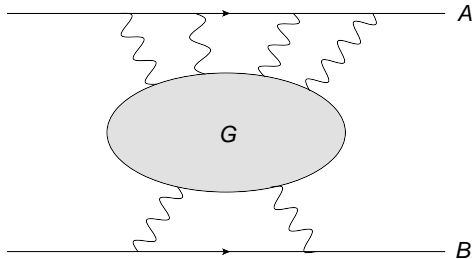}
  \caption{A process involving two eikonal lines A and B, interacting
   through the exchange of soft gluons forming diagram 
   $G$.}\label{fig_eik_exp}
  \end{center}
\end{figure}
Having constructed the effective Feynman rules, one may proceed 
to demonstrate the exponentiation of soft photon corrections
as follows. As an example, we consider graphs of the form shown 
in \fig{fig_eik_exp}, at a fixed order in the perturbative expansion.
\Fig{fig_eik_exp} consists of two eikonal lines, labelled $A$ and 
$B$, each of which emits a number of soft photons. Notice that
although we consider only two eikonal lines here, the analysis 
below generalises to any number of hard charged particles.
One may envisage lines $A$ and $B$ as emerging from a hard 
interaction, and one may consider the graph $G$ either as a 
contribution to an amplitude, or to a squared amplitude (in which 
case some of the propagators in $G$ will be cut). 

Diagram $G$ can be taken as consisting only of soft photons 
and fermion loops: in fact, at leading power in the soft momenta, 
soft photons originating from the hard scattering give no 
contribution (a statement which will have to be revisited when 
including NE corrections). Photons originating from one of the 
two eikonal lines must land on the other one, or on a fermion loop 
inside $G$. Indeed, a photon cannot land on the same eikonal line, 
as in that case the diagram is proportional to $p^\mu p_\mu = 0$. 

Using eikonal Feynman rules, one finds that graphs of the form of
\fig{fig_eik_exp} contribute to the corresponding (squared) 
amplitude a factor
\beq
  {\cal F}_{AB} \, = \, \sum_G \, \left[ \prod_i 
  \frac{p_A^{\mu_i}}{p_A \cdot k_i} \right]
  \left[\prod_j \frac{p_B^{\nu_j}}{p_B \cdot l_j} \right]
  G_{\mu_1 \ldots \mu_n ; \nu_1\ldots \nu_m} (k_i, l_j) \, ,
\label{amp3}
\eeq
where $k_i$, $l_j$ are the momenta of the photons emitted 
from lines $A$ and $B$ respectively, with $i = 1, \ldots, n$ 
and $j = 1, \ldots,  m$. 

\begin{figure}
  \begin{center}
  \scalebox{0.6}{\includegraphics{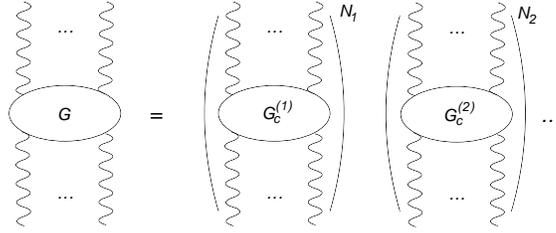}}
  \caption{Decomposition of a soft photon graph into 
  connected subdiagrams $G_c^{(i)}$, each of which 
  occurs $N_i$ times.}
  \label{decompcon}
  \end{center}
\end{figure}

Given that we have already summed over permutations in order 
to obtain the eikonal Feynman rules, each diagram $G$ can be 
uniquely specified by the set of connected subdiagrams it 
contains, as indicated schematically in \fig{decompcon}, where 
each possible connected subdiagram $G_c^{(i)}$ occurs $N_i$ 
times. According to the standard rules of perturbation theory, 
diagram $G$ has a symmetry factor corresponding to the number 
of permutations of internal lines which leave the diagram invariant. 
This symmetry factor is given by
\beq
  S_G = \prod_i \, S_i^{N_i} \, (N_i)! \, ,
\label{sym}
\eeq
where $S_i$ is the symmetry factor associated with each 
connected subdiagram $G_c^{(i)}$, and the factorials account 
for permutations of identical connected subdiagrams, which 
must be divided out. Contracting Lorentz indices as in 
\eqn{amp3}, the eikonal factor ${\cal F}_{AB}$ may be 
written as
\beq
  {\cal F}_{AB} \, = \, \sum_{\{N_i\}}\, \prod_i \, 
  \frac{1}{N_i!} \, \left[ {\cal F}_c^{(i)} \right]^{N_i} \, ,
\label{amp4}
\eeq
where 
\beq
  {\cal F}_c^{(i)} \, = \, \frac{1}{S_i} \, \left( \prod_q 
  \frac{p_A^{\mu_q}}{p_A \cdot k_q} \right)
  \left(\prod_r \frac{p_B^{\nu_r}}{p_B \cdot l_r} \right)
  G^{(i)}_{\mu_1 \ldots \mu_{n_q} ; \nu_1\ldots \nu_{m_r}} 
  (k_q, l_r) 
\label{Gcdef}
\eeq
is the expression for each connected subdiagram, including the 
appropriate symmetry factor. Recognising \eqn{amp4}
as an exponential series, it follows that
\beq
  {\cal F}_{AB} \, = \, \exp\left[\sum_i {\cal F}_c^{(i)} \right] \, .
\label{ampexp}
\end{equation}
We conclude that soft photon corrections exponentiate in the 
eikonal approximation, and the exponent is given by the sum of 
all connected subdiagrams.

The same result was recently rederived in~\cite{Laenen:2008gt}, 
using path integral methods to derive  soft photon exponentiation
from the well-understood exponentiation of disconnected diagrams 
in quantum field theory.

\subsection{Non-abelian eikonal exponentiation}
\label{nonabelianexp}

We now consider the generalisation of eikonal exponentiation 
to non-abelian theories. The results are well 
known~\cite{Gatheral:1983cz,Frenkel:1984pz}, but we will
introduce methods and notation that will prove useful in the 
extension to NE order. We will examine the case of two 
incoming or outgoing hard emitting particles connected by a 
color-singlet hard interaction, as happens for example in 
Drell-Yan production, deep inelastic scattering or $e^+e^-$ 
annihilation.  We note that the results below extend to 
multiple hard colored emitters, as recently shown in 
Refs.~\cite{Gardi:2010rn,Mitov:2010rp}.

The proof of the abelian result relies crucially on the application 
of the eikonal identity after summing over the permutations of all 
photon momenta on each eikonal line. In the nonabelian case, this 
identity cannot be used, due to the presence of non-commutative 
color matrices associated with each emission. Exponentiation,
however, is still possible, but with a somewhat more complicated
structure.

\begin{figure}
  \begin{center}
  \scalebox{0.7}{\includegraphics{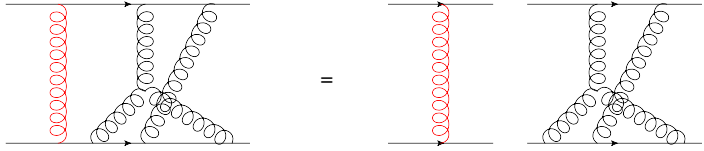}}
  \caption{Decomposition of a higher order non-Abelian diagram in 
  terms of webs.}
  \label{webdecomp}
  \end{center}
\end{figure}

First we introduce the concepts of {\it webs} and {\it groups}. 
A {\it web} is a two-eikonal irreducible diagram, {\it i.e.} a 
diagram that cannot be disconnected by cutting the two eikonal 
lines\footnote{As shown in Refs.~\cite{Gardi:2010rn,Mitov:2010rp},
this simple topological identification of webs must be suitably
modified when more than two eikonal lines are present.}. 
Higher order diagrams can be rewritten as sums of products of 
webs: an example is illustrated in \fig{webdecomp}, which shows 
a particular fourth order diagram that is not a web, and its 
subsequent decomposition into webs.

\begin{figure}
  \begin{center}
  \scalebox{0.7}{\includegraphics{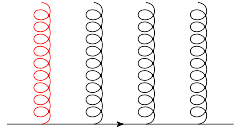}}
  \caption{Gluon groups obtained from the lower eikonal line of 
  \fig{webdecomp}, with the same color coding as in 
  that figure.}
  \label{group}
  \end{center}
\end{figure}

A {\it group} is the projection of a web onto a single eikonal line:
therefore, gluons emitted from an eikonal line belong to the same 
group if they belong to the same web. This is illustrated in 
\fig{group}, where we depict the groups that result on the lower 
eikonal line from the webs of \fig{webdecomp}.

Having introduced the concept of a group, one observes a useful 
generalisation of the eikonal identity, \eqn{eikonalid}, to sums
over permutations that do not affect the ordering of gluons within groups~\cite{Frenkel:1984pz,Berger:2003zh}. Let $\tilde{\pi}$ 
be a permutation of the gluons (with momenta $\{k_i\}$) emitted
by a given eikonal line of momentum $p$, with the restriction that 
the ordering of the gluons in each group $g$ be held fixed. Then 
one may write
\begin{align}
  \sum_{\tilde{\pi}} \, \frac{1}{p \cdot k_{\tilde{\pi}_1}}
  \frac{1}{p \cdot (k_{\tilde{\pi}_1} + k_{\tilde{\pi}_2})} \, 
   \ldots \, \frac{1}{p \cdot(k_{\tilde{\pi}_1} + \ldots + 
   k_{\tilde{\pi}_n})} & \, = \, \prod_{{\rm groups} \, \, g} \, \, 
   \frac{1}{p \cdot k_{g_1}} \frac{1}{p \cdot (k_{g_1} + 
   k_{g_2})} \notag \\
   & \qquad \ldots \, \frac{1}{p \cdot(k_{g_1} + \ldots + 
   k_{g_m})} \, .
\label{eikid2}
\end{align}
Here $k_{\tilde{\pi}_i}$ and $k_{g_i}$ are the momenta of the 
$i^\text{th}$ gluon in permutation $\tilde{\pi}$ and group $g$, 
respectively, where $g$ contains, say, $m$ gluons. As an example 
of this result, consider an eikonal line with 3 gluon emissions, where 
gluons 1 and 2 belong to one group, and gluon 3 to another one;
then, \eqn{eikid2} amounts to the statement
\begin{align}
  &\frac{1}{p \cdot k_1} \frac{1}{p \cdot(k_1 + k_2)}
  \frac{1}{p \cdot  (k_1 + k_2 + k_3)} \, + \, 
  \frac{1}{p \cdot k_1} \frac{1}{p \cdot(k_1 + k_3)} 
  \frac{1}{p \cdot(k_1 + k_3 + k_2)}  
  \notag\\
  & \qquad + \, \frac{1}{p \cdot k_3} \frac{1}{p\cdot(k_3 + 
  k_1)} \frac{1}{p \cdot(k_3 + k_1 + k_2)} \, = \,
  \frac{1}{p\cdot k_3} \, \left( \frac{1}{p \cdot k_2} 
  \frac{1}{p \cdot(k_1 + k_2)} 
  \right) \, .
\label{eikid2ex}
\end{align}
The RHS displays the factorization of gluons from different groups. 
Henceforth, unless otherwise stated, all permutations of gluon 
momenta involve fixed orderings of the gluons inside each group;
thus we drop the tilde on the permutation symbols $\tilde{\pi}$ 
for brevity. 

\begin{figure}
  \begin{center}
  \begin{equation*}
  \begin{array}{ccccc}
  \includegraphics{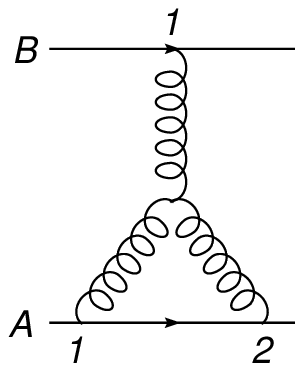}& 
  \bigcup_{132}^{12}&\includegraphics{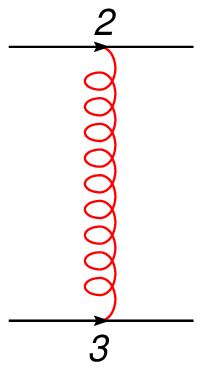} & = & 
  \includegraphics{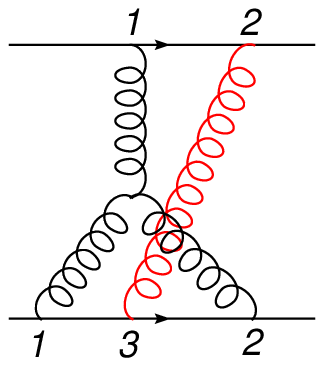}
  \end{array}
  \end{equation*}
  \caption{Illustration of the merging product 
  $\cup_{\pi_A}^{\pi_B}$ between two eikonal graphs, 
  where $\pi_A$ and $\pi_B$ are permutations of the 
  gluons on eikonal lines $A$ and $B$ such that ordering 
  within each group is held fixed.}
  \label{merge}
  \end{center}
\end{figure}

Next we examine contractions of soft gluons emanating from two
emitting lines connected by a color singlet hard interaction. A 
given diagram $G$ connecting the external lines then has two color 
indices in some representation of the non-abelian gauge group, one
index for each eikonal line. Color conservation forces the color 
structure of each diagram to be proportional to the identity matrix 
$\delta_{ij}$, where $i$ and $j$ are indices in the chosen 
representation. Denote by $\E(G)$ the eikonal amplitude arising 
from diagram $G$, not including color matrices. $\E(G)$ encodes 
the momentum information carried by a particular diagram, but not 
its color structure: we can then write the factor contributed by $G$ 
to the (squared) matrix element as
\beq
  {\cal F}_G \, = \, c_G \, \delta_{ij}\, \E(G) \, ,
\label{eiksubamp}
\eeq
where $c_G$ is the color factor associated with $G$, and for 
simplicity we are not displaying the internal Lorentz structure. 
This notation allows us to rewrite \eqn{eikid2} in a more formal 
and useful way. Consider two soft gluon diagrams $H_1$ and 
$H_2$, connecting the same two eikonal lines $A$ and $B$. One 
may define the product of these two eikonal diagrams, using 
\eqn{eikid2} in reverse, as
\beq
  \E (H_1) \, \E(H_2) \, = \, \sum_{\pi_A} \, \sum_{\pi_B} \,
  \E(H_1 \cup_{\pi_A}^{\pi_B} H_2) \, ,
\label{eikprod}
\eeq
where $H_1\cup_{\pi_A}^{\pi_B}H_2$ denotes the particular 
merging that results from combining $H_1$ and $H_2$ so that 
the gluon permutations on the eikonal lines are given by $\pi_A$ 
and $\pi_B$ respectively\footnote{This product endows the set 
of eikonal gluon amplitudes with the structure of a shuffle algebra
(for a definition, see for example Ref.~\cite{Weinzierl:2010ps}).}. 
Note that there may be a different number of gluons on each 
external line. An example of this is shown in \fig{merge}.

Each possible merging of $H_1$ and $H_2$ gives some diagram 
$G$. The same diagram $G$ could however result from different 
mergings that arise from different permutations $\pi_A$, $\pi_B$. 
One may then rewrite \eqn{eikprod} more formally as
\beq
  \E(H_1) \, \E(H_2) \, = \, \sum_{G} \, \E(G) 
  \, N_{G | H_1H_2} \, ,
\label{prod2}
\eeq
where the multiplicity $N_{G | H_1H_2}$ denotes the number of 
ways in which diagram $G$ can be generated from mergings of 
$H_1$ and $H_2$. \Eqn{prod2} generalises immediately to the 
product of any number of eikonal subdiagrams, as
\beq
  \E(H_1)^{s_1} \, \E(H_2)^{s_2} \, \ldots \, \E(H_n)^{s_n}
  \, = \, \sum_{G} \, \E(G) \, N_{G | H_1^{s_1} H_2^{s_2}
  \ldots H_n^{s_n}} \, ,
\label{prod3}
\eeq
where each diagram $H_i$ occurs $s_i$ times in the product. 

With these notations in hand, we are now ready to state the 
exponentiation theorem for nonabelian eikonal diagrams. Consider
the exponential
\beq
  \exp \left\{ \sum_{H} \, \bar{c}_{H} \, \E(H)\right\} \, ,
\label{eikexp1}
\eeq
where the sum is over all diagrams $H$, each with an accompanying
color factor $\bar{c}_H$ whose interpretation will become clear in
what follows (the bar is used to distinguish these color factors from 
those arising from conventional Feynman rules, $\{c_H\}$). Note 
that each $H$ may be decomposable in terms of products of smaller
subdiagrams. Using \eqn{prod3}, one may then write
\beq
  \exp \left\{ \sum_{H} \, \bar{c}_{H} \, \E(H)\right\}
  \, = \, \prod_H \left( \sum_n \frac{1}{n!} \left[ \bar{c}_H \, 
  \E(H) \right]^n \right) \, = \, \sum_G \, c_G \, \E(G) \, ,
\label{eikexp2}
\eeq
where the sum on the right-hand side is again over all possible
diagrams $G$ ({\it i.e.} we use a different label on the left- and 
right-hand sides), and $\{c_G\}$ are constants. Expanding the 
exponential, one generates all possible products of diagrams. 
Each such product is itself equal to a linear combination of diagrams, 
by \eqn{prod3}, and thus contributes to various terms on the 
right-hand side of \eqn{eikexp2}. Each diagram $H$ in the 
exponent thus appears on the right-hand side of \eqn{eikexp2} 
in two ways: either as itself ({\it i.e.} contributing only to the 
term where $G$ is equal to $H$), or as a component of larger
diagrams $G$. One may then equate coefficients on the left- 
and right- hand sides of \eqn{eikexp2}, so that the choice of 
constants $\{c_G\}$ uniquely fixes, at least in principle, the 
constants $\{\bar{c}_H\}$. Our choice is to require that the 
constants $\{c_G\}$ on the right-hand side of \eqn{eikexp2}
be the usual color factors of perturbation theory. We can then
show that the constants $\{\bar{c}_H\}$ are zero except for 
a subset of diagrams $H$ which have the property of being 
two-eikonal-line irreducible. These are the webs 
of~\cite{Gatheral:1983cz,Frenkel:1984pz}, that were 
referred to at the beginning of this section. 

\begin{figure}
\begin{equation*}
 \raisebox{-0.04\textheight}
 {\includegraphics{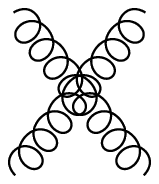}} \rightarrow \left\{ \left\{ 
 \raisebox{-0.04\textheight}
 {\includegraphics{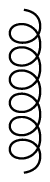}},
 \raisebox{-0.04\textheight}
 {\includegraphics{diagramI}} \right\} \, ,
 \left\{
 \raisebox{-0.04\textheight}{\includegraphics{diagramX}}
 \right\} \right\}
\end{equation*}
\caption{A diagram and its set of decompositions. The first decomposition has $m_{{\bf I}} = 2$ and $m_{\bf X} = 0$, 
the second decomposition has $m_{{\bf I}} = 0$ and 
$m_{\bf X} = 1$.}
\label{decompex}
\end{figure}

One proceeds as follows. First one notes, as we just discussed, 
that each diagram $G$ has a set of decompositions into 
subdiagrams $H$. This includes the case where $G$ cannot be 
decomposed into smaller parts, and thus has only the trivial 
decomposition where $H$ is equal to $G$ itself. An example
of how a simple diagram can be decomposed is shown in 
\fig{decompex}. Each decomposition can be uniquely labelled by 
a set of integers $\{m_H\}$ that specify how many times each
subdiagram $H$ occurs in the decomposition of $G$. In the simple
example shown in \fig{decompex}, there are two decompositions 
labelled by $\{ m_{\bf I},  m_{\bf X} \} = \{2,0\}$ and $\{ 0, 1\}$,
where the notation $m_{{\bf I, X}}$ is suggested by the 
figure. To make use of this decomposition, we note that one 
can rearrange the expansion of any product of exponentials, 
such as the one appearing in \eqn{eikexp2}, as
\beq
  \prod_H \left( \sum_n \frac{1}{n!} \, \big[ \bar{c}_H \, 
  \E(H) \big]^n \right) = \sum_{\{p_H\}} \prod_H 
  \frac{1}{p_H!} \, \big[ \bar{c}_H \, \E(H) \big]^{p_H} \, .
\label{eikexp3}
\eeq
In the present context, each dummy index $p_H$ can be 
interpreted precisely as the multiplicity $m_H$ of subdiagram
$H$ in graph $G$.

We can now exploit \eqn{prod3} to rewrite each product of 
subgraphs in \eqn{eikexp3} as a single sum over graphs. 
\Eqn{eikexp3} then becomes 
\beq
  \prod_H \left( \sum_n \frac{1}{n!} \, \big[ \bar{c}_H \, 
  \E(H)]^n \right) \, = \, \sum_G \, \E(G)\, 
  \sum_{ \{ m_H \} } N_{G | \{ m_H \} }
  \left( \prod_H \frac{1}{m_H !} \, \bar{c}_H^{\, m_H} \right) \, ,
\label{eikexp4}
\eeq
where we denote by $N_{ G | \{ m_H \} }$ the number of ways in
which diagram $G$ can be formed out of the given decomposition 
specified by  $\{m_H\}$. The expression now has the form of a 
single sum over all diagrams $G$ (weighted by non-trivial 
coefficients), and thus can be matched to the right-hand side of 
\eqn{eikexp2}. One finds
\beq
  c_G = \sum_{ \{ m_H \} } N_{G | \{ m_H \} } \left( \prod_H
  \frac{1}{m_H !} \,\bar{c}_H^{\, m_H} \right) \, .
\label{colfacts}
\eeq
This equation relates the coefficients $\{ \bar{c}_H \}$ to 
the color factors $\{ c_G \}$. A similar relation was given 
in~\cite{Berger:2003zh}, and an explicit solution (in which the 
$\{ \bar{c}_H \}$ are given in terms of the $\{ c_G \}$) 
was derived in~\cite{Laenen:2008gt}. One may interpret the 
coefficients $\{ \bar{c}_H \}$ as modified color factors for the 
diagrams appearing in the exponent of \eqn{eikexp2}. At 
present this equation appears to contain no information, as 
the full set of subdiagrams appears in both the exponent and 
on the right-hand side. Crucially however, the modified color 
factors are zero except for a subclass of diagrams that are 
two-eikonal-line irreducible, as we now show.

The proof proceeds inductively. The first step is to separate
from the sum over decompositions $\{ m_H \}$ on the right-hand 
side of \eqn{colfacts} the term involving the trivial decomposition. 
Those remaining involve proper decompositions, where $G$ 
genuinely reduces to a product of smaller subdiagrams. This
leads to writing \eqn{colfacts} as
\beq
  c_G = \bar{c}_G + \sum_{ \{ m'_H \} } N_{ G | \{ m'_H \} }
  \left( \prod_H \frac{1}{m'_H !} \, \bar{c}_H^{\, m'_H} \right) \, ,
\label{colfacts2}
\eeq
where the prime denotes a proper decomposition, and the first term 
comes from the trivial decomposition. For example, for the case depicted in fig.~\ref{decompex}, the proper
decomposition is only the leftmost one ($\{ m'_{\bf I} \} = 2$). Since
$c_{\bf X} = C_F^2 - C_F C_A/2$,  one finds  $\bar{c}_{\bf X} = - C_F C_A/2$.
One now assumes that the fact
that the modified color factors are zero for two-eikonal-line reducible 
diagrams has already been shown up to some given order, so that 
lower-order diagrams appearing on the right-hand side can be taken 
to have this property. This will then be used to show that two-eikonal-line irreducibility persists to higher orders.

\begin{figure}
  \begin{center}
  \scalebox{0.8}{\includegraphics{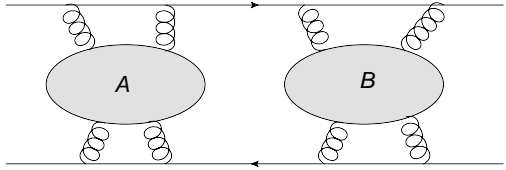}}
  \caption{General form of a diagram which is two-eikonal-line 
  reducible.}
\label{2eikred}
\end{center}
\end{figure}

Consider a general two-eikonal-line reducible diagram, shown in 
\fig{2eikred}.
The normal color factor for such a diagram is given by $c_{A B} = c_A \, c_B$.
The sum over proper decompositions for 
diagram $AB$ has the form given in \eqn{colfacts2},
\beq
  c_{A B} = c_A \, c_B = 
  \bar{c}_{A B} + \sum_{ \{ m'_H \} } N_{A B |
  \{ m'_H \} } \left( \prod_H \frac{1}{m'_H !} \, 
  \bar{c}_H^{\, m'_H} \right) \, ,
\label{multAB}
\eeq
where each subdiagram $H$ is now two-eikonal-line irreducible by
the induction hypothesis. This allows the multiplicity factor 
$N_{AB | \{ m'_H \}}$ to be factorised. Each proper 
decomposition $\{ m'_H \}$ can be split into two parts denoted 
by $\{ m_H^A \}$ and $\{ m_H^B \}$, where the subdiagrams in 
each case contribute solely to $A$ or $B$ respectively. That no 
subdiagram $H$ contributes to both $A$ and $B$ follows from 
the two-eikonal-line irreducibility of the former. For each 
subdiagram  $H$, one then has
\beq
  m'_H = m_H^A + m_H^B \, ,
\label{mhpsum}
\eeq
expressing the fact that the number of occurences of $H$ in 
diagram $AB$ is the sum of the number of occurrences of $H$ 
in $A$ ($m_H^A$) and the number of occurrences of $H$ in 
$B$ ($m_H^B$). The number of ways in which $H$ can occur 
in $AB$ is then given by
\beq
  \text{\bf{C}} (m_H^A, m_H^B) = \frac{m'_H !}{(m_H^{A}) ! 
  \, (m_H^{B}) ! } \, ,
\label{combfun}
\eeq
which gives the number of ways of choosing $m_H^A$ 
occurrences of $H$, out of a total of $m_H'$. This allows 
one to rewrite \eqn{multAB} as
\beq
   c_A \, c_B = 
  \bar{c}_{A B} + \sum_{ \{ m_H^A \} } 
  \sum_{ \{ m_H^B \} } N_{A | \{ m^A_H \} } \,
  N_{B | \{ m^B_H \} } \left( \prod_H \frac{1}{m'_H !} \,
  \text{\bf{C}} (m^A_H, m_H^B) \, \bar{c}_H^{\, m'_H} \right) \, ,
\label{multAB2}
\eeq
where the original multiplicity factor $N_{ AB | \{ m'_H \} }$ is 
replaced by separate multiplicity factors for $A$ and $B$, times 
the number of ways of partitioning diagram $H$ into $A$ and $B$, 
for each $H$. Substituting \eqn{combfun} into \eqn{multAB2}, 
the double sum over proper decompositions appearing on the 
{\it r.h.s} of \eqn{multAB2} factorizes into a product of sums 
depending separately on $A$ and $B$, and can be written as
\beq
  \left\{ \sum_{ \{ m_H^A \} } N_{A | \{ m^A_H \} }
  \left( \prod_H \frac{1}{(m_H^A) !} \, \bar{c}_H^{\, m_H^A} 
  \right) \right\} \left\{ \sum_{ \{ m_H^B \} } 
  N_{B | \{ m^B_H \} } \left( \prod_H \frac{1}{(m_H^B) !}
  \, \bar{c}_H^{\, m_H^B} \right) \right\} \, .
\label{multAB3}
\eeq
By \eqn{colfacts}, this is equal to $c_A c_B$, the product of 
the color factors of subdiagrams $A$ and $B$. We conclude that
$\bar{c}_{AB} = 0$, which is the desired result. The fact that the 
lowest order diagrams are two-eikonal line irreducible completes 
the inductive proof.

In this section we have derived the exponentiation of soft gluon  
corrections for two eikonal lines coupled by a color-singlet hard 
interaction, and in doing so have introduced methods and notations 
that will be useful in what follows. The nonabelian eikonal
exponentiation theorem~\cite{Gatheral:1983cz,Frenkel:1984pz} 
may then be summarised as follows.

\begin{quote}
{\it The sum of all diagrams involving soft-gluon emission from 
two eikonal lines connected in a color singlet exponentiates, and 
the exponent is a sum over two-eikonal line irreducible diagrams 
(webs, $H$) with modified color factors $\{\bar{c}_H \}$.}
\end{quote}

\noindent In the rest of this paper, we consider the extension of 
this result to next-to-eikonal order, for the scattering of two hard 
partons. Arguments towards the generalization of these results 
to the case of several hard partons, using the results of 
Ref.~\cite{Laenen:2008gt}, were given in Ref.~\cite{Gardi:2010rn}.

\section{NE corrections to Feynman diagrams}
\label{NErules}

In the previous section we reviewed the exponentiation of soft 
gluon contributions to scattering amplitudes. The proof of this 
result involves the use of effective Feynman rules in the eikonal 
approximation, valid at leading power in all gluon momentum 
components, as $k_i^\mu \rightarrow 0$. The proof of the 
existence of effective Feynman rules in the eikonal approximation 
relies crucially on the factorization of contributions from 
individual photons (in the abelian case) and from different 
gluon groups (in the nonabelian case), expressed via the 
eikonal identity~(\ref{eikonalid}) and the generalised eikonal 
identity~(\ref{eikid2}) respectively. 

We now wish to organize contributions to the amplitude that
are subleading by one power of a soft momentum, which we
call next-to-eikonal (NE) contributions. To be precise, as described
in the introduction, if we rescale all gluon momentum components 
by a scaling parameter $\eta$, as $k_i^\mu \to \eta k_i^\mu$, 
then the eikonal approximation to an amplitude with $n$ (real or 
virtual) gluons retains only terms that are ${\cal O} (\eta^{- n})$, 
before loop momentum integrations. Here we wish to organize terms
that are ${\cal O} (\eta^{- n +1})$. In order to do so, we
seek to identify effective Feynman rules which can be applied in calculating diagrams where one extra power of gluon momentum 
is retained for at most one of the available gluons. We will see 
that it is indeed possible to obtain effective rules, and that this 
result again relies on the factorization of contributions from 
different gluon groups. The method used here is analagous to 
that used in the previous section, and will lead to a confirmation 
of the results obtained using the path integral approach 
in~\cite{Laenen:2008gt}.

We will show that the effective Feynman rules for NE emissions 
include vertices for the correlated emission of gluon pairs, which
correct for the incomplete factorization of individual gluon 
contributions at NE level. In a nonabelian theory, these vertices
may couple gluons from different groups, and this will lead to
the introduction of further contributions to the logarithm of 
the amplitude, arising from pairs of correlated webs.

Beyond the eikonal approximation, it is no longer true that soft 
gauge boson emissions are insensitive to the spin of the emitting particle. For simplicity, we begin by studying the case of scalar 
particles. We will return to the spin $1/2$ case in 
Sect.~(\ref{spin1/2}).

\subsection{NE emission from a scalar particle}
\label{spin0}

In the case of scalar emitting particles, the eikonal approximation
to a scattering amplitude, given by \eqn{eq_spin0_most}, receives
NE corrections from three sources.

\begin{figure}
\begin{center}
  \begin{displaymath}
  \scalebox{0.7}{\includegraphics{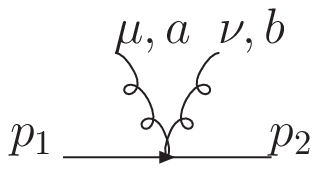}} = 
  g^{\mu\nu}\{T^a,T^b\}
  \end{displaymath}
  \caption{The 2 gluon vertex arising in scalar field theories, which is   
  absent for fermionic emitting particles.}
\label{2gluonscalar}
\end{center}
\end{figure}

\begin{itemize}
\item Subleading corrections to the gluon emission vertex, in which 
one of the momentum factors in the numerator of 
\eqn{eq_spin0_most} is replaced as
\beq
  p^{\mu_i} \rightarrow \frac{1}{2} \left( K_{i + 1} + 
  K_i \right)^{\mu_i} \, .
\label{NEvertex}
\eeq
\item Taylor expansion of a propagator factor, which leads to the  
replacement of an eikonal propagator in \eqn{eq_spin0_most} as
\beq
  \frac{1}{p \cdot K_i} \rightarrow - \, 
  \frac{K_i^2}{2 (p \cdot K_i)^2} \, .
\label{NEprop}
\eeq
\item For scalar emitting particles, the two-gluon (`seagull') vertex 
from the Lagrangian,  shown in \fig{2gluonscalar}, contributes to 
amplitudes at NE level. Indeed, a diagram with one seagull vertex has 
one propagator less then the corresponding diagram with only cubic 
vertices, so it is subleading by precisely one power of a soft 
momentum.
\end{itemize}

\noindent The full NE amplitude consists of a sum over all possible
insertions of the above replacements on each one of the hard lines. 
The seagull vertex has no analogue in the exact Feynman rules for 
emission from a fermion. We will see, however, that a vertex of this 
form arises in the effective Feynman rules for NE emissions from a 
fermion line, as shown in the following subsection.

In what follows, it will often be convenient to rewrite the NE 
emission vertex, using $K_i + K_{i+1} = 2 K_i - k_i$. 
\Eqn{NEvertex} then becomes
\beq
  p^{\mu_i} \rightarrow  K_i^{\mu_i} - \frac{1}{2} k_i^{\mu_i} \, .
\label{NEvertex2}
\eeq
This decomposes the vertex into a part which depends only on a 
single gluon momentum, and a part that depends on a single partial 
sum of gluon momenta. Such a decomposition is useful, since any 
vertex which explicitly depends only on local gluon momenta can be 
immediately interpreted as an effective Feynman rule, as the other 
gluon emissions will factorize upon application of the eikonal identity, 
as in \sect{nonabelianexp}.

To see this in more detail, note that a next-to-eikonal Feynman rule 
which depends on a single momentum in the numerator gives rise to 
a factor in the amplitude of the form
\beq
  k_i^{\mu_i} \, \prod_{j = 1}^n \frac{1}{p \cdot K_j} \, ,
\label{NEfactor1}
\eeq
where we have chosen the $i^{\text{th}}$ gluon to be NE, and
we did not display the remaining numerator factors of $p^{\mu_k}$ 
($k \neq i$) associated with the eikonal emissions. In evaluating the 
total NE contribution, by analogy with the eikonal case, one may 
replace the factor in \eqn{NEfactor1} with a sum over all permutations 
of the emitted gluons, holding the order in each group fixed. The 
total NE contribution the gives the factor
\beq
  k_i^{\mu_i} \sum_{\tilde{\pi}}
  \prod_{j = 1}^n \frac{1}{p \cdot K_{\tilde{\pi}_n}} \, ,
\label{NEfactor2}
\eeq
where, as in \eqn{eikid2}, $\tilde{\pi}$ is a permutation in which 
the order of the gluons in a group is preserved. The product 
of denominator factors now factorises by the generalised eikonal 
identity, \eqn{eikid2}, and one may extract the partial momentum 
sum $G_i$ corresponding to the NE gluon (assuming this to
come from group $G$) so that \eqn{NEfactor2} takes the form
\beq
  \frac{k_i^{\mu_i}}{2 p \cdot G_i} \times \big[ \ldots \big] \, ,
\label{NEfactor3}
\eeq
where the ellipsis denotes eikonal Feynman rules for all other gluons,
factorised into groups. The prefactor can then be interpreted as an 
effective Feynman rule.

A similar argument applies to the two-gluon vertex of 
fig.~\ref{2gluonscalar}. A diagram with such a Feynman rule 
contributes to the amplitude a factor
\beq
  g^{\mu_i \mu_{i + 1}} \prod_{j = 1}^n
  \frac{1}{p \cdot K_{\tilde{\pi}_n}} \, ,
\label{2gvfac1}
\eeq
where again we are not displaying the factors of $p^\mu_k$ 
($k \neq i, i + 1$) from the eikonal emissions. One may sum over
all gluon permutations $\tilde{\pi}$, and the generalised
eikonal identity then implies that the total contribution
from the particular two-gluon vertex of \eqn{2gvfac1} gives
the factor
\beq
  g^{\mu_i \mu_{i + 1}} \prod_{{\rm groups} \, \, g} \, \, 
  \frac{1}{p \cdot k_{g_1}} \frac{1}{p \cdot (k_{g_1} + 
   k_{g_2})} \, \ldots \, \frac{1}{p \cdot(k_{g_1} + \ldots + 
   k_{g_m})} \, .
\label{2gvfac2}
\eeq
If the two gluons entering the two-vertex come from the same 
group $G$, one may extract the relevant denominator factors from 
\eqn{2gvfac2} to yield an expression of the form
\begin{displaymath}
  \frac{g^{\mu_i \mu_{i + 1}}}{p \cdot(G_i + G_{i + 1})}
  \times \big[ \ldots \big] \, ,
\end{displaymath}
where the ellipsis denotes a product of eikonal Feynman rules for all 
other gluons, factorised into groups. The prefactor can be interpreted 
as an effective Feynman rule, coupling two gluons within a single 
group. If, on the other hand, the two gluons come from separate 
groups, then the seagull vertex couples together two groups: in other
words, the denominator of the effective Feynman rule contains
the sum of two partial momentum sums $G_i$, $H_j$ from different gluon groups. The two-gluon vertex therefore entangles two 
subdiagrams, each of which would be a web, were the seagull vertex 
not present. We return to this point in \sect{sec:exp}. Note also that 
the NE vertex still partly depends on sums over individual gluon 
momenta. This will be reconsidered  in \sect{factor}.

\subsection{NE emission from a spin-$\frac{1}{2}$ particle}
\label{spin1/2}

In the case of spin-$\frac{1}{2}$ fermions, minimally coupled to
massless gauge bosons, the lagrangian generates only three-point 
vertices.  At NE order, we find therefore only two types of 
contributions, arising from corrections to the vertices and to 
the propagators respectively. Taylor expansion of the propagators
in powers of the gluon  momenta proceeds exactly as in the 
scalar case. The vertex corrections, on the other hand, are more 
complicated, owing to the non-trivial Lorentz and color structure. 

Let us first examine the spinor structure. Consider the numerator 
of \eqn{spin1/2mat}, where we drop the spinor $u(p)$ and the 
hard matrix element ${\cal M}_0 (p)$ for simplicity. We write
\beq
  {\cal N}^{\mu_1, \ldots, \mu_n} (p, k_i) \equiv ({\slashed p} + 
  {\slashed K}_1) \gamma^{\mu_1}  \, \ldots \,  ({\slashed p} + 
  {\slashed K}_n) \gamma^{\mu_n} =  
  {\cal N}^{\mu_1, \ldots, \mu_n}_{\rm EIK} (p) +
  {\cal N}^{\mu_1, \ldots, \mu_n}_{\rm NE} (p, k_i)  \, + 
  \, \ldots\, ,
\label{num}
\eeq
where ${\cal N}^{\mu_1, \ldots, \mu_n}_{\rm NE}$ collects all 
terms linear in gluon momenta. In each such term, all factors of 
${\slashed p}$ to the right of the NE vertex insertion can be 
simplified using anticommutation and the Dirac equation (as in 
the eikonal case). We find then
\beqa
  {\cal N}^{\mu_1, \ldots, \mu_n}_{\rm NE} (p, k_i) & = & 
  \sum_{i = 1}^n \, \left[ \prod_{j = i + 1}^n \left( 2 p^{\mu_j}
  \right) \, \prod_{k = 1}^{i - 1} \, {\slashed p} \gamma^{\mu_k} 
  \, (2 K_i^{\mu_i} - \gamma^{\mu_i} {\slashed K}_i) \right] 
  \nonumber \\
  & = & {\cal N}^{\mu_1, \ldots, \mu_n}_{{\rm NE}, 1} (p, k_i) \, 
  + \, {\cal N}^{\mu_1, \ldots, \mu_n}_{{\rm NE}, 2} (p, k_i)  \, ,
\label{NEv1}
\eeqa
where we used anticommutation relations to move 
${\slashed K}_i^{\mu_i}$ to the right of $\gamma^{\mu_i}$ 
in the last factor of \eqn{NEv1}. The NE numerator is now 
written as the sum of two terms. The first term has no spinor 
structure, since one may anticommute all factors of 
${\slashed p}$ through to the right, where they annihilate 
the Dirac spinor as in the eikonal case. One may then rewrite 
this term using $2 K_i = K_i + K_{i + 1} + k_i$, as
\beq
  {\cal N}^{\mu_1, \ldots, \mu_n}_{{\rm NE}, 1} (p, k_i) = 
  \sum_{i = 1}^n \, \left[ \prod_{j \neq i} \left( 2 p^{\mu_j}
  \right) \, \left( K_i + K_{i + 1} \right)^{\mu_i} \right]
  + \sum_{i = 1}^n \, \left[ \prod_{j \neq i} \left( 2 p^{\mu_j}
  \right) \, k_i^{\mu_i} \right] \, .
\label{NEv2}
\eeq
The first summation on the right-hand side is the same as the 
sum over NE vertex corrections in the scalar case, as seen
from \eqn{NEvertex}. The second summation is over terms which 
only depend upon a single gluon momentum, and therefore can
be directly translated into Feynman rules. 

\begin{figure}
\begin{center}
  \scalebox{1.0}{\includegraphics{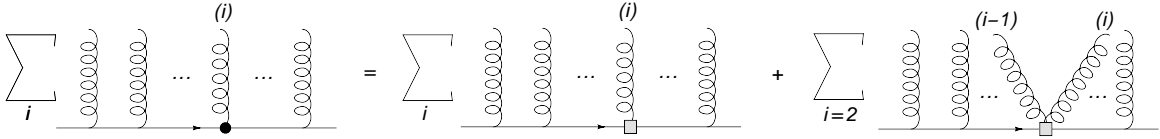}}
  \caption{Schematic representation of the theorem given in 
  \eqn{theor1}. Here $\bullet$ denotes a vertex which depends 
  upon a sum of gluon momenta, and $\Box$ a vertex depending 
  only upon the momenta entering the chosen vertex.}
\label{theor1fig}
\end{center}
\end{figure}

The second term in \eqn{NEv1}, which still has a spinor structure, 
requires more careful handling. Using only the Dirac algebra,
and a recursive argument, we prove in Appendix~\ref{proof} that the 
next-to-eikonal numerator ${\cal N}^{\mu_1, \ldots, 
\mu_n}_{{\rm NE}, 2} (p, k_i)$ can be written as
\beqa
  {\cal N}^{\mu_1, \ldots, \mu_n}_{{\rm NE}, 2} (p, k_i) 
  & \equiv & - \, 
  \sum_{i = 1}^n \, \left[ \prod_{j = i + 1}^n \left( 2 p^{\mu_j}
  \right) \, \prod_{k = 1}^{i - 1} \, {\slashed p} \gamma^{\mu_k} 
  \, \gamma^{\mu_i} {\slashed K}_i \right] 
  \label{theor1} \\
  & = & - \, 
  \sum_{i = 1}^n \, \left[ \prod_{j \neq i} \left( 2 p^{\mu_j}
  \right) \, \gamma^{\mu_i} {\slashed k}_i \, + \, 
  \prod_{j \neq i, i - 1} \left( 2 p^{\mu_j} \right) \,
  2 p \cdot K_i \, \gamma^{\mu_{i - 1}} \gamma^{\mu_i}
  \right]  \nonumber \, ,
\eeqa
where the sum in the second term of the last line starts at $i=2$.
\Eqn{theor1} shows that spin-dependent terms in the sum over 
all possible NE insertions on a hard fermion line can be expressed 
as the sum of two kinds of contributions. The first kind implements
single-gluon vertex corrections, each depending on the momentum 
of a single gluon. The second kind implements a new 2-gluon, `seagull'
vertex, as can be seen from the fact that the factor $2 p \cdot K_i$ 
in \eqn{theor1} is precisely such as to cancel the eikonal propagator
connecting the two neighbouring gluons $i$ and $i - 1$. The result
is shown in diagrammatic form in \fig{theor1fig}.

Combining \eqns{NEv2}{theor1}, one sees that the NE numerator
for a fermion line is built out of two kinds of of contributions. 
There are terms where single and double NE emissions manifestly
factorise ({\it i.e.} can be written in terms of the momenta and 
quantum numbers of the particles entering a specific vertex);
there are however also terms which still depend on sums of gluon
momenta along the eikonal line. We note that the latter, given by 
the first sum on the right-hand-side of \eqn{NEv2}, together with 
the corrections from Taylor expansion of the propagators, are the 
same as in the scalar case, as seen from \eqns{NEvertex}{NEprop}. 

Using arguments similar to those of \sect{spin0}, any term which
depends only upon single momenta factorises according to the
generalized eikonal identity, and thus can be interpreted as an 
effective next-to-eikonal Feynman rule. In particular, we find a 
one-gluon vertex resulting from the second sum in \eqn{NEv2} 
and from the first sum in \eqn{theor1}. This vertex has the form
\beq
  V_{\rm NE}^{\mu, a} (k) \equiv \left( k^\mu - 
  \gamma^\mu {\slashed k} \right) \, T^a \, .
  \label{1g}
\eeq
In practice this will be combined with an eikonal denominator 
factor (involving the appropriate partial momentum sum), as 
in \eqn{NEfactor3}. Here, however, we focus on the numerator 
structure only, and postpone a full presentation of the 
effective Feynman rules to \sect{NErulessum}. We also find a 
two-gluon emission vertex, arising from the second sum in 
\eqn{theor1}, which has the form
\beq
  V_{\rm NE}^{\mu \nu, \, a b} = 
  \gamma^\mu \gamma^\nu T^a T^b \, .
  \label{2g}
\eeq
Note that we have explicitly reinstated the color factors associated 
with each emission in the case of a non-abelian theory. The 
contributions arising from \eqns{1g}{2g} ultimately lead 
to a factorised expression for the amplitude. 
The same is not true, however, for the spin-independent contributions 
to the one-gluon emission vertex. We will see in the next section 
that the sum of these contributions can indeed be expressed in a 
factorised form, up to a remainder term expressing correlations 
between gluons in different groups. 

First, it is interesting to note some properties of the emission 
vertices in \eqns{1g}{2g}. Using
\beq
  \gamma^\mu {\slashed k} = \frac{1}{2} \left( [\gamma^\mu,
  {\slashed k}] + \{ \gamma^\mu, {\slashed k} \} \right) \, ,
\label{comk}
\end{equation}
we see that \eqn{1g} may be rewritten as
\beq
  V_{\rm NE}^{\mu, a} (k) = - 2 i \, k_\nu \sigma^{\mu\nu} \, ,
\label{1gb}
\eeq
where we recognise the generators of the Lorentz group 
$\sigma^{\mu \nu} \equiv - \frac{i}{4} [\gamma^\mu,
\gamma^\nu]$. The physical interpretation of this vertex, 
which does not occur for scalar emitting particles, is now clear: 
it is a (chromo-)magnetic moment vertex associated with the 
interaction between the spin of the emitting particle and the 
momentum of the emitted soft boson. This occurs for the first 
time at NE level, which is consistent with the fact that emitted 
radiation is insensitive to the spin of the emitter in the eikonal 
approximation.

One may similarly replace \eqn{2g}, using Bose symmetry, with
its symmetric part under the exchange of the two gluons. One
writes then
\beq
  V_{\rm NE}^{\mu \nu, \, a b} = 
  g^{\mu\nu} \{T^a,T^b\} + \frac{1}{2}\, [\gamma^\mu,
  \gamma^\nu] \, [T^a,T^b] \, .
\label{2gc}
\eeq
The first term has the same form as the scalar seagull vertex,
while the second term again involves the Lorentz generators 
$\sigma^{\mu\nu}$. The latter term vanishes in abelian gauge 
theories, thus it contributes a non-abelian component to the 
chromo-magnetic moment.

The results of this section can be translated into a set of effective
Feynman rules describing next-to-eikonal emissions. Some of these 
rules rely on individual momenta (they are `local' in the pictorial
representation of the Feynman graph), whereas others still rely 
upon partial sums of gluon momenta. Crucial to the eventual
exponentiation of soft gauge boson corrections is the factorization 
of gluon or photon emissions. In the eikonal approximation, this 
factorization results from the application of the eikonal identity 
\eqn{eikonalid} for abelian theories, and of the generalised eikonal
identity \eqn{eikid2} for nonabelian theories. We now consider the
extension of these same methods to NE order. Having derived 
effective Feynman rules, the next step is to show that contributions 
from different gluon groups factorise. As in the eikonal case, one 
may first separate the color structure of diagrams (as in 
\eqn{eiksubamp}), and then factorize the momentum structure 
arising from the NE Feynman rules. This factorization procedure 
is more complicated than at eikonal order (one lacks the relatively 
simple form of the generalised eikonal identity): in fact, it
is necessary to introduce extra effective Feynman rules, which 
correlate the emissions of pairs of gluons from different groups. 
This is the subject of the following subsection.

\subsection{Factorization of next-to-eikonal diagrams}
\label{factor}

In the previous section we have seen that there are two classes 
of effective Feynman rule at NE order: those depending only on 
the specific momentum associated with a particular gluon emission, 
and those that still rely on partial sums of gluon momenta. The first 
type of rule leads simply to the factorization of NE contributions 
from different gluon groups, using the same generalised eikonal 
identity, \eqn{eikid2}, as in the eikonal approximation. More work 
is needed to analyze the contributions from the second type of 
Feynman rules. We will see in this section that factorization for 
these contributions is weakly broken by a remainder term, which 
implements correlations between pairs of gluon emissions from 
different groups. The resulting structure, however, is still sufficient 
to achieve exponentiation of NE effects in terms of a subset of 
diagrams, as shown in Sect.~\ref{sec:exp}. 

Let us begin by introducing some notations and conventions 
that will be useful in the following. First of all, consistently with 
\eqn{eikfactfin}, we denote the combined eikonal vertex and 
propagator by
\beq
  \E^{\mu} (p, K) \, \equiv \, \frac{p^\mu}{p \cdot K} \, ,
\label{E}
\eeq
where one should keep in mind that in the following $K$ will be 
allowed to be a partial sum of gluon momenta.

Next, we introduce similar notations for NE corrections arising
from numerators and from the Taylor expansion of propagator 
denominators. In accordance with \eqn{NEvertex} and with
\eqn{NEprop}, we write
\beq
  \NE^{\mu}_P (p, K) \equiv - \, \frac{K^2}{2 (p \cdot K)^2} 
  \, p^{\mu} \, ,  \qquad \quad 
  \NE^{\mu}_V (p, K) \equiv \frac{K^{\mu}}{p \cdot K} \, ,
\label{NEdefform}
\eeq
where again $K$ will generically denote a partial momentum sum 
rather than a single momentum. One may also define the complete
NE correction factor
\beq
  \NE^{\mu} (p, K) \equiv \NE^{\mu}_P (p, K)  +
  \NE^{\mu}_V (p, K) \, ,
\label{NEtotform}
\eeq
however, given that each diagram at NE order has either a vertex 
or a propagator correction (but not both), we will mostly consider 
the two types of correction separately in what follows. When 
considering subgraphs built with a single hard line, we will omit
the hard momentum $p$ as an argument in all these definitions.

Consider next the organization of gluons into groups. As in the 
eikonal case, we partition the $n$ gluons emitted by any one of 
the hard lines into groups $g$: gluons in the same group belong to 
the same two-eikonal irreducible subgraph (web). Gluons are labelled
along a given eikonal line as in \fig{fig_basic_line}, {\it i.e.} from 
$1$ to $n$, where the index increases as one moves away from 
the hard interaction.  We will then need to sum over permutations 
$\pi$ such that the order of gluons in each group is held fixed. 
The order of gluons along the line in any such permutation is then 
given by $\pi_1, \ldots, \pi_n$, where $\pi_i$ labels the 
$i^{\text{th}}$ gluon insertion moving away from the hard 
interaction, in permutation $\pi$.

It will also be useful to introduce a label $l$ denoting those gluons 
which are the first of a group, in the order just stated. The group 
which contains this gluon we label by $g(l)$.  We then define
\beq
  \tilde{g} (l) \equiv \left\{\begin{array}{c} g, \quad  
  g \neq g(l) \\ g/\{l\}, \quad g = g(l) \end{array} \right. \, .
\label{gtilde}
\eeq
In words, a tilde over a group denotes that group without the first 
gluon, if that gluon is $l$. Finally, as before, we find it useful to 
denote by $G$ partial momentum sum of gluons in a group $g$ 
(while we denote by $K$ partial momentum sums in the entire set 
of available gluons). Thus $G_k$ will be the partial momentum sum 
from gluon $k$ up to gluon $m$, in a group $g$ containing $m$ 
gluons. We will also shorten $G_1$ to $G$, representing the sum 
of all momenta in group $g$.

\begin{figure}
\begin{center}
  \scalebox{1.0}{\includegraphics{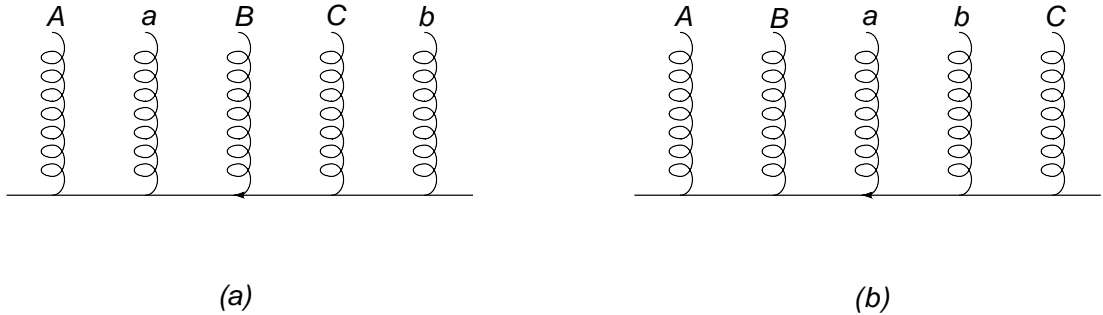}}
  \caption{Simple example of two groups of gluon emissions on a 
  single eikonal line. See the discussion in the text.}
\label{groupdef}
\end{center}
\end{figure}

To clarify the above definitions, consider the simple case of two
gluon groups, depicted in \fig{groupdef}, where we label the various 
gluon emissions by upper and lower case letters, corresponding to 
each group.
\Fig{groupdef}(a) shows a given hard line with two gluon groups 
$g = \{A,B,C\}$ and $h = \{a,b\}$. In this permutation, one 
has $\pi = (A,a,B,C,b)$. \Fig{groupdef}(b) corresponds to a different
permutation, $\pi = (A,B,a,b,C)$, where the order of gluons 
in each group is not changed. In this simple case, the label $l$ takes 
the values $A$ or $a$, and $g(A) = g$, while $g(a) = h$. The partial
momentum sums $G$ and $H$ in each case are given by $G = 
k_A + k_B + k_C$ and $H = k_a + k_b$. Finally, if $l = A$ one has 
$\tilde{g} (A) = \{B, C \}$ and $\tilde{h} (A) = \{a, b \}$; 
alternatively, if $l = a$ then one has $\tilde{g} (a) = \{A, B, C \}$ 
and $\tilde{h} (a) = \{b \}$. 

In constructing a (squared) amplitude at NE level we will encounter
expressions with the general structure of \eqn{amp3}, with two important differences: attachments of the gluon graphs to the hard 
lines will now include NE corrections, and, therefore, we will not be 
able to employ directly the generalized eikonal identity. As a 
consequence, it will be necessary to deal with sums over permutations, 
such as the one appearing in \eqn{eikfact2}. To handle these sums, 
we introduce a shorthand for the Lorentz tensors occurring in these
expressions. For every allowed permutation $\pi$, we define
\beqa
  \E(\pi) & = & \prod_{k = 1}^n 
  \E^{\mu_{\pi_k}} (K_{\pi_k}) \, , \nonumber  \\
  \NE(\pi) & = & \sum_{i = 1}^n \left[ \NE^{\mu_{\pi_i}}
  (K_{\pi_i}) \, \prod_{k \neq i} 
  \E^{\mu_{\pi_k}} (K_{\pi_k}) \right] \, , \nonumber \\
  \NE'(\pi) & = & \sum_{i = 2}^n \left[ \NE^{\mu_{\pi_i}}
  (K_{\pi_i}) \, \prod_{k \neq i} 
  \E^{\mu_{\pi_k}} (K_{\pi_k})  \right]  \, ,
  \label{NEdefs} \\
  \NE_1(\pi) & = & \NE^{\mu_{\pi_1}}
  (K_{\pi_1}) \, \prod_{k = 2}^n 
  \E^{\mu_{\pi_k}} (K_{\pi_k}) \, . \nonumber
\eeqa
which will be convenient in the following proof. In this shorthand 
notation Lorentz indices are not displayed on the left-hand sides
of \eqn{NEdefs}; also, one may, if needed, distinguish between
propagator or vertex corrections for the NE terms, by simply 
adding the appropriate index as in \eqn{NEdefform}. One easily 
sees that $\NE(\pi)$ is the sum over all possible NE insertions, for 
a given permutation $\pi$.  Similarly, $\NE'(\pi)$ is a sum over 
all insertions, where the first gluon is restricted to be eikonal, while
$\NE_1(\pi)$ is the term resulting when the first gluon
in permutation $\pi$ is next-to-eikonal. Thus one has
\beq
  \NE(\pi) = \NE_1(\pi) + \NE'(\pi) \, .
\label{NEeq}
\eeq
Finally, we note that we will employ the same notation as in 
\eqn{NEdefs} for Lorentz tensors constructed with gluons 
belonging to a given group. Thus, for example, $\E(g)$ will be 
the product of the eikonal factors associated with gluons belonging 
to group $g$ (whose order is fixed in any permutation $\pi$),
while $\NE(g)$ will be the sum over all possible NE insertions in
group $g$. Note that we also use the same notation for entire 
soft gluon diagrams, which were denoted by $E(G)$ in the eikonal
approximation already in \sect{eikonal}, and will be denoted by 
$\NE(G)$ at NE level. This does not lead to ambiguities, since soft 
gluon diagrams are just products of line factors, such as the ones 
described here, contracted with suitable color tensors and other 
factors related to internal soft gluon interactions. These factors 
do not affect the manipulations performed here and below, which 
are related to the combinatorics of soft gluon insertions on the 
hard lines.

We are now in a position to state the theorem governing the 
factorization of gluon groups at NE level. It can be written as
\beq
  \sum_\pi \NE(\pi) = \sum_h \left[ \NE(h) \prod_{g \neq h} 
  \E(g) \right] + \sum_{g \neq h} \left[ R (g, h) 
  \prod_{f \neq g, h} \E(f) \right] \, .
\label{theor2}
\eeq
This theorem can be seen as a further generalization, to NE level,
of the generalized eikonal identity~(\ref{eikid2}), which in the
present language would read simply $\sum_\pi \E(\pi) = \prod_g 
\E(g)$. The content of \eqn{theor2} can be summarised as follows.

\begin{quote}
{\it The NE contribution to soft gluon emissions from a hard line 
can be written in terms of two sums. In the first sum gluon
groups are factorised, and each term contains a sum over NE 
gluon insertions in a given group, times a product 
of eikonal factors for the remaining groups. The second sum 
organizes pairwise correlations between gluon groups, governed 
by a NE remainder function $R(g,h)$, times a product 
of eikonal factors for the remaining groups.}
\end{quote}

The proof is by induction, and it involves several steps. We must 
develop the induction argument separately for the NE vertex and propagator corrections: in each case, we will show a weak form
of factorization which leaves behind pairwise correlations of 
groups. The final step is to collect all these correlations and 
construct an explicit expression for the remainder function 
$R(g,h)$, which will generate its own Feynman rule.

We begin by noting that the theorem is trivially true for the 
emission of a single gluon. Next, we separate out the first 
gluon (the one emitted next to the hard interaction) from the 
sum over permutations. This gluon, being the first along the line, 
will also be the first gluon of a group. Thus, we can label it by
$l$, and denote the group it belongs to by $g(l)$. Furthermore, 
$l$ will either be eikonal or next-to-eikonal. Then, using the 
notations introduced above, we may write
\beq
  \sum_\pi \NE(\pi) = \sum_l \NE^{\mu_l} (K) 
  \sum_{\tilde{\pi}} \E(\tilde{\pi}) + \sum_l 
  \E^{\mu_l} (K) \sum_{\tilde{\pi}} \NE(\tilde{\pi}) \, ,
\label{theor2a}
\eeq
where $K = \sum_i k_i$ is the sum of all emitted gluon momenta. 
In this equation, $\tilde{\pi}$ represents a permutation of all but 
gluon $l$, where the order in each group has been held fixed. Thus, 
we have split the sum over permutations into a sum over the possible 
first gluons, and a sum over permutations of the remaining gluons. 
In the first term of \eqn{theor2a} one may use the generalised 
eikonal identity for the set of all gluons but the first,
\beq
  \sum_{\tilde{\pi}} E(\tilde{\pi}) = \prod_{\tilde{g}}
  E(\tilde{g}) \, .
\label{eikid}
\eeq
In the second term of \eqn{theor2a} one may use the validity of 
the conjecture (\ref{theor2}) for $n - 1$ gluons. One then gets
\beq
  \sum_\pi \NE(\pi) = \sum_l \NE^{\mu_l}(K) 
  \prod_{\tilde{g}} \E(\tilde{g}) + \sum_l \E^{\mu_l} (K) 
  \left( \sum_{\tilde{h}} \NE(\tilde{h}) \prod_{\tilde{g} 
  \neq \tilde{h}} \E (\tilde{g}) + \ldots \right) \, ,
\label{eq_basic_ind}
\eeq
where we did not explicitly write the remainder term, which will 
be dealt with in Sect.~\ref{remainder}. In the next two subsections
we consider separately the cases where $\NE^{\mu_l}$ represents
the propagator and vertex corrections of \eqn{NEdefform}, and 
we prove that for both of these \eqn{eq_basic_ind} can indeed 
be reduced to the form of \eqn{theor2}. This will also implicitly
determine the remainder term, which is then worked out in 
Sect.~\ref{remainder}.

\subsubsection{The NE propagator}
\label{prop}

In this subsection we consider the structure of \eqn{eq_basic_ind}
with $\NE^\mu (p, K) \to \NE^\mu_P (p, K)$, focusing on the
propagator correction given in \eqn{NEdefs}. For brevity, below we 
will omit the dependence on the hard momentum $p$.

Let us begin by considering the first term in \eqn{eq_basic_ind}. Here one may write
\beq
  \NE^{\mu_l}_P (K) = \frac{K^2}{2 (p \cdot K)^2} \, 
  p \cdot G_l \, \E^{\mu_l} (G_l) \, ,
\label{NE2}
\eeq
where the factor $p \cdot G_l$ cancels the eikonal propagator 
in $\E^{\mu_l} (G_l)$.
Substituting this into the first term of \eqn{eq_basic_ind} and 
using
\beq
  \E^{\mu_l} (G_l) \, \prod_{\tilde{g}} \E(\tilde{g})
  = \prod_g \E(g) \, ,
\label{theor2b}
\eeq
one has
\beq
  \sum_l \NE^{\mu_l}_P (K) \, \prod_{\tilde{g}} \E(\tilde{g}) =
  \left(\prod_g \E(g) \right) \, \frac{K^2}{2 (p \cdot K)^2} \,
  \sum_l p \cdot G_l \, .
\label{theor2c}
\eeq
Now we may use the fact that $\sum_l G_l = K$; indeed, note
that $G_l$ is the total momentum of gluons in group $g(l)$, 
and the sum over $l$ is equivalent to a sum over groups, since
the order of emissions within a group is never rearranged. We
can then write
\beq
  \sum_l \NE^{\mu_l}_P (K) \, \prod_{\tilde{g}} 
  \E(\tilde{g}) = \frac{K^2}{2 \, p \cdot K } \, \prod_g \E(g) \, .
\label{theor2d}
\eeq
We may now use the fact that
\beq
  K^2 = \sum_lG_l^2 \, + \, \sum_{l \neq r} \, G_l \cdot G_r \, .
\label{Ksq}
\eeq
The cross-terms involve correlated emissions of gluons in 
different groups, and thus enter the remainder term, to be
discussed in Sect.~\ref{remainder}. For the $G_l^2$ terms, 
one notes that
\beq
  \NE_{1, P} (g(l)) = \frac{G_l^2}{2 (p \cdot G_l)^2} 
  \, p^{\mu_l} \, E(\tilde{g}(l)) = \frac{G_l^2}{2\, p 
  \cdot G_l} \, E(g(l)) \, ,
\label{theor2e}
\eeq
where we absorbed an eikonal numerator and denominator in 
$\E(g(l))$. One  can use \eqn{theor2e} to obtain an expression
for $G_l^2$, which can then  be substituted into \eqn{theor2d}.
One finds
\beq
  \sum_l \NE^{\mu_l}_P (K) \, \prod_{\tilde{g}} 
  \E(\tilde{g}) = \sum_l \frac{p \cdot G_l}{ p \cdot K} \,\,
  \NE_{1, P} (g(l)) \prod_{g \neq g(l)} \E(g) \, .
\label{eq_prop_first}
\eeq

Next, consider the second term in \eqn{eq_basic_ind}. In this 
term, either $h = g(l)$ or $h \neq g(l)$, thus one may write
\beqa
  \sum_l \E^{\mu_l} (K) \sum_{\tilde{h}} \NE_P (\tilde{h})
  \prod_{\tilde{g} \neq \tilde{h}} \E(\tilde{g}) & = & 
  \sum_l \E^{\mu_l} (K)
  \, \NE_P (\tilde{g} (l)) \prod_{g \neq g(l)} \E(g) \nonumber \\
  && + \, \sum_l \E^{\mu_l}(K) \sum_{h \neq g(l)} \,
  \NE_P (h) \prod_{\tilde{g}  \neq h} \E(\tilde{g}) \, ,
\label{theor2f}
\eeqa
where in the second term we used the fact that $h$ does not 
contain gluon $l$ to replace $\tilde{h}$ with $h$. Next we use
the simple identity
\beq
  \E^{\mu_l} (K) = \frac{p \cdot G_l}{p \cdot K} \, 
  \E^{\mu_l} (G_l) \, ,
\label{theor2g}
\eeq
to find
\beqa
  \sum_l \E^{\mu_l} (K) \sum_{\tilde{h}} \NE_P (\tilde{h})
  \prod_{\tilde{g} \neq \tilde{h}} \E(\tilde{g}) & = &
  \sum_l \frac{p \cdot G_l}{p \cdot K} \, \NE'_P (g(l))
  \prod_{g \neq g(l)} \E(g) \nonumber \\
  && \, + \, \sum_l \sum_{h \neq g(l)} 
  \frac{p \cdot G_l}{p \cdot K} \, \NE_P (h) 
  \prod_{g \neq h} \E(g) \, .
\label{theor2h}
\eeqa
In the first term on the right-hand side, we have used the 
definition in \eqn{NEdefs} for $\NE'(g(l))$ to recognise that
\beq
  \NE'_P (g(l)) = \E^{\mu_l} (G_l) \, \NE_P (\tilde{g}(l)) \, ,
\label{NEprime2}
\eeq
while in the second term we have used the fact that
\beq
  \E^{\mu_l} (G_l) \prod_{\tilde{g} \neq \tilde{h}} \E(\tilde{g})
  = \prod_{g \neq h} \E(g) \, .
\label{Ecomb}
\eeq
We may now recombine \eqn{theor2h} with \eqn{eq_prop_first}. 
In doing so, the first term on the right-hand side of \eqn{theor2h} combines with the right-hand side of \eqn{eq_prop_first} using 
\eqn{NEeq}, and one finds
\beqa
  \sum_\pi \NE_P (\pi) & = & \sum_l \frac{p \cdot G_l}{p \cdot K}
  \, \NE_P (g(l)) \prod_{g \neq g(l)} \E(g) + 
  \sum_l \sum_{h \neq g(l)} \frac{p \cdot G_l}{p \cdot K} \,
  \NE_P (h) \prod_{g \neq h} \E(g) + \ldots 
  \nonumber \\
  & = & \sum_l \frac{p \cdot G_l}{p \cdot K} \, \sum_h
  \NE_P (h) \, \prod_{g \neq h} \E(g) + \ldots  \\
  & = & \sum_h \NE_P (h) \, \prod_{g \neq h} \E(g) 
  + \ldots  \, , \nonumber
\label{recomb}
\eeqa
where the ellipsis again denotes the remainder term. In the second 
line, we have used the fact that $\sum G_l = K$. The result gives 
the contribution of the NE propagator corrections to the right-hand 
side of \eqn{theor2}, as expected.

\subsubsection{The NE vertex}
\label{vertex}

We turn now to the NE vertex correction $\NE_V^\mu (p,K)$.
Using
\beq
  K^\mu = G^\mu_l + \sum_{k \neq l} G^\mu_k \, ,
\label{theor2av}
\eeq
we can rewrite the first term of \eqn{eq_basic_ind} as
\beqa
  \sum_l \NE^{\mu_l}_V (K) \prod_{\tilde{g}} \E(\tilde{g})
  & = & \sum_l \frac{p \cdot G_l}{p \cdot K} \, 
  \NE_{1, V} (g(l)) \prod_{\tilde{g} \neq g(l)} \E(\tilde{g})
  \nonumber \\
  && \, + \, \sum_l \sum_{k \neq l}
  \frac{G_k^{\mu_l}}{p \cdot K} \prod_{\tilde{g}}
  \E(\tilde{g}) \, ,
\label{theor2bv}
\eeqa
where we have used the fact that
\beq
  \NE_{1,V} (g(l)) = \frac{G_l^{\mu_l}}{p \cdot G_l} \, 
  \E(\tilde{g}(l)) \, .
\label{theor2bv2}
\eeq
The second term in \eqn{theor2bv} involves correlations between 
gluons in different groups, and enters the remainder term to be discussed in Sect.~\ref{remainder}. The first term is of the same 
form as the analagous result for the propagator correction in
\eqn{eq_prop_first}.

For the second term in \eqn{eq_basic_ind}, one proceeds as
in Sect.~\ref{prop}. Eqs.~(\ref{theor2f} - \ref{theor2h}) apply 
also in the case of the vertex correction, given that the form of 
$\NE_P (g)$ is not used there. Given that the forms of
\eqn{eq_prop_first} and of the first term in \eqn{theor2bv} 
are the same, combining the latter with \eqn{theor2h} leads 
(as in the previous subsection) to the right-hand side of 
\eqn{theor2}, which is the desired result.

\subsection{The remainder term}
\label{remainder}

We now turn to the determination of the explicit form of the 
two-group correlation function $R(g,h)$ in \eqn{theor2}.
As a first step, we note that we may again consider separately
the propagator and vertex corrections, given that each NE 
diagram has at most one of these. Thus, one may write
\beq
  R(g, h) = R_P (g, h) + R_V (g, h) \, ,
\label{Rsum}
\eeq
by analogy with \eqn{NEtotform}:  the right-hand side contains 
the sum of the contributions due to propagator and vertex 
corrections respectively. 

Let us now rewrite \eqn{eq_basic_ind}, including explicitly 
the contribution of the remainder term. One finds
\beqa
  \sum_\pi \NE(\pi) & = & \sum_l \NE^{\mu_l}(K)
  \prod_{\tilde{g}} \E(\tilde{g}) + \sum_l \E^{\mu_l} (K) 
  \sum_{\tilde{h}} \NE(\tilde{h}) \prod_{\tilde{g} \neq \tilde{h}} 
  \E(\tilde{g}) \nonumber \\
  &&  \, + \, \sum_l \E^{\mu_l} (K) \sum_{ \tilde{g} \neq 
  \tilde{h}} R \left( \tilde{g}, \tilde{h} \right) 
  \prod_{\tilde{f} \neq \tilde{g}, \tilde{h}} \E(\tilde{f}) \, .
\label{theor3a}
\eeqa

The derivation in Sects.~ \ref{prop} and \ref{vertex} has shown 
that the terms in the first line of \eqn{theor3a} reconstruct
the first sum on the right-hand side of our theorem, \eqn{theor2},
leaving behind two-group correlations which were dropped in
\eqn{Ksq} and in \eqn{theor2av}. Reinstating those terms, we
see that what we have been able to prove so far takes the form
\beqa
  \sum_\pi \NE(\pi) & = & \sum_h \NE(h) \prod_{g \neq h} E(g) 
  \, + \, \sum_l \E^{\mu_l} (K) \sum_{\tilde{g} \neq \tilde{h}}
  R \left( \tilde{g}, \tilde{h} \right) \prod_{\tilde{f} \neq \tilde{g},
  \tilde{h}} \E (\tilde{f}) \nonumber \\
  && \, + \, \sum_{l \neq r} \, \frac{G_l \cdot G_r}{2 \, p \cdot K}
  \, \prod_g E(g) + \sum_{l \neq r} \, \frac{G_r^{\mu_l}}{p \cdot K}
  \, \prod_{\tilde{g}}  E(\tilde{g}) \, .
\label{theor3a2}
\eeqa
Separating the remainder function $R(g,h)$ into its propagator and 
vertex components, according to \eqn{Rsum}, we note that the first 
term in the last line of \eqn{theor3a2} contributes (by definition) 
to $R_P$, whereas the second term contributes to $R_V$. Equating
\eqns{theor3a2}{theor2} then gives the relations
\beqa
  \sum_{g \neq h} R_P (g, h) \prod_{f \neq g, h} E(f) & = & 
  \sum_l \E^{\mu_l} (K) \sum_{\tilde{g} \neq \tilde{h} }
  R_P (\tilde{g}, \tilde{h}) \prod_{\tilde{f} \neq 
  \tilde{g}, \tilde{h} } \E(\tilde{f}) \nonumber \\
  && \, + \, \sum_{g \neq h} \, \frac{G \cdot H}{2 p \cdot K} \,
  \E(g) \, \E(h) \prod_{f \neq g, h} \E(g) \, ; 
  \label{rp} \\
  \sum_{g \neq h} R_V (g, h) \prod_{f \neq g, h} E(f) & = & 
  \sum_l \E^{\mu_l} (K) \sum_{\tilde{g} \neq \tilde{h}} 
  R_V (\tilde{g}, \tilde{h}) \prod_{\tilde{f} \neq 
  \tilde{g}, \tilde{h} } \E(\tilde{f}) \nonumber \\
  && \, + \, \sum_{g \neq h} \frac{H^{\mu_{g_1}}}{p \cdot K}
  \, \, \E(\bar{g}) \, \E(h) \, \prod_{f \neq g, h} \E(f) \, .
  \label{rv}
\eeqa
which may be solved for the remainder terms $R_{P, V}$. Notice
that we have relabeled the sums in the second terms on the 
right-hand sides of \eqns{rp}{rv} replacing $\sum_{r \neq l}
\to \sum_{g \neq h}$, {\it i.e.} replacing the sum over all first 
gluons with a sum over all groups; also, we denoted by $G$ and 
$H$ the sum of all momenta in groups $g$ and $h$ respectively, 
and we introduced the notation $\bar{g}$, which designates the 
group $g$ without its first gluon (which implies that $\tilde{g} =
\bar{g}$ if $l \in g$). 

We will proceed by writing down putative solutions to \eqns{rp}{rv}, 
and then showing that indeed the equations are satisfied. Our
proposed solution to \eqn{rv} can be written as 
\beq
  R_V (g, h) = \sum_\pi^{(g, h)} \, \sum_{i, \pi_i \in g} \, \, 
  \frac{H_j^{\mu_{g_i}}}{p\cdot(G_i+H_j)} \, \tilde{E}(\pi) \, .
\label{remvsol}
\eeq
Let us explain the notations we have introduced. In the first 
sum, $\pi$ is any permutation of the gluons in groups $g$ and $h$
such that, as usual, the order of gluons in each group is held fixed. 
There is then a sum over all gluons in group $g$, where for each 
gluon one implements a NE emission vertex involving the partial
momentum sum $H_j$ for the gluon $j$ from group $h$ which 
lies nearest to the right of the gluon from $g$. That is
\beq
  H_j^{\mu_{g_i}}=\sum_{j>i,\pi_j\in h}h_j^{\mu_{g_i}},
\label{Hjdef}
\eeq
where the individual gluon momenta $h_j$ all carry the Lorentz 
index of the gluon from $g$, so that the remainder term 
correlates this gluon with all the gluons in group $h$ that lie to 
the right of the gluon from $g$. Finally, in \eqn{remvsol}, $\tilde{E}
(\pi)$ denotes a product of eikonal Feynman rules for all gluons 
except the gluon from $g$.

To prove that \eqn{remvsol} solves \eqn{rv}, let us begin by rewriting 
the latter as
\beqa
  \sum_{g \neq h} R_V (g, h) \prod_{f \neq g, h} \E(f) & = &
  \sum_{g \neq h} \frac{H^{\mu_{g_1}}}{p \cdot(G + H)}
  \frac{p \cdot (G + H)}{p \cdot K} \, \, \E(\bar{g}) \, \E(h)
  \prod_{f \neq g, h} E(f) \nonumber \\
  && \, + \, \sum_l \E^{\mu_l} (K) \sum_{\tilde{g} \neq 
  \tilde{h}} \, R_V \left( \tilde{g}, \tilde{h} \right) 
  \prod_{\tilde{f} \neq \tilde{g}, \tilde{h}} E(\tilde{f}) \, ,
\label{rvsola}
\eeqa
In the second term, one may distinguish the cases in which gluon 
$l$ is neither in $g$ nor in $h$ ($l \notin \{ g, h \}$) from those in 
which $l \in g$ or $l \in h$. The second line in \eqn{rvsola} can 
correspondingly be split into three separate terms, and one gets
\beqa
  \sum_{g \neq h} R_V (g, h) \prod_{f \neq g, h} \E(f) & = & 
  \sum_{g \neq h} \frac{H^{\mu_{g_1}}}{p \cdot (G + H)}
  \frac{p \cdot (G + H)}{p \cdot K} \, \, \E(\bar{g}) \, \E(h)
  \prod_{f \neq g, h} \E(f) \nonumber \\
  && \, + \, \sum_{g \neq h} \, R_V (g, h) \sum_{l \notin \{g, h \}} 
  \frac{p \cdot G_l}{p \cdot K} \prod_{f \neq g, h} \E(f) 
  \label{rvsolb} \\
  && \, + \, \sum_{g \neq h} \, \frac{p \cdot (G + H)}{p \cdot K}
  \, \E^{\mu_{g_1}} (G + H) \, R_V (\bar{g}, h )
  \prod_{f \neq g, h} \E(f) \nonumber \\
  && \, + \, \sum_{g \neq h} \, \frac{p \cdot (G + H)}{p \cdot K}
  \, \E^{\mu_{h_1}} (G + H) \, R_V (g, \bar{h} )
  \prod_{f \neq g, h} \E(f) \, . \nonumber
\eeqa
Note that we have repeatedly used \eqn{theor2g} to rewrite 
$E^{\mu_l} (K)$ in terms of group partial momentum sums. 
The fourth line contains all mergings of $g$ and $h$ such that 
the first gluon along the line (next to the hard interaction) comes 
from group $h$. This gluon does not couple to any of the gluons 
in $g$, since $g$ and $h$  are assumed to be distinct; this 
is reflected in the fact that gluons from $g$ couple only to gluons 
from $h$ to their right-hand side, as described above. 

Similarly, the third line contains all mergings such that the first 
gluon along the line is a gluon from $g$, which does not couple 
to any of the gluons in $h$. Note finally that, in the first line, the generalised eikonal identity implies
\beq
  \E(\bar{g}) \, \E(h) = \sum_{\pi}^{(\bar{g}, h)} \, \E(\pi) \, ,
\label{geneikidim}
\eeq
where the right-hand side contains a sum over all mergings of 
$\bar{g}$ and $h$. The validity of \eqn{remvsol} can now be
demonstrated (after some straightforward but tedious algebra) 
as follows. Substituting \eqn{remvsol} into \eqn{rvsolb}, the first 
and third lines combine to give all contributions where the first 
gluon along the line comes from group $g$. This then combines 
with the fourth line to give the sum over all possible mergings, 
with a prefactor in the sum of $p \cdot (G + H)/p \cdot K$. 
Finally, this combines with the second term to give a total 
prefactor
\beq
  \frac{p \cdot (G + H)}{p \cdot K} + \sum_{l \notin \{ g, h \}}
  \frac{p \cdot G_l}{p \cdot K} = 1 \, .
\label{rvsolc}
\eeq
Equating the result to the left-hand side of \eqn{rvsolb}, one finds 
that $R_V(g, h)$ is given by \eqn{remvsol}, which is then, as expected, 
the solution to \eqn{rv}.  

We now proceed in the same way for the contribution of the
propagator correction, $R_P (g,h)$. Our proposed solution
to \eqn{rp} is
\beq
  R_P (g, h) = \sum_\pi^{(g,h)} \sum_{i, \pi_i \in g} 
 \frac{G_i \cdot H_j}{2 \, p \cdot (G_i + 
  H_j)} E(\pi) \, .
\label{rempsol}
\eeq
Again there is a sum over all gluons in group $g$, where $G_i$ is 
the partial momentum sum associated with this gluon. As 
in eq.~(\ref{remvsol}), $H_j$ is the partial momentum sum for
the gluon in group $h$ which lies nearest to the right of the 
gluon from $g$. There is then a product of eikonal Feynman rules
for all gluons in $g$ and $h$, including those which are correlated
by the two-gluon vertex, as is consistent with the first term
in the last line of eq.~(\ref{theor3a2}). 

The proof is directly analagous to the vertex case. 
One first rewrites \eqn{rp} as
\beqa
  \sum_{g \neq h} R_P (g, h) \prod_{f \neq g, h} \E(f) & = &
  \sum_{g \neq h} \frac{G \cdot H}{2 \, p \cdot (G + H)} \,
  \frac{ 2\, p \cdot (G + H)}{2 \, p \cdot K} \, \,  \E(g) \, \E(h)
  \prod_{f \neq g, h} \E(f) \nonumber \\
  && \, + \, \sum_l \E^{\mu_l} (K) \sum_{\tilde{g} \neq \tilde{h}} 
  R_P (\tilde{g},\tilde{h}) \prod_{\tilde{f}} \E(\tilde{f}) \, ,
\label{rpsola}
\eeqa
and again considers the separate cases in which $l \in \{ g, h \}$, 
$l \in g$ and $l \in h$, which allows one to separate the second 
line of \eqn{rpsola} into three sums. Carrying out manipulations 
similar to those applied to \eqn{rvsolb}, substituting \eqn{rempsol}
into the right-hand side, and combining terms, one finds as
expected that $R_P (g,h)$ on the left-hand side of \eqn{rpsola} 
is indeed given by \eqn{rempsol}. 

This concludes our proof of \eqn{theor2}. We have shown that 
the sum over all possible NE gluon insertions exhibits a partial
factorization into contributions arising from distinct groups. 
At NE level, a two-gluon vertex arises which correlates gluons at 
different positions along the line, including in general the case of 
non-adjacent gluons. This two-gluon vertex may either correlate 
pairs of gluons within the same group (in which case this remains a 
group), or it may correlate gluons in different groups (in which 
case the merging of $g$ and $h$ yields a single group). 

Having shown in this section that contributions from different groups
factorise, one may proceed to show that next-to-eikonal corrections
exponentiate. This is the subject of the following section.

\section{Exponentiation for NE matrix elements}
\label{sec:exp}

In \sect{nonabelianexp} we showed that soft gluon corrections 
exponentiate in the eikonal approximation for non-abelian theories. 
Crucial to that derivation was the generalised eikonal identity, 
\eqn{eikid2}, which states that, in summing over eikonal emissions,
contributions from different groups factorise on each external line. 
In this section we extend the argument to next-to-eikonal order, 
using the NE generalisation of \eqn{eikid2}, given by \eqn{theor2}.

By exponentiation at NE order, we mean the generalisation of
\eqn{eikexp2} to give
\beqa
  \sum_G \Big[ c_G^{\E}\E(G) + c_G^{\NE}\NE(G) \Big] & = & 
  \exp \left[ \sum_H \bar{c}_H \left( \E(H) + \NE(H) \right) 
  \right] \nonumber \\
  & = & \exp \left[ \sum_H \bar{c}_H \E(H) \right]
  \left[ 1 + \sum_K \bar{c}_K \NE(K) \right] \, .
\label{NEexptheorem}
\eeqa
The left-hand side of \eqn{NEexptheorem} consists of diagrams
$G$, spanning the external lines, and evaluated up to NE order, 
accompanied by color factors $c_G^{\E}$ ($c_G^{\NE}$) at
eikonal (next-to-eikonal) order. On the right-hand side, one 
has an exponent involving diagrams $H$, where the notations 
$\E(G)$ and $\NE(G)$ have the same meaning as in 
\sect{nonabelianexp}: they represent the momentum part of a 
given soft gluon diagram, with the hard interaction factored off. 
Care is needed in interpreting the sum over diagrams $H$. Those
diagrams involving the next-to-eikonal single gluon and nonlocal
two gluon vertices have topologies which are related to those already
occuring at eikonal order (i.e. for the two-gluon case, we may think 
of this vertex as correlating two single gluon emissions). 
However, diagrams involving the local (seagull-like) two gluon
vertex have topologies which have no counterpart at eikonal order.
For such diagrams $H$, one has $\E(H) = 0$ in the above sum, 
so that the form of the above result is still correct. Note that 
\eqn{NEexptheorem} implies that the modified colour factors for 
the NE diagrams are the same as those for the eikonal diagrams, when
both give a non-vanishing contribution. We will see in what follows 
that this is indeed true.

In the simple case we are considering (two hard partons coupled
by a color singlet interaction), the color factors for all subdiagrams 
commute, since they must all be proportional to the identity 
matrix in the chosen color representation. As in the eikonal case, 
\eqn{NEexptheorem} would contain no information if it were not 
for the fact that the modified color factors $\bar{c}_H$ are zero 
except for a subset of diagrams, which are two-eikonal-line
irreducible. The diagrams on the right-hand side can then be
interpreted as eikonal and next-to-eikonal webs. Note that, up to 
NE order, one may rewrite the NE exponentiation theorem as shown 
in the second line, which consists of a Taylor expansion of the NE 
part of the exponential. Higher order terms in this expansion are 
NNE and so on, and thus one may choose whether to place NE 
webs in the exponent or not. We return to this point later. First, 
we must show that the form of \eqn{NEexptheorem} is indeed 
correct.

To simplify the argument, we begin by neglecting the remainder 
term in  \eqn{theor2}. We also neglect the local two-gluon vertex discussed in sections \ref{spin0} and \ref{spin1/2}, consisting of 
a seagull vertex, plus a spin-dependent correction for fermions. 
Thus, we consider only NE vertices involving the emission of a 
single gluon. By analogy with \eqn{eikprod}, we may then use 
\eqn{theor2} to write, for example,
\beq
  \NE ( H_1H_2 ) = \NE (H_1) \, \E (H_2) + \E (H_1) \, \NE (H_2)
  = \sum_{\pi_A} \sum_{\pi_B} \NE(H_1 \cup_{\pi_A}^{\pi_B}
  H_2) \, ,
\label{NEprod}
\eeq
where $\NE(H_i)$ denotes the next-to-eikonal contribution to 
a subdiagram $H_i$ (which is a sum over propagator and vertex 
corrections). We also introduce the notation $\NE(H_1H_2)$ for
the sum over all possible next-to-eikonal insertions in the diagram
formed by a product of subdiagrams $H_1$ and $H_2$. As in 
\sect{nonabelianexp}, we use $\pi_I$ to represent a permutation 
of the gluons on eikonal line $I$, where the ordering of gluons in 
each group is held fixed. As stated above, we have ignored the 
remainder term $R(H_1,H_2)$ in \eqn{NEprod}, so that only 
one-gluon vertices are present. Each merging of diagrams $H_1$ 
and $H_2$ leads to a new diagram $G$, but this diagram can be 
formed in different ways. As in the eikonal case, one may then write
\beq
  \E(H_1) \, \NE(H_2) + \NE(H_1) \, \E(H_2) = \sum_G \NE(G)
  \, N_{G | H_1 H_2} \, ,
\label{NEprod2}
\end{equation}
where $N_{G | H_1 H_2}$ is a multiplicity factor representing the 
number of ways that $G$ can result from the merging of $H_1$ 
and $H_2$. Note that this is the same factor as in the eikonal case, 
given that we are so far considering only one-gluon vertices, 
and the topology of the merged diagrams is independent of whether 
gluons are next-to-eikonal. Eq.~(\ref{NEprod2}) is easily generalised
to the product of any number of diagrams $H_i$, where each 
diagram may occur $m_{H_i}$ times. Consider first the case in 
which one has $m_H$ copies of a single diagram $H$. Then one 
may write
\beqa
  \NE(H^{m_H}) & = & \sum_{n = 0}^{m_H - 1} \E(H)^n \, 
  \NE(H) \, \E(H)^{m_H - 1- n} \nonumber \\
  & = & m_H \, \NE(H) \, \E(H)^{m_H - 1} \, ,
\label{NEprod3}
\eeqa
where the notation $\NE(H^{m_H})$ naturally generalizes the one
introduced above for $\NE(H_1H_2)$. Here the first line follows 
directly from \eqn{theor2} (in which the remainder term is also 
neglected), and in the second line we have used the fact that the 
factors $\E(H)$ and $\NE(H)$ commute with each other. One may 
then write a general NE product of many diagrams, by analogy with
\eqn{prod3}, as
\beq
  \sum_j \Bigg[ m_{H_j} \, \NE (H_j) \, \E (H_j)^{m_{H_j} - 1}
  \prod_{i \neq j} \E(H_i)^{m_{H_i}} \Bigg] \, = \, \sum_G 
  \NE(G) \, N_{G | H_1^{m_{H_1}} \ldots \, H_n^{m_{H_n}}} \, ,
\label{NEprod4}
\eeq
where $N_{G | H_1^{m_{H_1}} \ldots H_n^{m_{H_n}}}$ is the
multiplicity for forming diagram $G$ from the set of sub-diagrams 
$H_i$, each occurring $m_{H_i}$ times. As discussed above, this 
is the same factor as in the eikonal case.

Following the reasoning of \sect{nonabelianexp}, we now consider 
the expansion of the quantity
\beq
  \exp \left[ \sum_H \bar{c}_H \E(H) \right] \cdot
  \sum_K \bar{c}_K \, \NE(K) = \sum_K \bar{c}_K \, \NE(K)
  \cdot \prod_H \left\{ \sum_n \frac{1}{n!} \Big[ \bar{c}_H
  \E(H) \Big]^n \right\} \,.
\label{neamp1}
\eeq
This corresponds to the second term on the right-hand side of 
\eqn{NEexptheorem}, and we will now show that the modified color factors $\bar{c}_K$ indeed satisfy the same relation (in terms of the 
normal color factors $c_K$) as in the eikonal case. This allows us to
interpret the diagrams $K$ as next-to-eikonal webs, which must also
be two-eikonal line irreducible. 

In order to do this, we follow once again the reasoning of 
\sect{nonabelianexp}. Using \eqn{eikexp3}, we rewrite the product
over diagrams $H$ in \eqn{neamp1} as a sum over multiplicities
$m_H$. Then we examine a single term in the sum over $K$ and 
$\{m_H\}$ that results on the {\it r.h.s.} of \eqn{neamp1}. We
rewrite this term by extracting from the product over diagrams
$H$ the factor corresponding to $H = K$. We get 
\beqa
  \bar{c}_K \NE(K) \prod_H \frac{1}{m_H!} [\bar{c}_H 
  \E(H)]^{m_H} & = & 
  \bar{c}_K \NE(K) \bar{c}_K^{m_K - 1} 
  \frac{\E(K)^{m_K - 1}}{(m_K - 1)!} \prod_{H \neq K}
  \frac{[\bar{c}_H \E(H)]^{m_H}}{m_H!} 
  \label{neamp4} \\
  & = & m_K \, \NE(K) \E(K)^{m_K - 1} \left(\prod_{H \neq K}
  \E(H)^{m_H} \right) \, \prod_J \frac{\bar{c}_J^{m_J}}{m_J!} 
  \nonumber \, ,
\eeqa
where we also relabelled $m_K \rightarrow m_K - 1$, so that diagram
$K$ occurs a total of $m_K$ times.

Substituting this in \eqn{neamp1}, and using \eqn{NEprod4}, one 
finds finally
\beq
  \exp \left[ \sum_H \bar{c}_H \E(H) \right] \cdot 
  \sum_K \bar{c}_K \NE(K) = \sum_G \, \NE(G)
  \sum_{\{m_H\}} N_{G | \{H^{m_H}\}} \left(
  \prod_H \frac{\bar{c}_H^{m_H}}{m_H!} \right) \, .
\label{neamp5}
\eeq
Equating terms of NE order on both sides of \eqn{NEexptheorem}, 
one finds that
\beq
  c_G^{\NE} = \sum_{\{m^H\}} N_{G | \{H^{m_H}\}} \left(
  \prod_H \frac{\bar{c}_H^{m_H}}{m_H!} \right) \, .
\label{colfactNE}
\eeq
This relation between the normal color factors $\{c_G\}$ and the
modified color factors $\{\bar{c}_G\}$ is exactly the same as in 
the eikonal case, given by \eqn{colfacts}. Thus, the modified color 
factors for NE webs are the same as those obtained for the
corresponding eikonal webs. In particular, the modified color 
factors for two-eikonal-line reducible diagrams vanish also at
NE level. This is not surprising, given that the color factor carried 
by a NE one-gluon vertex is the same as that carried by the eikonal 
vertex. The derivation of \eqn{NEexptheorem}, however, relies 
crucially on the non-trivial fact that the momentum-dependent 
contributions from different groups factorise. 

Having demonstrated the argument for the simplest case in which 
correlations between groups (and the seagull vertex) are neglected, 
we now prove the result of \eqn{NEexptheorem} in full. As shown 
in \sect{remainder}, the complete form of the NE generalised eikonal 
identity is given by \eqn{theor2}. The corresponding generalisation 
of \eqn{NEprod4} is then 
\beqa
  && \sum_G \NE(G) \, N_{G | \{H_1^{m_{H_1}} \ldots \, 
  H_n^{m_{H_n}}\}} \, = \,
  \sum_j \Bigg[ m_{H_j} \, \NE (H_j) \, \E (H_j)^{m_{H_j} - 1}
  \prod_{i \neq j} \E(H_i)^{m_{H_i}} \Bigg] \nonumber \\
  && \qquad \quad + \,
  \sum_j \Bigg[ m_{H_j} (m_{H_j} - 1) \, R(H_j, H_j) \, 
  \E(H_j)^{m_{H_j} - 2} \prod_{i \neq j} \E(H_j)^{m_{H_j}} 
  \Bigg] \label{NEprod4b} \\
  && \qquad \quad +\, \sum_{i \neq j} \Bigg[ 
  m_{H_i} m_{H_j} \, R(H_i, H_j) \, 
  \E(H_i)^{m_{H_i} - 1} \, \E(H_j)^{m_{H_j} - 1}
  \prod_{k \neq i, j} E(H_k)^{m_{H_k}} \Bigg] \, .
  \nonumber
\eeqa
This formula includes three types of NE contribution:
\begin{itemize}
\item Single gluon vertices: as before, these reside in the first 
term on the right-hand side, contributing to the factor $\NE(H_j)$.
\item The remainder term of \sect{remainder}: this contributes to 
the second and third terms on the right-hand side, via factors of
$R(H_j, H_i)$.
\item The local NE two-vertex of fig. 9 (in the scalar case) and
\eqn{theor1} (in the fermion case). As discussed in \sect{spin0}, 
this vertex may either couple gluons from the same group $G$, or 
it may couple gluons from two different groups $G$ and $H$. In the 
former case, one includes it in the first term on the right-hand side 
of \eqn{NEprod4b}; in the latter case, one regards it as part of the 
remainder term, since it couples together two groups, which would 
remain separate if the local two-vertex were not present. As 
remarked already above, $\E(H) = 0$ for such diagrams.   
\end{itemize}
In those terms which involve the remainder function $R$, we explicitly 
separated the contributions from correlations between two instances 
of the same subdiagram (involving $R(H_j, H_j)$), and correlations
between different subdiagrams (involving $R(H_i, H_j)$). The 
factors in each sum come from the number of ways in which these 
combinations occur. Now, to proceed along the lines of the discussion 
leading to \eqn{colfactNE}, we consider the expansion of the 
quantity
\beq
  {\cal E} \equiv \exp \left[ \sum_H \bar{c}_H \E(H)^{m_H} 
  \right] \left\{ \sum_K \bar{c}_K \NE(K) + \sum_{K,L} 
  R(K, L) \bar{c}_K \bar{c}_L \right\} \equiv 
  {\cal E}_1 + {\cal E}_2 \, .
\label{neamp1b}
\eeq
The first term in square brackets leads to \eqn{neamp4}. The 
second term contains correlations between groups, and one may 
again separate out the term with $K = L$ in the sum, to obtain
\beq
  {\cal E}_2 = \left[ \sum_K \bar{c}_K^2 R(K, K) + 
  \sum_{K \neq L} \bar{c}_K \bar{c}_L R(K,L) \right]
  \prod_H \left\{ \sum_n \frac{1}{n!} \Big[ \bar{c}_H
  \E(H) \Big]^n \right\} \, .
\label{neamp3b}
\eeq
As above, one may rewrite the expanded exponential in terms of a 
sum over the possible multiplicities  $\{m_H\}$. In doing so, one 
may take the terms involving $K$ and $L$ out of the product over 
diagrams $H$. It is also convenient to relabel $m_K$ and $m_L$, 
where appropriate, so that each group $H$ occurs $m_H$ times
in each sum. Eq.~(\ref{neamp3b}) then becomes
\beqa
  {\cal E}_2 & = & \sum_{\{m_H\}} \left[ \sum_K \bar{c}_K^2
  R(K,K) \, \frac{ [ \bar{c}_K \E(K) ]^{m_K - 2}}{(m_K - 2)!}
  \prod_{H \neq K} \frac{ [ \bar{c}_H \E(H) ]^{m_H}}{m_H!}
  \right. \nonumber \\
  & + & \left. \, \sum_{L \neq K} \bar{c}_K \bar{c}_L
  R(K,L) \, \frac{ [ \bar{c}_K \E(K) ]^{m_K - 1} 
  [ \bar{c}_L \E(L)]^{m_L - 1}}{(m_K - 1)! (m_L - 1)!}
  \prod_{H \neq K,L} \frac{[\bar{c}_H \E(H)]^{m_H}}{m_H!}
  \right] \, .
\label{neamp4b}
\eeqa
Combining \eqn{neamp4b} with \eqn{neamp4} gives
\beqa
  {\cal E} & = & \sum_{\{m_H\}} \left[\sum_K m_K \, \NE(K)
  \, \E(K)^{m_K - 1} \prod_{H \neq K} \E(H)^{m_K} 
  \right. \nonumber \\
  & + & \sum_K m_K (m_K - 1) R(K,K) \, \E(K)^{m_K - 2} 
  \prod_{H \neq K} \E(H)^{m_H} \nonumber \\
  & + & \left. \, \sum_{K,L} m_K m_L R(K,L) \, 
  \E(K)^{m_K - 1} \, \E(L)^{m_L - 1} \prod_{H \neq K, L}
  \E(H)^{m_H} \right] \prod_J \frac{\bar{c}_J ^{m_J}}{m_J!} \, .
\label{neamp5b}
\eeqa
The prefactor in brackets has precisely the form required by 
\eqn{NEprod4b}, and one finds
\beqa
  {\cal E} & = & \exp \left[ \sum_H \bar{c}_H \, \E(H)^{m_H} 
  \right] \left\{ \sum_K \bar{c}_K \, \NE(K) + \sum_{K, L} 
  \bar{c}_K \bar{c}_L \, R(K,L) \right\} \nonumber \\
  & = & \sum_G \NE(G) \sum_{\{m_H\}} N_{G | \{H^{m_H}\}}
  \left(\prod_H \frac{\bar{c}_H^{m_H}}{m_H!} \right) \, .
\label{neamp6b}
\eeqa
Comparing again with \eqn{NEexptheorem}, one finds precisely the 
same color relation already expressed in \eqn{colfactNE}. This 
completes the proof of the relation
\beq
  \sum_G c_G \Big[\E(G) + \NE(G) \Big] = 
  \exp \left[ \sum_H \bar{c}_H \, \E(H) \right] \left[1 + 
  \sum_K \bar{c}_K \, \NE(K) + \sum_{K, L}
  \bar{c}_K \bar{c}_L \, R(K,L) \right] \, ,
\label{NEexptheorem2}
\eeq
as given in the second line on \eqn{NEexptheorem}. Here however,
in a slight abuse of notation, we have explicitly extracted the 
remainder term in the next-to-eikonal contribution on the right-hand 
side. The modified color factors in this equation are the same as 
those obtained in the eikonal case (for those topologies that
occur already at eikonal order). 
As stated above, the NE 
terms can be moved into the exponent up to NE order. The content 
of this theorem can be summarised as follows.
\begin{quote}
{\it Matrix elements containing gluon emissions computed up 
to next-to-eikonal level exponentiate, and the exponent contains 
a sum over webs. The NE webs belong to two classes: (i) 
two-eikonal-line irreducible subgraphs including at most one NE 
emission vertex, and (ii) novel two-eikonal-line irreducible graphs 
resulting from joining together two eikonal webs with a two-gluon 
vertex.}
\end{quote}
We emphasize again that the exponentiation theorem of 
\eqn{NEexptheorem} applies only to contributions from soft 
gluon emissions which are factorizable from the hard interaction.
In order to gain complete control on NE logarithms in the presence 
of hard virtual gluons, one must still include the non-factorizable 
corrections denoted by ${\cal M}_r$ in \eqn{Mstruc}. Furthermore,
in order to gain control on the threshold expansion at NLO, one 
must also take into account hard collinear gluon emissions, which 
lead to threshold logarithms of the same order as NE terms. 
We leave the study of these further contributions to future work.

\subsection{Next-to-eikonal Feynman rules}
\label{NErulessum}

In the preceding sections we have classified all next-to-eikonal
contributions in diagrams for scattering amplitudes involving two 
colored lines, and shown that they exponentiate in terms of 
(next-to-)eikonal webs. In this section we provide a set of effective 
Feynman rules with which these webs can be efficiently computed.

Our rules are defined elaborating upon those of the eikonal 
approximation. As usual, one begins by identifying eikonal webs 
through the criterion of two-eikonal-line irreducibility. The 
momentum space part of each web is computed using the 
Feynman rules for the eikonal approximation, given in \eqn{eikrule}, 
and the corresponding color factor can be computed either 
recursively \cite{Gatheral:1983cz,Frenkel:1984pz,Berger:2003zh} 
or directly, using the replica trick \cite{Laenen:2008gt}. 

At next-to-eikonal order, three new types of webs appear. The first
type consists of diagrams obtained by replacing precisely one vertex 
or propagator in an eikonal web with a next-to-eikonal one, which 
must be done for each vertex or propagator in the eikonal web. 
The second type is found by constructing new two-eikonal-line
irreducible diagrams, using eikonal Feynman rules, and precisely 
one NE two-gluon vertex. Together, these two types of webs
constitute the second term in the last factor in \eqn{NEexptheorem2}. 
The third type of web corrsponds to the third term in the last 
factor in \eqn{NEexptheorem2}. At order $k$, one must consider 
all pairs of eikonal webs of lower orders $i, j$ such that $i + j = k$, 
and for each such pair implement the correlation in 
\eqns{remvsol}{rempsol}. The rules for the color factors 
of each next-to-eikonal type are given in the previous section.
Let us now treat each type in more detail, giving the effective
Feynman rules involved. Note that we will use a photon-like symbol
to represent a generic gauge boson, which may be abelian or non-abelian.

For the first type of webs, we can formulate a one-gluon vertex rule, 
combining the second term on the right-hand side of \eqn{NEvertex2}
with the propagator correction of \eqn{NEprop}. For scalar partons,
the result is
\beq
  \includegraphics[width=2cm]{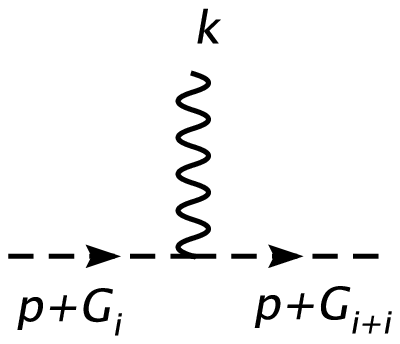}   \qquad
  {\raisebox{3ex}{\ensuremath{ = \, \, \, 
  t^A \Big( \frac{2 G_i^\mu - k^\mu}{2 p \cdot G_i} - 
  \frac{G_i^2 \, p^\mu}{2 (p \cdot G_i)^2} \Big) 
  \, = \, 
  \frac{t^A}{2 p \cdot G_i} \Big( G_i^\mu + G_{i + 1}^\mu 
  - \frac{G_i^2 \, p^\mu}{p \cdot G_i}  \Big) \, .}}}
\label{1gv}
\eeq
Note that because we provide rules that are only to be used in web
diagrams, NE insertions and the partial momentum sums that they
contain are automatically restricted to a single group, unless otherwise
stated. The spinor equivalent of \eqn{1gv} is  
\beq
  \includegraphics[width=2cm]{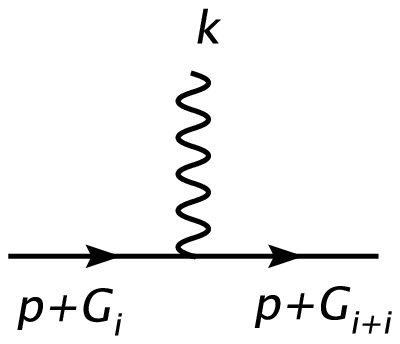}   \qquad
  {\raisebox{3ex}{\ensuremath{ = \, \, \, 
  t^A \Big( \frac{2 G_i^\mu + \gamma^\mu \slashed{k}}{2 p 
  \cdot G_i} - \frac{G_i^2 \, p^\mu}{2 (p \cdot G_i)^2} \Big) 
  \, = \, \frac{t^A}{2 p \cdot G_i}  \Big( G_i^\mu + 
  G_{i + 1}^\mu + \slashed{k} \, \gamma^\mu - k^\mu - 
  \frac{G_i^2 \, p^\mu}{p \cdot G_i} \Big) \, ,}}}
\label{1gvspin}
\eeq
which follows from \eqns{NEv2}{theor1}. On the right-hand-side, 
we have shown explicitly how the vertex part of this Feynman rule 
can be thought of as the sum of a spin-zero piece and a magnetic
moment vertex, as remarked in \sect{spin1/2}.

For the second type of web we merely need to give the two-gluon 
vertex rules. Any two-eikonal-line irreducible diagram one builds 
with one such vertex and any number of eikonal vertices and 
propagators is a next-to-eikonal web. For the case of a scalar 
emitter, one has the NE vertex rule
\beq
  \includegraphics[width=2cm]{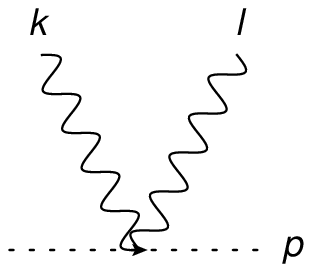}   \qquad
  {\raisebox{3ex}{\ensuremath{ = \, \, \, 
  \frac{g^{\mu \nu} \, \{ t^A, t^B \}}{2 p \cdot (k + l)} \, ,}}}
\label{2gv}
\eeq
where we have included the relevant eikonal propagator. 
Note that $t^A$ and $t^B$ are color generators in the 
appropriate representation and that the numerator corresponds 
in fact to the exact vertex of scalar perturbation theory, as given 
in \fig{2gluonscalar}. For the equivalent spinor vertex one has
\beq
  \includegraphics[width=2cm]{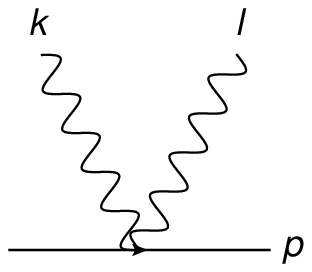}   \qquad
  {\raisebox{3ex}{\ensuremath{ = \, \, \,
  \frac{g^{\mu \nu} \, \{ t^A, t^B \}}{p \cdot (k + l)} \,
  + \, \frac{\left[ \gamma^\mu, \gamma^\nu \right] \, 
  \left[ t^A, t^B \right]}{2 p \cdot (k + l)} \, ,}}}
\label{2gvspinfig}
\eeq
where the second term is the chromomagnetic moment 
vertex referred to in \sect{spin1/2}. The momenta in 
\eqns{2gv}{2gvspinfig} are single gluon momenta in the 
abelian case. In the non-abelian case, one must use the appropriate
partial momentum sums corresponding to each gluon: if the gluons
both originate from group $G$ (with numbers $i$ and $i + 1$), then
one should use $k = G_i$ and $l = G_{i + 1}$. If the gluons come 
from different groups $G$ (gluon number $i$) and $H$ (gluon 
number $j$), one should evaluate the Feynman rule using the 
appropriate partial momentum sums $G_i$ and $H_j$. Note that 
in this case it is more natural to absorb this Feynman rule as part 
of the remainder vertex.

Finally, in the third type of web we must correlate any two gluons 
in different groups, by means of an additional two-gluon vertex
rule that corresponds to the remainder terms derived in
\sect{remainder}. This has the form
\beq
  \includegraphics[width=2cm]{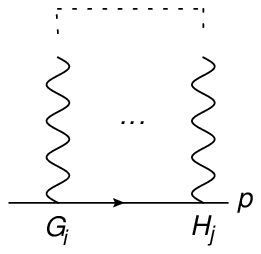}   \qquad
  {\raisebox{3ex}{\ensuremath{\, \, \, = \, \, \,
  t^A \otimes t^B \left( \frac{- \, G_i^\mu \, p^\nu \, 
  (p \cdot H_j) \, - \, H_j^\nu \, p^\mu \, (p \cdot G_i)\,
  + \, p^\mu p^\nu \, G_i \cdot H_j}{2 p \cdot(G_i + H_j) \, 
  p \cdot G_i \, p\cdot H_j} \right) \, .}}}
\label{2gvrem}
\eeq
Here the tensor product in the color factor indicates the fact that 
the color matrices correspond to gluons on different parts of the 
external line, belonging to sets that in the eikonal approximation 
would be different groups. Recall that $G_i$ and $H_j$ denote 
the partial momentum sums for gluons at position $i$ ($j$) in 
group $g$ ($h$), where it is understood that $j > i$, as described 
in \sect{remainder}. 

The above results match those obtained in~\cite{Laenen:2008gt}.  
It is useful to relate our results here with those of that paper where
they were derived using path integral methods. This comparison is 
the subject of the following section. As it is not immediately relevant 
to the demonstration of the application of the Feynman rules in 
\sect{sec:DY}, the reader may omit the following section without 
interrupting the flow of the paper.

\section{Relation to the path integral formalism}
\label{NErulespath}

In this section, we briefly review the method and results
of~\cite{Laenen:2008gt}, which first demonstrated the 
exponentiation of a class of NE corrections using a path integral 
approach, and derived effective Feynman rules for NE emissions. 
A comparison of the two approaches is useful for two reasons. 
First, it offers a non-trivial check of the effective Feynman rules 
derived here and in~\cite{Laenen:2008gt}; second, the present 
derivation reveals certain subtleties associated with NE Feynman 
rules, that were not fully addressed in the path integral approach. 
For the sake of completeness, we will briefly describe the path 
integral approach, referring the reader to Ref.~\cite{Laenen:2008gt} 
for a full exposition. We shall moreover restrict ourselves to the case 
of scalar emitting particles, which will be sufficient to clarify the 
correspondence between the two approaches.

The starting point in~\cite{Laenen:2008gt} was to separate the 
gauge field into hard and soft modes, and then consider (as we 
do here) a scattering amplitude for the production of $n$ final 
state hard particles, each of which may emit further soft radiation. 
For the simple case of abelian gauge theory with scalar emitting 
particles, such a scattering amplitude has the factorised form
\beq 
  {\cal M} (p_1, \ldots, p_n) = \int{\cal D} A^\mu_s
  \int \prod_{i = 1}^n d x_i \, \, H(x_1, \ldots, x_n)
  \prod_{k = 1}^n \langle p_k | (S_{\rm quad} - {\rm i}
  \varepsilon)^{-1} | x_k \rangle \, p_k^2 \, \,
  {\rm e}^{ {\rm i} S[A_s]} \, .
\label{amp1}
\eeq
Here $A^\mu_s$ is the soft gauge field (with action $S[A_s]$), 
and $H(x_1,\ldots, x_n)$ is the factor representing the hard 
interaction producing the emitting scalar particles at positions 
$x_i$. The factors $\langle p_k | (S_{\rm quad} - {\rm i} 
\varepsilon)^{ - 1} | x_k \rangle$ represent propagators for 
a scalar particle in a soft background gauge field (with
$S_{\rm quad}$ denoting the quadratic operator for the scalar field 
in the Lagrangian). These propagators are computed between 
states of given initial position (the point $x_k$ at which the particle 
is created by the hard interaction) and given final momentum $p_k$.
The explicit factors of $p_k^2$ in \eqn{amp1} truncate the free
propagators associated with the external legs. As noted above, 
the factorization represented by \eqn{amp1}, which is exact in the
eikonal approximation, does not encompass all NE corrections to 
the amplitude, since there are contributions arising from virtual 
diagrams containing hard gluons which do not factorize in the 
form of \eqn{amp1}. Such contributions can be traced and 
organized using an appropriate extension of Low's 
theorem~\cite{DelDuca:1990gz}, and their treatment is left to 
future work.

The propagator factors in \eqn{amp1} can be represented as 
first-quantised path integrals~\cite{Strassler:1992zr,vanHolten:1995ds}. One finds
\beq
  p_k^2 \, \langle p_k | (S_{\rm quad} - {\rm i} \varepsilon)^{- 1} 
  | x_k \rangle = e^{- {\rm i} p_k \cdot x_k} f_k (\infty) \, ,
\label{path1}
\end{equation}
where
\beqa
  f_k (\infty) & = & \int_{y_k(0) = 0}{\cal D} y_k \, \exp
  \left\{ {\rm i} \int_0^\infty d t \left[ \frac{1}{2} 
  \dot{y}_k^2 (t) + \left(p_k + \dot{y}_k (t) \right) \cdot A_s 
  \left(x_k + p_k t + y_k(t) \right) 
  \right. \right. \nonumber \\
  && \quad \left. \left. + \, \,  \frac{{\rm i}}{2} \, 
  \partial \cdot A_s \left( x_k + p_k t + y_k(t) \right) 
  \right] \right\} \, .
\label{path2}
\eeqa
Here $y_k(t) $ is the fluctuation about the classical straight-line
path associated with the $k$-th scalar particle, and the path integral 
is over all such fluctuations, subject to the boundary conditions of 
given initial position $x_k$ and final momentum $p_k$. The implementation of these boundary conditions is discussed in more 
detail in Ref.~\cite{Laenen:2008gt}. Substituting \eqn{path2}
into \eqn{amp1} yields
\beqa
  {\cal M} (p_1, \ldots, p_n) & = & \int{\cal D} A^\mu_s 
  \int \prod_{i = 1}^n d x_i \, \int \prod_{k = 1}^n {\cal D} y_k
  \, \, H(x_1, \ldots, x_n) \, {\rm e}^{ {\rm i} S [A_s]} \,
  {\rm e}^{- {\rm i} (x_1 \cdot p_1 + \ldots + x_n \cdot p_n)}
  \nonumber \\
  && \quad \times \, \prod_k \exp \left[ {\rm i} \int_0^\infty
  d t \left(\frac{1}{2} \dot{y}_k^2 (t) + \left(p_k + \dot{y}_k(t) 
  \right) \cdot A_s + \frac{{\rm i}}{2} \partial \cdot A_s
  \right)\right] \, .
\label{amp2}
\eeqa
We see that the scattering amplitude has taken the form of a 
generating functional for a quantum field theory of the soft gauge 
field $A^\mu_s$. The exponent in the second line contains terms 
linear in $A_s^\mu$, which act as sources. These sources 
are located on the external lines, and the path integral 
over $A_s^\mu$ generates all diagrams that connect these 
sources. Thus formulated, the exponentiation of soft photon
corrections (in terms of connected subdiagrams) is precisely
equivalent to the well-known exponentiation of connected 
diagrams in quantum field theory~\cite{Laenen:2008gt}.

One advantage of this approach is the clear physical interpretation 
in terms of the worldline trajectories of the emitting hard particles. 
To see which diagrams exponentiate, one must calculate the soft 
gauge field Feynman rules that result after carrying out the path
integrations over the fluctuations $y_k (t)$ in \eqn{amp2}. This 
was achieved in Ref.~\cite{Laenen:2008gt} by systematically 
expanding about the classical straight-line trajectory $x_k (t) = 
x_k + p_k t$, for each external particle moving in the direction
of $p^\mu_k$. Note that the classical trajectory corresponds to 
the eikonal approximation in which the emitting particles do not 
recoil. Following Ref.~\cite{Laenen:2008gt}, we rescale the hard
momenta $p_k$ by means of a scaling variable $\lambda$, so 
that, for large $\lambda$, terms of order $\lambda^0$ and 
$\lambda^{-1}$ constitute the eikonal and the next-to-eikonal 
approximations, respectively. With this choice, the expansion of 
\eqn{amp2} up to NE order reads
\beq
  {\cal M} (p_1, \ldots, p_n) = \int{\cal D} A^\mu_s 
  \int \prod_{i = 1}^n d x_i \, H (x_1, \ldots, x_n) \, 
  {\rm e}^{ {\rm i} S [A_s]} \, {\rm e}^{- {\rm i} (x_1 \cdot p_1
  + \ldots + x_n \cdot p_n)} \, \prod_k F_k (n_k, A_s) \, ,
\label{amp3b}
\eeq
where $n^\mu$ is the direction of the hard line carrying 
momentum $p^\mu$, and
\beqa
  && F_k (n_k, A_s) = \exp \Bigg\{ {\rm i} \int_0^\infty d t
  \left[ n_k \cdot A_s (n_k t) + \frac{{\rm i}}{2 \lambda}
  \partial \cdot A_s (n_k t) + \frac{{\rm i}}{2 \lambda} t \,
  n_{k \mu} \Box A^\mu_s (n_k t) \right] \nonumber \\
  && \quad + \, \int_0^\infty d t \int_0^\infty d t' 
  \left[ - \frac{{\rm i}}{2 \lambda} \delta(t - t') \, g_{\mu \nu}
  \,A_s^\mu (n_k t) \, A_s^\nu (n_k t') - \frac{{\rm i}}{\lambda}
  \theta(t - t') \, n_{k \mu} \, g_{\sigma \nu} \left[
  \partial^\sigma A^\mu_s (n_k t) \right] \right. \nonumber \\
  && \quad \left. \times \, \left[A^\nu_s (n_k t') \right] -
  \frac{{\rm i}}{2 \lambda} \text{min} (t, t') \, n_{k \mu} \, 
  n_{k \nu} g^{\sigma \tau} \left[ \partial_\sigma
  A^\mu_s (n_k t) \right] \left[ \partial_\tau A^\nu_s (n_k t') 
  \right] \right] \Bigg\} \, . 
\label{amp3bf}
\eeqa
Eq.~(\ref{amp3b}) now has the form of a path integral over 
$A^\mu_s$ only, and the Feynman rules that couple the soft 
gauge field to the external lines can in principle be read off from 
the exponent. To that end, it is conventional to first perform a 
Fourier transform to momentum space. The practical procedure
for this, in the present context, depends upon whether the gauge 
field is abelian or not.

Consider first briefly the case where the gauge field is abelian, 
so that the source terms for $A_s^\mu$ in \eqn{amp3b} commute. 
One may perform the Fourier transform in the exponent in the second 
line,  as was done in Ref.~\cite{Laenen:2008gt}\footnote{We note 
some misprints in~\cite{Laenen:2008gt}, involving the wrong 
correlators being used in eqs.~(B.26) and (B.28) of that paper. 
The results as given here are correct.}. For a line with direction 
$n^\mu$, one obtains
\beqa
  && F (n, A_s) = \exp \Bigg\{ \int \frac{d^d k}{(2 \pi)^d} \, 
  \tilde{A}_s^\mu (k) \left[ - \frac{n_\mu}{n \cdot k} +
  \frac{1}{\lambda} \left(\frac{k_\mu}{2 n \cdot k} - 
  k^2 \frac{n_\mu}{2 (n \cdot k)^2} \right) \right]
  \nonumber \\
  && \quad + \int \frac{d^d k}{(2 \pi)^d} \int 
  \frac{d^d l}{(2 \pi)^d} \, \tilde{A}_s^\mu (k) \, 
  \tilde{A}_s^\nu (l) \, \frac{1}{\lambda} \left( 
  \frac{g_{\mu\nu}}{2 n \cdot (k + l)} - 
  \frac{n_\nu l_\mu n \cdot k + n_\mu k_\nu n 
  \cdot l}{2 (n \cdot l)(n \cdot k) [n \cdot (k + l)]} \right.
  \nonumber \\
  && \left. \quad + \, \frac{(k \cdot l) n_\mu n_\nu}{2 (n 
  \cdot l)(n \cdot k) [n \cdot (k + l)]} \right) \Bigg\} \, ,
\label{pathrules2}
\eeqa
with $\tilde{A}_s^\mu (k)$ the soft gauge field in momentum 
space. The first line in \eqn{pathrules2} contains one-photon 
emission vertices at both eikonal and next-to-eikonal accuracy.  
When combined, these correspond exactly to the rule of \eqn{1gv}. 
The second and third lines of \eqn{pathrules2} contain contributions 
to a two-photon vertex. Combining these terms results precisely 
in \eqn{2gvrem}, after removing color matrices and with the 
replacement $G_i \rightarrow k$, $H_j \rightarrow l$. 

Having seen how the correspondence between the two approaches 
works in the case of abelian gauge fields, we now consider the 
non-abelian case. As in Ref.~\cite{Laenen:2008gt}, we assume a 
color singlet hard interaction with two outgoing lines.  Performing
the path integral over fluctuations $y_k(t)$, we get once again
\eqn{amp3b}, with $n = 2$. Now however the soft photon fields
$A^\mu_s$ in \eqn{amp3bf} must be replaced with matrix valued
soft gluon fields ${\bf A}^\mu_s \equiv A^\mu_{s, a} T^a$, which
do not commute; as a consequence, the exponential in 
\eqn{amp3bf} must be replaced with a path ordered exponential.

\begin{figure}
  \begin{center}
    \scalebox{1.0}{\includegraphics{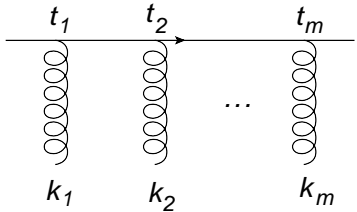}}
    \caption{A series of emissions from an external line in position
      space. In a non-abelian theory, emissions are ordered, and path
      ordering of the color matrices is equivalent to time ordering of 
      the emissions, with $t_1 < t_2 < \ldots < t_m$.}
    \label{Tordered}
  \end{center}
\end{figure}

To illustrate the consequences of this fact, consider for example
fig.~\ref{Tordered}, which shows a number of eikonal emissions 
from the same external line, which would be generated by the first 
term in the exponent of \eqn{amp3bf}. 
For an abelian gauge field emitted from a hard particle moving in
the direction $n^\mu$, a line dressed as in fig.~\ref{Tordered} 
would lead, after a Fourier transform to momentum space, to a 
factor in the amplitude of the form
\beq
  L_{\rm eik} (n, k_i) \, = \,
  \frac{1}{m!} \left[ {\rm i} \int_0^\infty d t_1 n^{\mu_1}
  {\rm e}^{{\rm i} n \cdot k_1 t_1}
  \ldots {\rm i} \int_0^\infty d t_m n^{\mu_m} 
  {\rm e}^{{\rm i} n \cdot k_m t_m} \right] 
  \tilde{A}_{\mu_1} (k_1) \ldots \tilde{A}_{\mu_m} (k_m) \, ,
\label{torder1}
\eeq
where henceforth for brevity we omit the subscript $s$ on the
soft gauge fields. The exponential factors in \eqn{torder1} arise from 
the Fourier transform. Carrying out the $t_i$ integrals\footnote{We 
implicitly assume the prescription $n \cdot k_i + i \varepsilon$ to 
ensure convergence of the $t_i$ integral.}, the coefficient of the 
product of photon fields $\tilde{A}_{\mu_1}(k_1) \ldots 
\tilde{A}_{\mu_n} (k_n)$ is
\beq
  C_{\rm eik}^{\mu_1, \ldots, \mu_m} (n, k_i) 
  \, = \, \frac{1}{m!}  \,
  \prod_{i = 1}^m \left( - \frac{n^{\mu_i}}{n \cdot k_i} \right) \, ,
\label{torder2}
\eeq
which is the familiar product of eikonal Feynman rules for $m$ 
photon emissions, with uncorrelated momenta. Recall that in the
diagrammatic approach the same result arises from the application 
of the eikonal identity, \eqn{eikonalid}, after relabelling photon 
momenta and summing over permutations. In the non-abelian case, 
the path ordering prescription modifies \eqn{torder1} to yield
\beqa
  {\bf L}_{\rm eik} (n, k_i) & = &
  \left[ {\rm i} \int_0^\infty d t_1 n^{\mu_1} {\rm e}^{{\rm i} 
  n \cdot k_1 t_1} \, {\rm i} \int_{t_1}^\infty d t_2 n^{\mu_2}
  {\rm e}^{ {\rm i} n \cdot k_2 t_2} \ldots  {\rm i} 
  \int_{t_{m - 1}}^\infty d t_m n^{\mu_m} {\rm e}^{{\rm i} 
  n \cdot k_m t_m} \right] \nonumber \\
  && \qquad \times \, \tilde{{\bf A}}_{\mu_1}(k_1) \ldots
  \tilde{{\bf A}}_{\mu_m}(k_m) \, ,
\label{torder3}
\eeqa
where the limits of integration have been modified to reflect the 
time-ordering of the gluon fields. Carrying out the $t_i$ integrals, 
we find that the coefficient of $\tilde{{\bf A}}_{\mu_1}(k_1)
\ldots \tilde{{\bf A}}_{\mu_n}(k_n)$ becomes
\beq
  {\cal C}_{\rm eik}^{\mu_1 \ldots \mu_m} (n, k_i) \, = \,
  \frac{(- 1)^m \, n^{\mu_1} \ldots n^{\mu_m}}{n \cdot k_m \,
  n \cdot(k_m + k_{m - 1}) \, \ldots \, n \cdot (k_1 + 
  \ldots  + k_m)} \equiv \prod_i^m \left(- \frac{n^{\mu_i}}{n 
  \cdot K_i} \right) \, .
\label{torder4}
\eeq
One observes that the effect of path-ordering is to replace single
gluon momenta in the denominators of the eikonal Feynman
rules with partial momentum sums. The sums are restricted to 
gluon groups since, as shown in \sect{nonabelianexp}, contributions
from different groups of gluon emissions factorise. This indeed 
corresponds to the effective eikonal Feynman rules summarized in 
the previous section.

\begin{figure}
  \begin{center}
    \scalebox{1.0}{\includegraphics{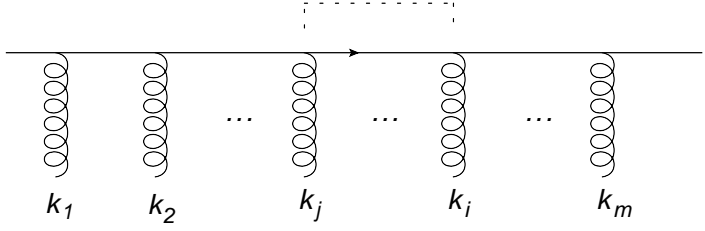}}
    \caption{Schematic form of diagrams involving a two-gluon
      vertex. Additional eikonal emissions may occur between the two
      correlated gluons. }
    \label{1gvjoin}
  \end{center}
\end{figure}

Let us now analyse the remaining terms in the exponent of
the non-abelian version of \eqn{amp3bf} along similar lines, in
order to recover the NE Feynman rules. Recall that the rules are 
different depending on whether gluons are in the same group or in 
different groups. Let us therefore first consider the case of gluon 
emissions along an external line (as in fig.~\ref{Tordered}), where 
the gluons all belong to the same group. We will see that different 
subsets of terms in the exponent of \eqn{amp3bf} combine to give 
the Feynman rules obtained in the diagrammatic analysis of the 
previous sections. Let's begin by considering, in coordinate space, 
the second term on the first line and the second term on the second 
line of \eqn{amp3bf}, in the nonabelian case. They read
\beqa
  {\bf E}_{\rm NE}^{(1+2)} (n, x_i) & \equiv &  
  \int_0^\infty d t \left[ - \frac{1}{2} \partial \cdot 
  {\bf A}(n t) \right] \nonumber \\
  && - \, {\rm i} \int_0^\infty d t \int_0^\infty 
  d t' \, \theta(t - t') \, n^\mu \, g_{\sigma \nu} 
  \left[ \partial^\sigma {\bf A}^\mu(n t) \right] 
  \left[{\bf A}^\nu(n t') \right] \, ,
\label{torder5}
\eeqa
where we have set $\lambda = 1$ in \eqn{amp3bf}. When the
exponent is expanded (including the eikonal terms we discussed
above), the first term in \eqn{torder5} generates diagrams of the 
form shown in fig.~\ref{Tordered}, where one of the emitted gluons 
has one NE one-gluon vertex. If this is the $j^{\rm th}$ gluon, the corresponding momentum space expression is
\beqa
  {\bf L}_{\rm NE}^{(1, j)} (n, k_i) & = &
  \left[ {\rm i} \int_0^\infty d t_1n^{\mu_1} {\rm e}^{{\rm i} 
  n \cdot k_1 t_1} \ldots \left(- \frac{\rm i}{2} \right)
  \int_{t_{j - 1}}^\infty d t_j k_j^{\mu_j} {\rm e}^{{\rm i} 
  n \cdot k_j t_j} \ldots {\rm i} \int_{t_{m - 1}}^\infty 
  d t_m n^{\mu_m} {\rm e}^{ {\rm i} n \cdot k_m t_m}
  \right] \nonumber \\
  && \quad \times \, \tilde{{\bf A}}_{\mu_1}(k_1)
  \ldots \tilde{{\bf A}}_{\mu_m}(k_m) \, .
\label{1gvpath2}
\eeqa
Carrying out the $t_i$ integrals, the coefficient of the product
of gauge fields in \eqn{1gvpath2} becomes
\beq
  {\cal C}_{{\rm NE}, (1,j)}^{\mu_1, \ldots, \mu_m} (n, k_i) 
  \, = \, \frac{k^{\mu_j}_j}{2 n \cdot K_j} \prod_{k \neq j}
  \left(- \frac{n^{\mu_k}}{n \cdot K_k} \right) \, .
\label{1gvpath3}
\end{equation}
The second term in \eqn{torder5}, on the other hand, leads to 
diagrams of the type shown in fig.~\ref{1gvjoin}.
In such diagrams, a pair of gluons is correlated through the
appropriate two-gluon vertex. Note that the two gluons need not 
be adjacent on the external line, as indeed \eqn{torder5} allows:
path ordering restricts the second field merely to lie to the right of 
the first in the ordered gauge field product, and further eikonal 
emissions may lie in between. More specifically, taking the left-most
correlated gluon to be the $j^{th}$ one, fig.~\ref{1gvjoin} leads to 
the momentum space expression
\beqa
  {\bf L}_{\rm NE}^{(2,j)} (n, k_i) & = &
  - \sum_{i > j} \Bigg[ {\rm i} \int_0^\infty d t_1 n^{\mu_1}
  {\rm e}^{{\rm i} n \cdot k_1t_1} \ldots {\rm i} 
  \int_{t_{j - 1}}^\infty d t_j {\rm e}^{{\rm i} n
  \cdot k_j t_j} \ldots {\rm i} \int_{t_{i - 1}}^\infty d t_i 
  n^{\mu_i} k_i^{\mu_j} {\rm e}^{{\rm i} n \cdot k_i t_i}
  \ldots  \nonumber  \\
  && \times  \, {\rm i} \int_{t_{m - 1}}^\infty d t_m 
  n^{\mu_m} {\rm e}^{{\rm i} n \cdot k_m t_m} \Bigg]
  \, \tilde{{\bf A}}_{\mu_1} (k_1) \ldots 
  \tilde{{\bf  A}}_{\mu_m}(k_m) \, .
\label{1gvpath5}
\eeqa
After carrying out the $t_i$ integrals, the coefficient of the gluon 
fields in \eqn{1gvpath5} is
\beq
  {\cal C}_{{\rm NE}, (2,j)}^{\mu_1, \ldots, \mu_m} (n, k_i) \, = \,
  \sum_{i > j} \, \frac{k_i^{\mu_j}}{n \cdot K_i} \,,
  \prod_{l \neq j} \left(- \frac{n^{\mu_l}}{n \cdot K_l} \right) \, .
\label{1gvpath6}
\eeq
The total contribution from the $j^{th}$ gluon, arising from the
terms in \eqn{torder5}, is then given by the sum of \eqn{1gvpath3} 
and \eqn{1gvpath6}, which gives
\beqa
  {\cal C}_{{\rm NE}, (1 + 2, j)}^{\mu_1, \ldots, \mu_m} 
  (n, k_i) & = & \frac{1}{2 n \cdot K_j} \left(k_j^{\mu_j} + 
  2 \sum_{i > j} k_i^{\mu_j} \right) \prod_{k \neq j} 
  \left(- \frac{n^{\mu_k}}{n \cdot K_k} \right) 
  \nonumber \\ & = & \frac{K_j^{\mu_j} + 
  K_{j + 1}^{\mu_j}}{2 n \cdot K_j} \, \prod_{k \neq j}
  \left(- \frac{n^{\mu_k}}{n \cdot K_k} \right) \, .
\label{1gvpathtot}
\eeqa
As expected, this result has the form of a product of eikonal factors 
times a NE Feynman rule for gluon $j$, and is precisely equal to the 
first part of the expression for effective NE Feynman rule
in \eqn{1gv}.

Next, we consider the last term in the first line and the term on
the third line in the exponent of \eqn{amp3bf}. In the non-abelian
case, and for $\lambda = 1$, they read
\beqa
  {\bf E}_{\rm NE}^{(3+4)} (n, x_i) & \equiv &  
  - \frac{1}{2} \, \int_0^\infty d t \, t \, n^\mu \, \Box 
  \mathbf{A}_\mu (n t) 
  \label{termsprop} \\
  && - \, \frac{\rm i}{2} \int_0^\infty d t \int_0^\infty d t' 
  \text{min} (t, t') \, n^\mu \, n^\nu \, g^{\sigma\tau} 
  \left[\partial_\sigma {\bf A}_\mu (n t) \right] 
  \left[\partial_\tau {\bf A}_\nu (n t') \right] \, . \nonumber
\eeqa
The first term leads again to diagrams of the form of 
fig.~\ref{Tordered}, but with one of the eikonal emissions replaced 
by a NE one-gluon vertex, this time of the type generated by the
expansion of a propagator. Without loss of generality we can take
this gluon to be the$j^{th}$ gluon. The corresponding momentum 
space expression is
\beqa
  {\bf L}_{\rm NE}^{(3,j)} (n, k_i) & = &
  \left[{\rm i} \int_0^\infty d t_1 n^{\mu_1} 
  {\rm e}^{{\rm i} n \cdot k_1t_1} \,\, {\rm i} \int_{t_1}^\infty 
  d t_2 n^{\mu_2} {\rm e}^{{\rm i} n\cdot k_2 t_2} 
  \ldots {\rm i} \int_{t_{j - 1}}^\infty d t_j t_j 
  \left(\frac{n^{\mu_j}}{2} k_j^2 \right) {\rm e}^{{\rm i} 
  n \cdot k_j t_j} \ldots \right. \nonumber \\
  && \quad \left.
  \ldots \, {\rm i} \int_{t_{m - 1}}^\infty d t_m 
  n^{\mu_m} {\rm e}^{{\rm i} n \cdot k_m t_m} \right]
  \tilde{{\bf A}}_{\mu_1}(k_1) \ldots 
  \tilde{{\bf A}}_{\mu_m}(k_m) \, .
\label{proppath2}
\eeqa
Carrying out the $t_i$ integrals to the right of the NE vertex, the
coefficient of the product of gluon fields for this term reads
\beq
  {\cal C}_{{\rm NE}, (3, j)}^{\mu_1, \ldots, \mu_m} (n, k_i) 
  \, = \, {\rm i}^{n - 1} n^{\mu_1} \ldots n^{\mu_n} 
  \left(\prod_{k > j} \frac{\rm i}{n \cdot K_k} \right)
  \int_0^\infty d t_1 {\rm e}^{{\rm i} n \cdot k_1t_1}
  \ldots \int_{t_{j - 1}}^\infty d t_j t_j \, \frac{k_j^2}{2} \,  
  {\rm e}^{{\rm i} n \cdot K_j t_j} \, .
\label{proppath3}
\eeq
The right-most integral in this expression may be performed 
through integration-by-parts, and we find
\beqa
  {\cal C}_{{\rm NE}, (3, j)}^{\mu_1, \ldots, \mu_m} (n, k_i) & = &
  {\rm i}^{n - 1} n^{\mu_1} \ldots n^{\mu_n}  
  \left(\prod_{k >{j - 1}} \frac{\rm i}{n \cdot K_k} \right)
  \int_0^\infty d t_1 {\rm e}^{{\rm i} n \cdot k_1t_1}
  \ldots \int_{t_{j - 2}}^\infty d t_{j - 1} \nonumber \\ 
  && \times \, \left( t_{j - 1} + \frac{\rm i}{n \cdot K_j} \right) \, 
  \frac{k_j^2}{2} \, {\rm e}^{{\rm i} n \cdot K_{j - 1} t_{j - 1}} \, .
\label{proppath4}
\eeqa
This procedure can be iterated for the remaining integrations, and 
we find the expression
\beq
  {\cal C}_{{\rm NE}, (3, j)}^{\mu_1, \ldots, \mu_m} (n, k_i) \, = \,
  (- 1)^n \Bigg[ \frac{k_j^2 \, n^{\mu_j}}{2 (n \cdot K_j)^2}
  \prod_{k \neq j} \frac{n^{\mu_k}}{n \cdot K_k} + 
  \sum_{i < j} \frac{k_j^2 \, n^{\mu_i}}{2 (n \cdot K_i)^2}
  \prod_{k \neq i} \frac{n^{\mu_k}}{n \cdot K_k} \Bigg] \, .
\label{proppath5}
\eeq
The first term in \eqn{proppath5} has the form of a product of 
eikonal emissions times a NE Feynman rule for gluon $j$, as expected. 
The second term implements correlations between gluon $j$ and all 
gluons to the left of it.  We may conveniently rewrite this to also 
involve all gluons to the right of gluon $j$, using the fact that one 
must ultimately sum over all gluons along the line. One may then
use the identity
\beqa
  \sum_j \sum_{i < j} \frac{k_j^2 \, n^{\mu_i}}{2 (n \cdot K_i)^2}
  \prod_{k \neq i} \frac{n^{\mu_k}}{n \cdot K_k} & = &
  \sum_i \sum_{j > i} \frac{k_j^2 \, n^{\mu_i}}{2 (n \cdot K_i)^2}
  \prod_{k \neq i} \frac{n^{\mu_k}}{n \cdot K_k} \nonumber \\
  & = & \sum_j \sum_{i > j} \frac{k_i^2 \, n^{\mu_j}}{2 (n \cdot
  K_j)^2} \prod_{k \neq j} \frac{n^{\mu_k}}{n \cdot K_k} \, ,
\label{summanip}
\eeqa
where we relabelled $j \leftrightarrow i$ in the second line. 
In this way we collect terms from all diagrams where $j$ is either 
a single NE emission, or is correlated with gluons to its right. We
can then replace \eqn{proppath5} with
\beq
  {\cal C}_{{\rm NE}, (3, j)}^{\mu_1, \ldots, \mu_m} (n, k_i) \, = \,
  (- 1)^n \Bigg[ \frac{k_j^2 \, n^{\mu_j}}{2 (n \cdot K_j)^2}
  \prod_{k \neq j} \frac{n^{\mu_k}}{n \cdot K_k}
  - \sum_{i > j} \frac{k_i^2 \, n^{\mu_j}}{2 (n \cdot K_j)^2}
  \prod_{k \neq j} \frac{n^{\mu_k}}{n \cdot K_k} \Bigg] \, .
\label{proppath6}
\eeq
This must be combined with the contributions arising from the 
second term in \eqn{termsprop}, which generates diagrams of 
the form shown in fig.~\ref{1gvjoin}. This term correlates gluons 
at different positions along the line and, as before, the paired gluons 
are not necessarily adjacent, but may have additional eikonal 
emissions in between. Considering diagrams where the left-most 
gluon of the pair is the $j^{th}$, and performing manipulations 
similar to those leading to \eqn{proppath6}, we are led to
\beq
  {\cal C}_{{\rm NE}, (4, j)}^{\mu_1, \ldots, \mu_m} (n, k_i) \, = \,
  (- 1)^n \Bigg[ \sum_{i > j} \frac{k_i \cdot k_j \, n^{\mu_j}}{(n
  \cdot K_j)^2} \prod_{k \neq j} \frac{n^{\mu_k}}{n \cdot K_k}
  + \sum_{i > j} \sum_{l < j} \frac{k_i \cdot k_j \, n^{\mu_l}}{(n
  \cdot K_l)^2} \prod_{k \neq l} \frac{n^{\mu_k}}{n \cdot K_k}
  \Bigg] \, .
\label{proppath8}
\eeq
The structure of this result is analagous to \eqn{proppath5}, and 
follows from iterated integrations by parts. The second term 
implements correlations involving one gluon to the left of gluon 
$j$ (gluon $l$), and one to the right of it (gluon $i$).

As before, it is convenient to regroup terms in order to consider
all diagrams where gluon $j$ is the left-most one. One may do
this after summation over all gluons $j$ using the identity
\beq
  \sum_j \sum_{l < j} \sum_{i > j} \frac{k_i \cdot k_j \, 
  n^{\mu_l}}{(n \cdot K_l)^2} \, = \, \sum_j \sum_{j < l < i}
  \frac{k_i \cdot k_l \, n^{\mu_j}}{(n \cdot K_j)^2} \, ,
\label{summanip2}
\eeq
where we have interchanged the order of summation over $j$ and 
$l$, and then relabelled $j \leftrightarrow l$. We can now replace 
\eqn{proppath8} with
\beqa
  {\cal C}_{{\rm NE}, (4, j)}^{\mu_1, \ldots, \mu_m} (n, k_i) & = &
  (- 1)^m \Bigg[ \sum_{i > j} \frac{k_i \cdot k_j \, n^{\mu_j}}{(n
  \cdot K_j)^2} \prod_{k \neq j} \frac{n^{\mu_k}}{n \cdot K_k}
  + \sum_{j < l < i} \frac{k_i \cdot k_l \, n^{\mu_j}}{(n 
  \cdot K_j)^2} \prod_{k \neq j} \frac{n^{\mu_k}}{n \cdot K_k}
  \Bigg] \nonumber \\
  & = & (- 1)^m \sum_{j \leq l < i} \frac{k_i \cdot k_l
  \, n^{\mu_j}}{(n \cdot K_j)^2} \prod_{k \neq j}
  \frac{n^{\mu_k}}{n \cdot K_k} \, .
\label{proppath9}
\eeqa
The total contribution from the terms of \eqn{termsprop}, for
each gluon $j$ and its correlations with those to its right,
as obtained from the sum of \eqns{proppath6}{proppath8}, is
given by
\beqa
  {\cal C}_{{\rm NE}, (3 + 4, j)}^{\mu_1, \ldots, \mu_m} (n, k_i) 
  & = & (- 1)^m \frac{n^{\mu_j}}{2 (n \cdot K_j)^2}
  \left(k_j^2 + \sum_{i > j} k_i^2 + 2 \sum_{j \leq l < i}
  k_l \cdot k_i \right) \prod_{k \neq j} \frac{n^{\mu_k}}{n
  \cdot K_k} \nonumber \\
  & = & (- 1)^m \, \frac{K_j^2 \, n^{\mu_j}}{2 n \cdot K_j}
  \prod_{k \neq j} \frac{n^{\mu_k}}{n \cdot K_k} \, .
\label{propresult}
\eeqa
This term has the form of a product of eikonal emissions, multiplied
times a NE emission vertex for the emission of gluon $j$. This vertex 
has precisely the form of the second term in the effective NE 
Feynman rule in \eqn{1gv}. This is a non-trivial check on the structure 
of this term, since the various scalar products $k_i \cdot k_l$ 
coming from \eqn{termsprop} must  combine to produce precisely
the squared partial momentum sum $K_j^2$.

The last term to be considered is the second term on the second 
line in the exponent of \eqn{amp3bf}. In the non-abelian theory, 
and for $\lambda = 1$, it gives
\beqa
  {\bf E}_{\rm NE}^{(5)} (n, x_i) & \equiv &  
  - \frac{\rm i}{2} \int_0^\infty d t \int_0^\infty d t' \,
  \delta(t - t') \, g^{\mu \nu}{\bf A}_\mu (n t) 
  {\bf A}_\nu (n t') \nonumber \\ 
  & = & - \frac{\rm i}{2} \int_0^\infty d t \, g^{\mu \nu}
  {\bf A}_\mu (n t) {\bf A}_\nu (n t) \, .
\label{gmunuterm}
\eeqa
The delta function implies that the two gluons are emitted from 
the same position on the eikonal line. Thus, this term correlates
adjacent gluons, and indeed is easily seen to generate the NE 
two-gluon vertex Feynman rule of \eqn{2gv}.

For gluons in the same group, we have now shown that 
the NE Feynman rules derived in this paper precisely correspond 
to those that arise from the path integral formalism of 
Ref.~\cite{Laenen:2008gt}. To complete the demonstration 
of this correspondence, some remarks are in order concerning 
what happens when gluons belong to different groups. We will see 
that this analysis gives a useful interpretation of the remainder 
term found in eqs.~(\ref{remvsol}) and~(\ref{rempsol}).

\begin{figure}
  \begin{center}
    \scalebox{1.0}{\includegraphics{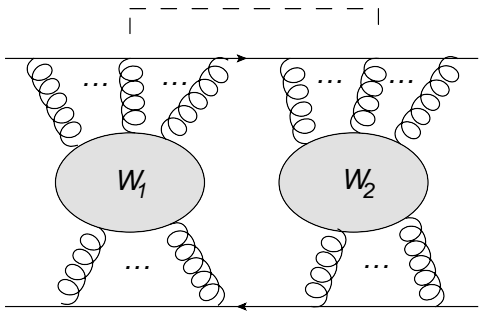}}
    \caption{A diagram composed of two eikonal webs $W_1$ and 
    $W_2$, correlated by a next-to-eikonal two-gluon vertex. Such 
    diagrams contribute to the NE exponent, and thus can be 
    described as NE webs.}
    \label{twoweb}
  \end{center}
\end{figure}

In the path integral formalism, it can be proven that diagrams 
that are two-eikonal-line reducible do not contribute to 
the exponent. This is well-known for eikonal webs
\cite{Gatheral:1983cz,Frenkel:1984pz,Sterman:1981jc}: 
the non-trivial statement proven here and in~\cite{Laenen:2008gt} 
is that the same is also true for next-to-eikonal webs. One must 
however keep in mind that there are NE webs arising from diagrams 
such as the one shown in fig.~\ref{twoweb}, consisting of what would 
be a product of two eikonal webs at eikonal order, but where two 
gluons (one from each web) become correlated by a NE two-gluon 
vertex acting at different points on the external line.
These diagrams are then two-eikonal-line irreducible, and thus
must contribute to the exponent.  Corresponding diagrams where 
the two-gluon vertex correlates gluons within one of the two webs 
are not present in the exponent, since they remain two-eikonal-line 
reducible. As a consequence, the contributions from all possible 
two-gluon vertex insertions no longer combine to form vertices 
depending on a single partial momentum sum, as they did above 
in the case of diagrams containing a single web. Instead, the 
leftover contributions form a remainder term, which has the 
same form as that discussed previously, in \sect{remainder}.

In more detail, consider again \eqn{1gvpathtot}, and assume for 
simplicity that the line being considered has contributions from two 
groups $g$ and $h$ only (the following discussion easily generalises 
to the case of more groups). Let us also take gluon $j$ as coming 
from group $g$. One may then split the sum on the left-hand side 
of \eqn{1gvpathtot} into two parts, involving gluons from $g$ and 
$h$ respectively, to get
\beq
  {\cal C}_{{\rm NE}, (1 + 2, j)}^{\mu_1, \ldots, \mu_m} (n, k_i) 
  \, = \, \frac{1}{2 n \cdot K_j} \left(g_j^{\mu_j} + 
  2 \sum_{i > j,\, i \in g} g_i^{\mu_j} + 2 \sum_{i > j, \, i \in h}
  h_i^{\mu_j} \right) \prod_{k \neq j} \left(- 
  \frac{n^{\mu_k}}{n \cdot K_k} \right) \, ,
\label{1gvrem}
\eeq
where as usual $g_i$ is the momentum of the $i^{\text{th}}$ gluon
from $g$. The first two terms in brackets combine to give the 
combination of partial momentum sums $G_j + G_{j + 1}$, as already
discussed in \eqn{1gvrem}, and lead to a next-to-eikonal emission 
vertex for gluon $j$ with eikonal Feynman rules for the remaining 
gluons. The second term can be rewritten as
\beq
  \frac{H_i^{\mu_j}}{n \cdot (H_i + G_j)} \prod_{k \neq j}
  \left(- \frac{n^{\mu_k}}{n \cdot K_k} \right) \, ,
\label{1gvrem2}
\eeq
where $i$ labels the gluon from $h$ that is nearest (on the 
right-hand side) to the gluon from $g$. This has the form of a 
two-gluon correlation, with eikonal Feynman rules for the remaining 
gluons, and is the same as that given in \eqn{remvsol}. Note that the 
previous result is expressed as a sum over all possible permutations 
of the gluons along the line, whereas here we consider only one such 
permutation. Also, we have yet to sum over the gluons $j$ from 
group $g$.

One may also consider \eqn{propresult}, again in the case in which
two groups are present along the line. Considering the bracketed
combination in the first line, all terms involving gluons from the 
same group combine to form a single squared partial momentum 
sum $G_j^2$ or $H_j^2$, as already discussed. There remain the 
terms in the double sum, in which the two gluons come from different 
groups. Considering the case where gluon $j$ comes from group 
$g$, the leftover terms are
\beq
  2 \sum_{l = j}^m \sum_{k > l} g_l \cdot h_k \, = \, 2 
  G_j \cdot H_i \, ,
\label{1gvrem3}
\eeq
where again $i$ labels the gluon from $h$ that is nearest (on 
the right-hand side) to the gluon from $g$. Combining with the 
other factors in \eqn{propresult}, one has a contribution to the 
remainder term of
\beq
  \frac{G_j \cdot H_i}{2 n \cdot(G_j + H_i)} \,
  \prod_{k} \left(- \frac{n^{\mu_k}}{n \cdot K_i} \right) \, .
\label{1gvrem4}
\eeq
After summing over all gluons from $g$, and all permutations 
of gluons on the line, this agrees with the result obtained in 
\eqn{rempsol}. We have therefore demonstrated the full equivalence 
between the results of this paper and the path integral approach of 
Ref.~\cite{Laenen:2008gt}.

The above discussion can be summarised as follows. In the 
diagrammatic approach of this paper, we find that contributions 
from different groups (derived from webs at eikonal and NE order 
containing local vertices only) factorise, up to a remainder term 
that correlates gluons between different groups. The NE Feynman 
rules thus obtained are precisely those that one finds through the 
path integral formalism, where the remainder term can be interpreted 
as correlating pairs of eikonal webs to form a new class of NE web. 
That the same structure and effective Feynman rules emerge 
from both the diagrammatic and path-integral viewpoints, is a 
highly non-trivial check of both approaches.

\section{An illustrative example: the Drell-Yan process}
\label{sec:DY}

In order to illustrate the results obtained in the previous sections, 
showing how they can be implemented in a concrete calculation,
we consider in this section soft gluon corrections to the Drell-Yan
production of a virtual photon in quark-antiquark annihilation. 
Since our focus here is just to give an example of the implementation
of our NE rules, we will restrict our attention to abelian-like contributions 
to the cross section, which build up the $C_F^p$ color structure at
$N^pLO$. Furthermore, we will consider only real radiation, since
when including mixed real-virtual corrections we will need to account
for Low's theorem \cite{Low:1958sn,Burnett:1967km,DelDuca:1990gz} contributions, which are not within the scope of 
the present paper. With these restrictions, we will compute soft
gluon corrections up to NE level, and up to NNLO, using the rules
developed in the previous sections. This is the simplest situation in 
which one can explicitly test NE exponentiation (as some of the 
two-loop NE corrections arise by expanding the exponential of the 
NE one-loop result). Furthermore, this calculation can be viewed as
a preliminary step towards the generalisation of existing soft-gluon 
resummation methods to NLO in the threshold expansion, as well as 
a check of the validity of our analysis. 

The NLO corrections to the Drell-Yan process were computed 
in~\cite{Altarelli:1979ub}, while the NNLO results was obtained
in~\cite{Hamberg:1991np} (eikonal NNLO results were first presented
in~\cite{Matsuura:1989sm,Matsuura:1988nd}). We begin with
a very brief summary of the LO calculation in order to introduce 
notation. 

\begin{figure}
\begin{center}
\scalebox{0.8}{\includegraphics{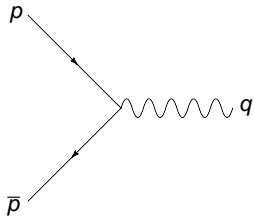}}
\caption{Born level diagram for the Drell-Yan production of a 
an off-shell photon.}
\label{DYBorn}
\end{center}
\end{figure}

The LO Drell-Yan process~\cite{Drell:1970wh}, shown at parton level
in fig.~\ref{DYBorn}, proceeds through the channel
\beq
  {\cal Q}(p) + \bar{{\cal Q}} (\bar{p}) \rightarrow 
  \gamma^*(q^\alpha) \, ,
\label{DYdef}
\eeq
where ${\cal Q}$ denotes a quark, and $q^\alpha$ is the momentum 
of the off-shell photon.
Since at higher orders (or at hadron level) the center-of-mass 
energy $s$ and the photon mass $Q^2 = q^2$ do not coincide, 
it is customary to parametrize the Drell-Yan total cross section in
terms of the variable $z = Q^2/s$, $0 < z < 1$. One then normalises
the higher order cross-section to the Born result defining the 
$n$-loop $K$ factor $K^{(n)}(z)$ as
\beq
  K^{(n)}(z) \, = \, \frac{1}{\sigma^{(0)}} \frac{d 
  \sigma^{(n)}(z)}{dz} \, ,
\label{Kfacdef}
\eeq
where $\sigma^{(n)}$ is the cross-section at ${\cal O}
(\alpha_s^n)$. The Born amplitude is simply
\beq
  {\cal M}_{(0)}^\alpha \, = \, \bar{u}(p) \gamma^\alpha 
  v(\bar{p}) \, ,
\label{Bornamp}
\eeq
where we drop coupling factors and spin averaging, as these will 
ultimately cancel upon taking ratios to compute the $K$ factor. 
Up to inessential factors, the Born cross section is given by
\beqa
  \sigma^{(0)} & \propto & - \int \frac{d^d q}{(2 \pi)^{d - 1}} \,
  (2 \pi)^n \, \delta^{(n)} (q - p - \bar{p}) \,
  \delta_+(q^2 - Q^2) \, \text{Tr} \left[{\slashed p} 
  \gamma^\alpha \bar{\slashed p} \gamma_\alpha \right] 
  \nonumber \\
  & = & 2 \pi \delta_+ (q^2 - Q^2) \,4 p \cdot \bar{p} \, (d - 2) \, ,
\label{Borncross}
\eeqa
where we will henceforth take $d = 4 - 2 \epsilon$. Clearly, the
$K$ factor at leading order is simply given by
\beq
  K^{(0)} (z) = \delta(1 - z) \, .
\label{sig0}
\eeq
We now move on to the NLO, next-to eikonal calculation.

\subsection{Next-to-eikonal contributions at NLO}
\label{DYNLO}

So far in this paper, we have only considered NE contributions to 
matrix elements. In calculating soft gluon contributions to 
cross-sections, one must also consider phase space integration. 
In the eikonal approximation (as is well known), the $n$-gluon 
phase space can be written as a product of decoupled one-gluon 
phase spaces. This is a necessary condition for the exponentiation 
of the cross-section which underlies soft-gluon resummations. 
At NE order the exact factorization of phase space breaks down, 
and it becomes necessary to study the unfactorized corrections
to all orders. This is done, for the inclusive cross section, in 
Appendix B. 
Preliminarily, we note that at NE level one
expects the schematic structure
\beq
  \sigma_{\NE} \, = \, \int \text{dPS}_{\text{eik}} \, 
  |{\cal M}_{\NE}|^2 + \int \text{dPS}_{\NE} 
  |{\cal M}_{\text{eik}}|^2 \, .
\label{PSstruc}
\eeq
Here ${\cal M}_{\text{eik}}$ and ${\cal M}_{\NE}$ are the 
eikonal and next-to-eikonal contributions to the matrix element,
extensively discussed in the previous sections, and we have
introduced an analogous expansion for the multi-gluon phase 
space measure of integration $\text{dPS}$. The NE correction 
to the phase space measure will also induce correlations between 
pairs of soft gluons, with a sum over all possible correlations. Up
to NNLO, these corrections are easy to implement since the
exact phase-space integration can be performed analytically, as 
we will do below.

\begin{figure}
  \begin{center}
  \scalebox{0.8}{\includegraphics{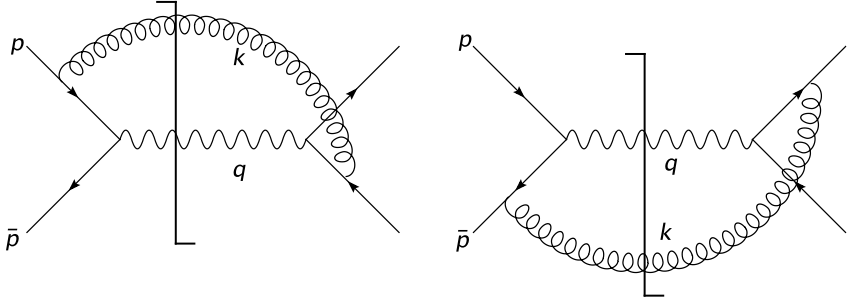}}
  \caption{Squared matrix elements contributing at NLO in the 
  eikonal limit. The cut denotes the final state.}
  \label{DYeik}
  \end{center}
\end{figure}

We begin by computing the eikonal contributions to the NLO 
$K$-factor. Eikonal Feynman rules easily yield
\beqa
  {\cal M}_{{\rm eik}, \alpha}^{(1)} \left[
  {\cal M}_{\rm eik}^{(1) \, \alpha} \right]^\dag
  & = & 2 g_s^2 C_F \, \frac{p \cdot \bar{p}}{(p \cdot k)
  (\bar{p} \cdot k)} \, \text{Tr} \left[{\slashed p} \gamma^\alpha
  \bar{\slashed p} \gamma_\alpha \right] \nonumber \\
  & =  & 2 g_s^2 C_F \, \frac{8 s^2}{u t} (1 - \epsilon) \, .
\label{A1E}
\eeqa
where the factor of 2 in the first line comes from combining the two 
diagrams shown in fig.~\ref{DYeik}, and we have introduced the Mandelstam invariants
\beq
t = - 2 k\cdot p \, ; \qquad u = - 2 k \cdot\bar{p} \, .
\label{mandies}
\end{equation}
To carry out the phase space integration, one may parametrise 
the four-vectors as
\beq
  p = \frac{\sqrt{s}}{2} \left(1, 1, {\bf 0}_\perp \right) \, ;
  \quad 
  \bar{p} = \frac{\sqrt{s}}{2} \left(1, -1, {\bf 0}_\perp \right) \, ;
  \quad 
  k = \frac{(1 - z) \sqrt{s}}{2} \left(1, \cos\theta, \sin\theta, 
  0\right) \, .
\label{4vecs}
\eeq
Defining then 
\beq
  y = \frac{1 + \cos \theta}{2} \, ,
\label{y}
\eeq
one finds
\beq
  t = - 2 s (1 - y)(1 - z) \, ; \qquad u = - 2 s y (1 - z) \, .
\label{mandies2}
\eeq
The NLO $K$-factor, defined in \eqn{Kfacdef}, is computed by 
including a factor $\delta(k^0 - (1 - z) p^0)$ in the two-particle
phase space, and introducing the appropriate normalization, according
to \eqn{Borncross}. Setting the renormalization scale $\mu^2 = 
Q^2$ one finds, at the eikonal level
\beq
  K^{(1)}_{\rm eik} (z) \, = \, 
  \frac{1}{16 \pi^2} \, 
  \frac{(4 \pi)^{\epsilon}}{\Gamma(2 - \epsilon)} \, 
  z^\epsilon (1 - z)^{1 - 2 \epsilon} \int_0^1 
  d y \, \big[y (1 - y) \big]^{- \epsilon} 
  {\cal M}_{{\rm eik}, \alpha}^{(1)} \left[
  {\cal M}_{\rm eik}^{(1) \, \alpha} \right]^\dag  \, .
\label{K1a}
\eeq
Substituting the squared eikonal matrix element given in \eqn{A1E},
performing the integration and Taylor expanding in $\epsilon$ 
this yields
\beqa
  K^{(1)}_{\rm eik} (z) & = & \frac{\alpha_s}{4 \pi} \, C_F \,  
  \Bigg\{ - \frac{8}{\epsilon} {\cal D}_0 (z) + 16 {\cal D}_1 (z)
  - \frac{8 \log(z)}{1 - z} \nonumber \\
  && - \, 4 \epsilon \left[4 {\cal D}_2(z) - 3 \zeta_2 
  {\cal D}_0 (z) - \frac{4 \log{z} \log(1 - z)}{(1 - z)}
  + \frac{\log^2 z}{1 - z} \right] \Bigg\} \, ,
\label{K1E}
\eeqa
where, following~\cite{Hamberg:1991np}, we use the notation
\beq
  {\cal D}_p (z) \, = \, \left[ \frac{\log^p (1 - z)}{1 - z}
  \right]_+ 
\label{plusdistdef}
\eeq
to denote plus distributions. Note that we are not displaying
terms proportional to $\delta(1 - z)$, which mix with virtual
contributions not considered here. Note also that \eqn{K1E} 
contains terms that do not behave as $(1- z)^{-1}$: since we 
are using eikonal  matrix elements, such terms arise from 
subleading corrections to the phase space measure.

\begin{figure}
  \begin{center}
  \scalebox{0.8}{\includegraphics{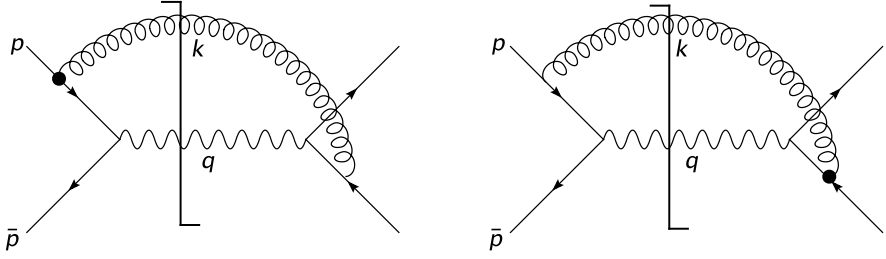}}
  \caption{Diagrams contributing at NE level to the squared amplitude 
  for NLO real emission. Two additional diagrams, arising from complex 
  conjugation, are not shown.}
  \label{NEdiags}
  \end{center}
\end{figure}

Let's now consider NE corrections to the matrix element. According 
to the results of \sect{NErules}, the effective Feynman rule for single gluon emission at NE order is
\beq
  \frac{\gamma^\mu {\slashed k}}{2 p \cdot k} - k^2
  \frac{p^\mu}{2 (p \cdot k)^2} \, .
\label{NErule}
\eeq
For real emission, the second term in \eqn{NErule} vanishes since
$k^2 = 0$. The diagrams contributing at NE level are shown in 
fig.~\ref{NEdiags}, and consist of all possible insertions of at most  
one NE vertex in the eikonal diagrams of figure \ref{DYeik}.
The corresponding contribution to the squared matrix element is
\beqa
  {\cal M}_{\NE, \alpha}^{(1)} 
  \left[{\cal M}_{\rm eik}^{(1) \, \alpha} \right]^\dag
  + {\cal M}_{{\rm eik}, \alpha}^{(1)}
  \left[ {\cal M}_{\NE}^{(1) \, \alpha} \right]^\dag
  & = &\frac{\bar{p}_\mu}{\bar{p} \cdot k}
  \frac{ \text{Tr} \left[{\slashed p} \gamma^\mu
  {\slashed k} \gamma^\alpha \bar{\slashed p} 
  \gamma_\alpha \right]}{2 p \cdot k} + 
  \frac{p_\mu}{p \cdot k} \frac{ \text{Tr} \left[{\slashed p}
  \gamma^\alpha \bar{\slashed p} \gamma^\mu {\slashed k}
  \gamma_\alpha \right]}{2 \bar{p} \cdot k} 
  \nonumber \\
  & = & 8 \, (1 - \epsilon) \left(\frac{s}{t} + \frac{s}{u} \right).
\label{NEamp}
\eeqa
Carrying out the phase space integration, dividing by the Born normalisation and expanding in $\epsilon$ leads to
\beqa
  K_{\NE}^{(1)} (z) & = & \frac{\alpha_s}{4\pi} \, C_F \, 
  \Bigg\{ \frac{8}{\epsilon} - 16 \log (1 - z) + 8 \log z 
  \nonumber \\
  && \quad - 4 \epsilon \left[ - 4 \log^2 (1 - z) + 4 \log z 
  \log(1 - z) - \log^2 z + 3 \zeta_2 \right] \Bigg\} \, .
\label{KNE}
\eeqa
Combining \eqns{K1E}{KNE} and retaining only terms with a
logarithmic dependence on $z$, up to NE order, we obtain 
\beqa
  K^{(1)} (z) & = & \frac{\alpha_s}{4\pi} \, C_F \, 
  \Bigg\{ - \frac{8}{\epsilon} {\cal D}_0 (z) +16 {\cal D}_1(z)
  - 16 \log(1 - z) -  4 \epsilon \Big[ 4 \, {\cal D}_2 (z) - 
  3 \zeta_2 {\cal D}_0 (z) \nonumber \\
  && \quad - 4 \log^2 (1 - z) + 4 \log(1 - z) \Big] \Bigg\} \, .
\label{K1total}
\eeqa
Note that to obtain this result we expanded around $z = 1$, including 
for example the replacement $\log z \to - (1- z)$.

One may compare this result with the corresponding exact NLO
result including the complete $z$ dependence~\cite{Altarelli:1979ub}.
One easily verifies that expanding the exact result around $z=1$ 
gives \eqn{K1total}. Thus, the effective Feynman rules derived 
in the previous sections pass a first simple test.

\subsection{Next-to-eikonal contributions at NNLO}
\label{DYNNLO}

In this section we compute the contribution proportional to 
$C_F^2$ to the NNLO $K$-factor for DY production of a virtual 
photon, and, as in the previous section, we restrict our attention 
to real gluon emission from a $q \bar{q}$ initial state. This is the 
simplest possible example illustrating the analysis performed in
sections~\ref{eikonal}-\ref{sec:exp}.

The structure of NE corrections found in the preceding sections 
implies that the amplitude for the double-real-emission graphs can
be extracted, up to NE accuracy, from the expansion of an exponential 
with the schematic structure
\beq
  {\cal M} \, = \, {\cal M}^{(0)} \, \exp \left[
  \bar{\cal M}_{\rm eik}^{(1)} + \bar{\cal M}_{\NE}^{(1)}
  + \bar{\cal M}_{\NE}^{(2)} \right] \, ,
\label{calastruc}
\eeq
where ${\cal M}^{(0)} \propto \text{Tr} \left[\slsh{p} 
\gamma^\alpha \pb \gamma_\alpha \right]$, and the bar over 
an amplitude represents the expression for that amplitude with Born 
result factored out. Note there is no ${\cal O}(\alpha_s^2)$ eikonal 
amplitude in the exponent. This is because we are considering only
abelian graphs, and in that set there is no connected subdiagram 
involving two eikonal real emissions. In this language, the single and
double real radiation matrix elements, up to NE accuracy, can be
written as
\beqa
  {\cal M}^{(1)} & = & {\cal M}^{(0)} \left[ 
  \bar{\cal M}_{\rm eik}^{(1)} + \bar{\cal M}_{\NE}^{(1)} \right]
  \, , \nonumber \\
  {\cal M}^{(2)} & = & {\cal M}^{(0)} \left[ 
  \frac{1}{2} \left(\bar{\cal M}_{\rm eik}^{(1)} \right)^2 + 
  \bar{\cal M}_{\rm eik}^{(1)} \bar{\cal M}_{\NE}^{(1)} 
  + \bar{\cal M}_{\NE}^{(2)} \right] \, ,
\label{enematr}
\eeqa
Since we are not considering virtual corrections, the only 
contributions to the squared matrix element that we will need 
at ${\cal O} (\alpha_s^2)$ and at NE accuracy are given by
\beqa
  {\cal M}^{(2)} \left[ {\cal M}^{(2)} \right]^\dag & =  &
  \left| {\cal M}^{(0)} \right|^2 \Bigg\{
  \left| \frac{1}{2} \left(\bar{\cal M}_{\rm eik}^{(1)} 
  \right)^2 \right|^2 +
  2 \, {\bf Re} \left[ \frac{1}{2} \left(\bar{\cal M}_{\rm eik}^{(1)} 
  \right)^2 \left( \bar{\cal M}_{\rm eik}^{(1)} 
  \bar{\cal M}_{\NE}^{(1)} \right)^\dag \right]
  \nonumber \\
  && \qquad + \, 2 \, {\bf Re} \left[ \frac{1}{2} 
  \left(\bar{\cal M}_{\rm eik}^{(1)} 
  \right)^2 \left( \bar{\cal M}_{\NE}^{(2)} \right)^\dag
  \right] \Bigg\} \nonumber \\
  & \equiv & 
  \left| {\cal M}^{(2)} \right|^2_{\rm eik} + 
  \left| {\cal M}^{(2)} \right|^2_{\NE, {\rm 1g}} + 
  \left| {\cal M}^{(2)} \right|^2_{\NE, {\rm 2g}} \, ,
\label{expoterms}
\eeqa
where, as the notation suggests, the first term is the eikonal approximation to the double real radiation matrix element squared,
the second term is the NE correction arising from single-gluon
NE vertices, and the third term is the NE correction arising
from double-gluon NE vertices. We will treat each contribution
in turn.

Let's begin with the eikonal result, which has the explicit expression
\beq
  \left| {\cal M}^{(2)} \right|^2_{\rm eik} \, = \, {\cal N} \, 
  \frac{p \cdot \bar{p}}{p \cdot k_1\, p \cdot k_2 \,
  \bar{p} \cdot k_1 \, \bar{p} \cdot k_2} \,
  \text{Tr} \left[\slsh{p} \gamma^\alpha \pb \gamma_\alpha
  \right] \, ,
\label{ME}
\eeq
where, for simplicity, we have included in the normalization 
${\cal N}$ all coupling, color and spin averaging factors, most of 
which drop out when constructing the $K$ factor. 

The integration of \eqn{ME} over the three-body phase space is 
non-trivial, and we briefly discuss it here, since it will be needed 
also for the evaluaton of the NE corrections. We follow
Ref.~\cite{Hamberg:1991np}, and we evaluate the integrals
in the center-of-mass frame of the two outgoing gluons. 
In order to perform the integral, we parametrise the momenta 
in $d = 4 - 2 \epsilon$ dimensions as in~\cite{Hamberg:1991np}.
We write
\beqa
  k_1 & = & \frac{\sqrt{s_{12}}}{2} \left( 1, 0, \ldots, 
  \sin \theta_2 \sin \theta_1, \cos \theta_2 \sin\theta_1,
  \cos \theta_1\right) \, , \nonumber \\
  k_2 & = & \frac{\sqrt{s_{12}}}{2} \left( 1, 0, \ldots,
  - \sin \theta_2 \sin \theta_1, - \cos \theta_2 \sin \theta_1,
  - \cos \theta_1\right) \, ,  \nonumber \\
  p & = & \frac{(s - \tilde{t})}{2 \sqrt{s_{12}}} \left(1, 0,
  \ldots, 0, 1\right) \, , \label{p} \\
  q & = & \left( \frac{s - Q^2 - s_{12}}{2 \sqrt{s_{12}}}, 0, 
  \ldots, 0, | \vec{q} | \sin \psi, | \vec{q} | \cos\psi \right) \, ,
  \nonumber \\
  \bar{p} & = & \left( \frac{\tilde{t} + s_{12} - Q^2}{2 
  \sqrt{s_{12}}}, 0, \ldots, 0, | \vec{q} | \sin \psi, | \vec{q} |
  \cos \psi - \frac{(s - \tilde{t})}{2 \sqrt{s_{12}}} 
  \right) \, . \nonumber
\eeqa
Here we have introduced the quantities
\beqa
  \tilde{t} & \equiv & 2 p \cdot q \, = \, (p + q)^2 - Q^2 \, ,
  \nonumber \\
  \tilde{u} & = & 2 \bar{p} \cdot q \, = \, (\bar{p} + q)^2 - Q^2
  \, , \nonumber \\
  s_{12} & = & 2 k_1\cdot k_2 \, = \, s - \tilde{t} - \tilde{u}
  + Q^2 \,  , \label{s12} \\
  \cos \psi & = & \frac{(s - Q^2)(\tilde{u} - Q^2) - 
  s_{12} (\tilde{t} + Q^2)}{(s - \tilde{t})
  \sqrt{\lambda(s, Q^2, s_{12})}} \, , \nonumber \\
  | \vec{q} | & = & \frac{\sqrt{\lambda(s, Q^2, s_{12})}}{2
  \sqrt{s_{12}}} \, , \nonumber
\eeqa
where $\lambda$ is the K\"{a}llen function, $\lambda(a, b, c)
= a^2 + b^2 + c^2 - 2 a b - 2 a c - 2 b c$.
The Mandelstam invariants $\tilde{t}$ and $\tilde{u}$ can in turn 
be expressed as functions of the photon energy fraction $z = 
Q^2/s$ and of two further variables $x$ and $y$, normalized so 
that $0 < x < 1$ and $0 < y < 1$, with the definitions
\beqa
  \tilde{u} & = & s \, \big[ 1 - y (1 - z) \big] \, , \nonumber \\
  \tilde{t} & = & s \left[ z + y (1 - z) - 
  \frac{y (1 - y) x (1 - z)^2}{1 - y(1 - z)} \right] \, .
  \label{z}
\eeqa
The $d$-dimensional three-body phase space can then be written as
\beqa
  \int \text{dPS}_3 & = & \frac{1}{(4 \pi)^d} 
  \frac{s^{d - 3}}{\Gamma(d - 3)} (1 - x)^{2 d - 5}
  \int_0^\pi d \theta_1 \int_0^\pi d \theta_2 \,
  (\sin \theta_1)^{d - 3} \, (\sin \theta_2)^{d - 4}
  \nonumber \\
  && \, \times \, \int_0^1d y \int_0^1d x
  \big[y (1 - y) \big]^{d - 3} \big[x (1 - x) \big]^{d/2 - 2}
  \big[1 - y (1 - z) \big]^{1 - d/2} \, .
\label{PS}
\eeqa
To integrate the matrix element squared in \eqn{ME} we still 
need expressions for the propagator factors in the current
parametrization. They are
\beqa
  p \cdot k_1 & = & \frac{s - \tilde{t}}{4} \, (1 - \cos \theta_1)
  \, \nonumber \\
  p \cdot k_2 & = & \frac{s - \tilde{t}}{4} \, (1 + \cos \theta_1)
  \, \label{proppar} \\
  \bar{p} \cdot k_1 & = & A - B \cos \theta_1 - C \sin\theta_1
  \cos \theta_2 \nonumber \\
  \bar{p} \cdot k_2 & = & A + B \cos \theta_1 + C \sin \theta_1
  \cos \theta_2 \nonumber \, ,
\eeqa
where
\beqa
  A & = & \frac{\tilde{t} + s_{12} - Q^2}{4} \, , \nonumber \\
  B & = & \frac{\sqrt{s_{12}}}{2} \, | \vec{q} | \cos \psi - 
  \frac{(s - \tilde{t})}{4} \, , \label{B} \\
  C & = & - \frac{\sqrt{s_{12}}}{2} \, | \vec{q} | \sin \psi \, .
  \nonumber
\eeqa
It can be verified that $A^2 = B^2 + C^2$. Using this fact and
putting together \eqn{ME} and \eqn{PS} one may verify that all 
the angular integrals to be performed can be reduced to the 
standard form
\beqa
  I^{(p,q)}_d (\chi) & = & \int_0^\pi d \theta_1 
  \int_0^\pi d \theta_2 \frac{\sin^{d - 3} \theta_1
  \sin^{d - 4} \theta_2}{(1 - \cos \theta_1)^p 
  (1 - \cos \chi \cos \theta_1 - \sin \chi \sin \theta_1
  \cos \theta_2 )^q} \label{int2} \\
  & = & 2^{1 - p - q} \pi \, 
  \frac{\Gamma(\frac{d}{2} - 1 - q) \Gamma(\frac{d}{2} - 1 - p)
  \Gamma(d - 3)}{\Gamma(d - 2 - p - q)
  \Gamma^2(\frac{d}{2} - 1)} \, 
  \phantom{!}_2F_1\left[p, q; \frac{d}{2} - 1; 
  \cos^2 \left(\frac{\chi}{2} \right) \right] \, , \nonumber 
\eeqa
where $\phantom{!}_2F_1$ is the standard hypergeometric 
function. Identifying $\cos \chi = - B/A$, using $\cos^2
(\chi/2) = (1 + \cos \chi)/2$, and expanding in powers of
$\epsilon = 2 - d/2$, one gets, for example,
\beq
  I_{4 - 2 \epsilon}^{(1,1)} (\chi) \, = \, 
  - \frac{\pi}{\epsilon} \left[ \frac{A + B}{2 A} 
  \right]^{- 1 - \epsilon} \left\{1 + 
  \epsilon^2 \, \text{Li}_2 \left[ \frac{A - B}{2 A} \right]
  + {\cal O} \left( \epsilon^3 \right) \right\} \, .
\label{In11}
\eeq

After performing the angular integrals in this way, the remaining 
integrations over $x$ and $y$ are easily carried out, after expanding 
the integrand in powers of $(1 - z)$. This is consistent here since,
for real emission contributions to the Drell-Yan cross section, the 
soft gluon expansion and the threshold expansion coincide.
The resulting eikonal contribution to the NNLO $K$ factor reads
\beqa
  K^{(2)}_{\rm eik} (z) & = & \left(\frac{\alpha_s}{4\pi} \,  
  C_F \right)^2 \, \left[ - \, \frac{32}{\epsilon^3} \, {\cal D}_0 (z)
  + \frac{128}{\epsilon^2} \, {\cal D}_1 (z) - 
  \frac{256}{\epsilon} \, {\cal D}_2 (z) - 
  \frac{320}{\epsilon} \log(1 - z) \right. \nonumber \\
  && \qquad \left. + \, \frac{1024}{3} \, 
  {\cal D}_3 (z) + 640 \log^2 (1 - z) + \ldots \right] \, .
\label{etot}
\eeqa
Note that for the sake of illustration we display only terms with a 
logarithmic dependence on $(1 - z)$, and specifically those with
rational coefficients, that are easily extracted from the exact 
calculation for comparison.

We now turn to NE corrections to the matrix element, and specifically
we consider the second term in the last line of \eqn{expoterms}, 
which arises from the insertion of single-gluon NE vertices. The 
application of the rules derived in the previous sections yields
\beq
  \left| {\cal M}^{(2)} \right|^2_{\NE, {\rm 1g}}  \, = \, 
  - \, {\cal N} \, \left[ \frac{s}{p \cdot k_1 \, \bar{p} \cdot k_1}
  \left(\frac{1}{p \cdot k_2} + \frac{1}{\bar{p} \cdot k_2}
  \right) + \left( k_1 \leftrightarrow k_2 \right) \right] \, 
  \text{Tr} \left[\slsh{p} \gamma^\alpha \pb \gamma_\alpha
  \right] \, .
\label{NEterm}
\eeq
The integration over phase space can be carried out as for the 
eikonal term, and one finds a contribution to the $K$ factor 
given by
\beqa
  \hspace{-2mm} 
  K^{(2)}_{\NE, {\rm 1g}} (z) & = & - \,  
  \left(\frac{\alpha_s}{4\pi} \, C_F \right)^2
  \left[ \frac{128}{\epsilon^2} \, \log(1 - z) 
  - \frac{256}{\epsilon} \, \log^2(1 - z) 
  \right. \nonumber \\ 
  && \left. \qquad + \, 
  \frac{1024}{3} \log^3(1 - z) + \ldots \right] \, ,
\label{NEresult}
\eeqa
where again for simplicity we display only logarithmic terms with 
rational coefficients.

\begin{figure}
  \begin{center}
  \scalebox{0.6}{\includegraphics{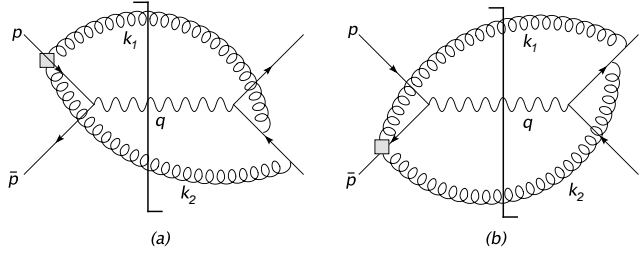}}
  \caption{Cut Feynman diagrams involving the next-to-eikonal 
  two-gluon vertices summarized by the tensor $R^{\mu\nu}$. 
  Note that also the complex conjugate diagrams must be included.}
  \label{Rdiags}
  \end{center}
\end{figure}

Finally, we need the contribution given by the final term in 
\eqn{expoterms}, which involves NE contributions to two-gluon 
emission that don't already arise from the exponentiation of the 
single emission. That is, one must include the contribution of the 
two-gluon vertices discussed in \sect{NErules}. The sum of the
relevant vertices on, say, the quark line carrying momentum 
$p$ gives the tensor
\beq
  R^{\mu \nu} (k_1, k_2; p) \, = \, - \, \frac{(p \cdot k_2)
  p^\mu k_1^\nu + (p \cdot k_1) k_2^\mu p^\nu - 
  (p \cdot k_1)(p \cdot k_2) \, g^{\mu \nu} - 
  (k_1 \cdot k_2) p^{\mu} p^{\nu}}{p \cdot(k_1 + k_2)} \, ,
\label{rmunu}
\eeq
where an additional factor of two arises upon summing over 
the different orderings of the gluon momenta. Note that,
given that this vertex joins two adjacent gluons, it includes
the seagull vertex (i.e. the term in $g^{\mu\nu}$), as discussed
in section~\ref{NErulessum}. It is easily checked 
that this gives zero if either gluon lands on a fermion line carrying
the same momentum $p$ as the emitter. Thus, the only possible 
contributions come from diagrams such as those shown in 
fig.~\ref{Rdiags}.
In fact, these diagrams also give a vanishing contribution, as can 
be seen from the following argument. Consider for example
the diagram in fig.~\ref{Rdiags}(a). This involves the factor
\beqa
  \frac{R^{\mu \nu} \bar{p}_\mu \bar{p}_\nu}{\bar{p}
  \cdot k_1 \, \bar{p} \cdot k_2 \, p \cdot k_1 \, p \cdot k_2}
  & = &  - \, \frac{p \cdot \bar{p}}{p \cdot (k_1 + k_2) \,
  \bar{p} \cdot k_1\, \bar{p} \cdot k_2 \, p \cdot k_1 \,
  p \cdot k_2} \nonumber \\
  && \qquad \times \, \Big[ p \cdot k_2 \, \bar{p} \cdot k_1 
  + \bar{p} \cdot k_2 \, p \cdot k_1 - k_1 \cdot k_2 \,
  p \cdot \bar{p} \Big] \, .
\label{rmunua}
\eeqa
Note that the term in $R^{\mu \nu}$ involving $g_{\mu \nu}$ 
vanishes, since $\bar{p}^2 = 0$. If we now pick light-cone 
coordinates, such that $p^\mu = (p^+,0^-,{\bf 0}_\perp)$ 
and $\bar{p}^\mu = (0^+,\bar{p}^-,{\bf 0}_\perp)$, with 
$p^\pm \equiv (p^0 \pm p^3)/\sqrt{2}$, we observe that the 
factor in brackets on the {\it r.h.s.} of \eqn{rmunua} becomes
\beq
  p \cdot k_2 \, \bar{p} \cdot k_1 + \bar{p} \cdot k_2 \, 
  p \cdot k_1 - k_1 \cdot k_2 \, p \cdot \bar{p}
  \, = \, - \, \frac{s}{2} \, {\bf k}_{1, \perp} \cdot 
  {\bf k}_{2, \perp} \, .
\label{bterm}
\eeq
The prefactor of this term depends only on longitudinal momentum 
components. As a consequence, the diagram involving the two-gluon
vertex gives a vanishing contribution upon integration over the gluon 
transverse momenta.

We conclude that the NE contribution to the NNLO $K$ factor is 
given just by the sum of \eqns{etot}{NEresult}. This, as expected, 
agrees with the result obtained from a full calculation of the double
real emission contribution, followed by an expansion to NE order. 
The latter result is not available directly in the literature, but we 
include a short summary of the calculation in appendix~\ref{tworeal}.

\section{Conclusion}
\label{sec:conclusion}

In this paper we have considered the exponentiation of soft 
gluon contributions to gauge theory amplitudes and cross sections, 
from a Feynman diagram viewpoint, classifying the structure of 
next-to-eikonal corrections. We showed that the notion of 
exponentiation can indeed be extended to NE order, as given 
by \eqn{NEexptheorem2}. We note however that at NE level there
are non-factorizable contributions which must be separately considered,
and which are related to the radiationless amplitude by Low's theorem
and its generalizations. Factorizable contributions exponentiate
and fall into two classes. First, there are conventional (eikonal) webs, 
modified by including effective NE Feyman rules; then, there are new 
webs formed by taking pairs of eikonal webs, and correlating a pair
of gluons (one from each original web) with a two-gluon vertex. The 
derivation of this result relies upon establishing the appropriate 
generalisation of the eikonal identity, which is given in \eqn{theor2} 
and expresses the fact that contributions from different groups 
factorise on a given external line, up to a remainder term which gives 
rise to the two-gluon vertex contributions mentioned above.

We performed a detailed comparison of the effective NE Feynman 
rules derived in this paper with the results obtained using the path 
integral formalism of~\cite{Laenen:2008gt}, finding complete 
agreement between the two formalisms. This is a highly non-trivial 
cross-check of our classification of NE contributions to matrix elements, 
and also provides further insight on how to apply them in practice. In
particular, a diagrammatic formalism makes contact with conventional
factorization proofs, and is likely to be amenable to generalizations
to hard collinear emissions which would be needed in order to perform
a complete analysis of the threshold expansion beyond leading order.

The complete structure of next-to-eikonal corrections to QCD cross 
sections receives contributions not only from matrix elements but also
from the integration over phase space for multiple soft-gluon emission.
The discussion in this paper (see appendix B) shows how to treat these corrections for 
any number of gluons in the case of the Drell-Yan process; we expect 
that it should be possible to generalize these arguments to other 
processes.

Having classified the structure of NE corrections, it is interesting to 
ponder what might happen at NNE order (and beyond). Terms at this 
order in the expansion in gluon momenta are no longer divergent, but 
may nevertheless be numerically significant. The path-integral arguments 
of~\cite{Laenen:2008gt}, as well as the recursive arguments presented 
in this paper, make clear that there is always a subset of external 
emission graphs which exponentiates at any given order in the 
momentum expansion. At higher orders, new vertices involving multiple gluon emissions will arise. More serious problems are bound to emerge, 
however, from non-factorizable contributions, which are governed
by Low's theorem at NE level. It is difficult to generalise the 
Low-Burnett-Kroll argument to higher orders in momentum, as 
gauge invariance does not then completely fix the structure of 
non-exponentiating contributions. Nevertheless, it might be the case 
that factorizable contributions dominate in practice, allowing for at 
least a partial resummation of NNE terms and beyond. It should also 
be noted that, as more finite orders terms are generated through exponentiation, it would become increasingly important to accurately
match the resummed result to fixed-order perturbation theory, since
a simple expansion of the exponential is bound to generate an 
increasing number of subleading terms which are beyond the accuracy
guaranteed by resummation.

The results of this paper should be seen as a step towards the generalization of Sudakov resummation beyond leading order in the threshold expansion. In order to achieve the goal several further
steps are still needed: the simple organization of phase space proposed 
in this paper must be brought to bear in a more formal way and in more 
general cross sections; non-factorizable contributions must be better understood; finally modifications of collinear evolution for hard 
collinear emissions must be fully understood at next-to-leading order 
in the threshold expansion. In order to show that our work is a step
in the right direction, and as a first illustration of how to apply our
effective Feynman rules at NE order, we reproduced a subset of NE 
logarithms in the real emission contribution to the Drell-Yan $K$-factor 
up to NNLO (specifically, those associated with the $C_F^n$ color 
structures). This is in a sense the simplest possible example, given 
that the limit $z \rightarrow 1$ for this process corresponds to soft 
singularities only, with no mixing from hard collinear divergences. 
One must also note that we have only discussed the simple case of 
a color-singlet hard interaction linking two partonic lines. We expect however that the bulk of our considerations will extend to more 
general color configurations, as indicated in ~\cite{Gardi:2010rn}. 
Work towards understanding these various generalizations is ongoing.

\subsection*{Acknowledgements}

This research has been supported by the Foundation for Fundamental Research of Matter (FOM), as part of the programme TPP (Theoretical Particle Physics in the era of the LHC, FP 104); by the National 
Organization for Scientific Research (NWO); by MIUR (Italy) under 
contract 2006020509$\_$004;  and by the European Community's 
Marie-Curie Research Training Network `Tools and Precision Calculations 
for Physics Discoveries at Colliders'  (`HEPTOOLS'), under contract 
MRTN-CT-2006-035505. CDW is supported by the STFC postdoctoral fellowship ``Collider Physics at the LHC'', and thanks the Nikhef theory group for hospitality. LM thanks the CERN Theory Division for hospitality
and support during crucial stages of this work.

\appendix

\section{Rearranging NE numerators for spin-$\frac{1}{2}$
particles}
\label{proof}

We wish to prove that, when acting on a massless spinor, the
following identity holds:
\beqa
  N^{\mu_1 \ldots \mu_n} & \equiv &
  \sum_{i = 1}^n \, \left[ \prod_{j = i + 1}^n \left( 2 p^{\mu_j}
  \right) \, \prod_{k = 1}^{i - 1} \left( {\slashed p} 
  \gamma^{\mu_k} \right) \,
  \gamma^{\mu_i} {\slashed K}_i \right]
  \label{theorapp} \\
  & = &  
  \sum_{i = 1}^n \, \left[ \prod_{j \neq i} \left( 2 p^{\mu_j}
  \right) \, \gamma^{\mu_i} {\slashed k}_i \, +
  \prod_{j \neq i, i - 1} \left( 2 p^{\mu_j} \right) \,
  2 p \cdot K_i \, \gamma^{\mu_{i - 1}} \gamma^{\mu_i}
  \right] \nonumber \, ,
\eeqa
where we use $N^{\mu_1 \ldots \mu_n}$ as a shorthand for
$- {\cal N}_{\NE, 2}^{\mu_1 \ldots \mu_n} (p, k_i)$, and
where $K_i = \sum_{p = i}^n k_p$. Note that, as the notation 
suggests, the second term in the second line vanishes for $i = 1$.
We begin by observing that \eqn{theorapp} can easily be verified for 
low values of $n$, and indeed is trivial for $n = 1$. This suggests 
setting up a recursive argument. In order to proceed, we first isolate 
the contribution of the first gluon in the {\it l.h.s.} of \eqn{theorapp}, 
writing
\beq
  N^{\mu_1 \ldots \mu_n} \, = \, \prod_{j = 2}^n 
  \left( 2 p^{\mu_j} \right) \, \gamma^{\mu_1} 
  {\slashed K}_1 + \, {\slashed p} \gamma^{\mu_1} \,
  \sum_{i = 2}^n \, \left[ \prod_{j = i + 1}^n \left( 2 p^{\mu_j}
  \right) \, \prod_{k = 2}^{i - 1} \left({\slashed p} 
  \gamma^{\mu_k} \right) \,
  \gamma^{\mu_i} {\slashed K}_i \right] \, ,
\label{isoone}
\eeq  
We can now use the induction hypothesis, assuming that 
\eqn{theorapp} holds for $n-1$ emissions, and using it to replace
the sum in the second term of \eqn{isoone}. We get 
\beqa
  N^{\mu_1 \ldots \mu_n} & = & \prod_{j = 2}^n 
  \left( 2 p^{\mu_j} \right) \, \gamma^{\mu_1} 
  {\slashed K}_1  \label{induc}  \\
  && + \, \, {\slashed p} \gamma^{\mu_1} \,
  \sum_{i = 2}^n \, \left[ \prod_{j \neq i} \left( 2 p^{\mu_j}
  \right) \, \gamma^{\mu_i} {\slashed k}_i \, +
  \prod_{j \neq i, i - 1} \left( 2 p^{\mu_j} \right) \,
  2 p \cdot K_i \, \gamma^{\mu_{i - 1}} \gamma^{\mu_i}
  \right] \, . \nonumber
\eeqa
Observe that the factor $(2 p^{\mu_1})$ never appears 
on the second line, where the sum starts with the second emission,
and for the same reason the last term starts contributing only 
for $i = 3$. We now operate on the first term on the second line 
of \eqn{induc}, using the commutation relation
\beq
  \left[ {\slashed p} \gamma^\mu, \gamma^\nu {\slashed k} 
  \right] \, = \, 2 \, p \cdot k \, \gamma^\mu \gamma^\nu - 
  2\, p^\nu \, \gamma^\mu {\slashed k} 
  \label{com1}
\eeq
in order to bring the operator ${\slashed p}$ to the right of the 
expression, where it annihilates the massless spinor. We find
\beqa
  N^{\mu_1 \ldots \mu_n} & = & \prod_{j = 2}^n 
  \left( 2 p^{\mu_j} \right) \, \gamma^{\mu_1} 
  {\slashed K}_1 + \sum_{i = 2}^n \, \left[ 
  \prod_{j \neq i} \left( 2 p^{\mu_j} \right) \, 
  \left( \gamma^{\mu_i} 
  {\slashed k}_i \, {\slashed p} \gamma^{\mu_1} 
  +  2 \, p \cdot k_i \gamma^{\mu_1} \gamma^{\mu_i} -
  \gamma^{\mu_1} {\slashed k}_i \, 2 p^{\mu_i} \right)
  \right] \nonumber \\
  && + \, \, {\slashed p} \gamma^{\mu_1} \sum_{i = 3}^n
  \left[ \prod_{j \neq i, i - 1} \left( 2 p^{\mu_j} \right) \,
  2 \, p \cdot K_i \, \gamma^{\mu_{i - 1}} \gamma^{\mu_i}
  \right] \, .
\label{commlong}
\eeqa
The last term in the sum on the first line has a factor 
$\prod_{j = 2}^n \left( 2 p^{\mu_j} \right)$ for every value of 
the summation index $i$: one can then perform the sum getting
simply a factor of ${\slashed K}_2$. This can be combined with
the first term in \eqn{commlong}, using the fact that $K_1 - K_2 
= k_1$. The result is simply $\prod_{j = 2}^n \left( 2 p^{\mu_j} 
\right) \gamma^{\mu_1} {\slashed k}_1$. This can, in turn, 
be combined with the first term in the sum on the first line, 
where the factor of ${\slashed p}$ can be commuted across 
$\gamma^{\mu_1}$ to yield $2 p^{\mu_1}$. One now has 
precisely all terms needed to reconstruct the sum of the first
terms on the {\it r.h.s.} of \eqn{theorapp}. One finds at this 
stage
\beqa
  N^{\mu_1 \ldots \mu_n} & = &
  \sum_{i = 1}^n \, \left[ \prod_{j \neq i} \left( 2 p^{\mu_j}
  \right) \, \gamma^{\mu_i} {\slashed k}_i \right] +
  \sum_{i = 2}^n \left[ \prod_{j \neq i} \left( 2 p^{\mu_j} \right) \, 
  2 \, p \cdot k_i \gamma^{\mu_1} \gamma^{\mu_i} \right]
  \nonumber \\
  && \, + \, \, {\slashed p} \gamma^{\mu_1} \sum_{i = 3}^n
  \left[ \prod_{j \neq i, i - 1} \left( 2 p^{\mu_j} \right) \,
  2 \, p \cdot K_i \, \gamma^{\mu_{i - 1}} \gamma^{\mu_i}
  \right] \, .
\label{stageone}
\eeqa
Next, we act on the second line in \eqn{stageone} applying the
commutation relation
\beq
  \left[ {\slashed p} \gamma^\rho, \gamma^\mu \gamma^\nu
  \right] \, = \, \gamma^\rho \gamma^\mu \, \, 2 \, p^\nu
  - \gamma^\rho \gamma^\nu \, \, 2 \, p^\mu \, ,
\label{com2}
\eeq
also in order to bring the factor ${\slashed p}$ to act on the
massless spinor to the right. This leads to
\beqa
  N^{\mu_1 \ldots \mu_n} & = &
  \sum_{i = 1}^n \, \left[ \prod_{j \neq i} \left( 2 p^{\mu_j}
  \right) \, \gamma^{\mu_i} {\slashed k}_i \right] +
  \sum_{i = 2}^n \left[ \prod_{j \neq i} \left( 2 p^{\mu_j} \right) \, 
  2 \, p \cdot k_i \gamma^{\mu_1} \gamma^{\mu_i} \right]
  \label{stagetwo} \\
  && \hspace{-5mm} 
  + \, \sum_{i = 3}^n \left[ \prod_{j \neq i, i - 1} 
  \left( 2 p^{\mu_j} \right) \, 2 \, p \cdot K_i \, \Big(
  \gamma^{\mu_{i - 1}} \gamma^{\mu_i} \, 2 \, p^{\mu_1} +
  \gamma^{\mu_1} \gamma^{\mu_{i - 1}} \, 2 \, p^{\mu_i} -
  \gamma^{\mu_1} \gamma^{\mu_i} \, 2 \, p^{\mu_{i - 1}}
  \Big) \right] \, . \nonumber
\eeqa
At this stage, one just needs to properly recombine the terms in
in the second and in the third sum in \eqn{stagetwo} in order to 
recover \eqn{theorapp}. We do this in the following way.
\begin{itemize}
\item Combining the term with $i = 2$ in the second sum with
the term $i = 3$ for the second term of the third sum, and using
the fact that $k_2 + K_3 = K_2$, one gets from these two terms
a contribution
\beq
  \prod_{j \neq 2} \left( 2 p^{\mu_j} \right) \, 2 \, p \cdot K_2 \,
  \gamma^{\mu_1} \gamma^{\mu_2} \, ,
\label{contri}
\eeq
which is the first non-vanishing contribution (the one with $i = 2$) to
the last term in \eqn{theorapp}.
\item One easily recognizes that all remaining terms in the last sum in
\eqn{theorapp} are given precisely by the first term in the last sum in 
\eqn{stagetwo}, when the sum is performed in the range $3 \leq 
i \leq n$), as written. Eq.~(\ref{theorapp}) is thus already complete:
it remains to be shown that the leftover terms in \eqn{stagetwo}
cancel.
\item Consider indeed the remaining terms in the second sum in 
\eqn{stagetwo} (those with $3 \leq i \leq n$). Using the fact that
$k_i - K_i = - K_{i + 1}$, they can be combined with the third term 
of the third sum, which is summed over the same range, to get
\beq
  \sum_{i = 3}^n \left[ \prod_{j \neq i} \left( 2 p^{\mu_j} \right)
  \, \left( - 2 \, p \cdot K_{i + 1} \right) \, \gamma^{\mu_1}
  \gamma^{\mu_i} \right] \, .
\label{alcontri}
\eeq
It is easy to see that this sum precisely cancels the leftover 
contributions from second term in the last sum in \eqn{stagetwo}, 
which are in the range $4 \leq i \leq n$, by just shifting $i + 1 \to i$
in \eqn{alcontri}. This completes the proof of \eqn{theorapp}.
\end{itemize}

\section{Next-to-eikonal phase space at arbitrary order}
\label{NEphasespace}

In the \sect{DYNNLO} we have illustrated the use of next-to-eikonal
Feynman rules by calculating certain NNLO contributions to the 
Drell-Yan cross section. In order to correctly calculate logarithmic 
contributions which become dominant near partonic threshold, one 
must include corrections to the phase space measure of integration, 
as well as to matrix elements. We have seen that, at one and two 
loops, these corrections have a simple form and are easily taken 
into account. In order to perform a resummation, however, we will
need to study NE corrections to the phase space measure to all
orders in perturbation theory, that is for any number of emitted 
gluons. In this appendix we show, for the Drell-Yan process and
for observables where the soft gluon phase space is integrated 
over, that the phase space measure remains very simple at NE
order and for any number of emitted gluons. Indeed, it turns out
to be possible to express this measure assigning identical factors
to each emitted gluon, in a way that generalizes the known
factorization properties of multi-gluon phase space at eikonal 
level. This will be instrumental in extending resummation techniques
for physical cross sections to subleading orders in the threshold 
expansion.

Let us briefly recall the effects of subleading phase space corrections 
at one and two loops. In the case of one final state gluon (\eqn{K1a}),
dependence on the energy fraction $z$ factors out of the integration 
over the angular parameter $y$, and subleading corrections 
straightforwardly result upon expanding the $z$-dependent prefactor 
in powers of $\epsilon$. When two real gluon emissions are 
present, as in \eqn{PS}, the integration measure contains the factor 
\beq
  \big[1 - y \, (1 - z) \big]^{- 1 + \epsilon} \, = \, 1 +
  \left(1 - \epsilon \right) (1 - z) + {\cal O}\left[(1 - z)^2 \right] \, .
\label{PSfac}
\eeq
Keeping only the first order term on the right-hand side corresponds 
to the NE approximation, which in the present case coincides with
the next-to-leading order in the threshold expansion. We will now 
show how \eqn{PSfac} can be simply generalized to $n$-gluon 
emission.

Our starting point is the $(n+1)$-body phase space for the
emission of $n$ gluons together with a final state vector boson.
Using standard factorization properties, we write this as
\beqa
  \int \text{dPS} \left(p + \bar{p} \rightarrow q + k_1 + 
  \ldots k_n \right) & = & \int_0^\infty \frac{d K^2}{2 \pi}
  \int \text{dPS} \left(p + \bar{p} \rightarrow q + K \right) 
  \nonumber \\ && \quad \times \, 
  \int \text{dPS} \left( K \rightarrow k_1 + 
  \ldots k_n \right) \, .
\label{PS1}
\eeqa
Here $\text{dPS}(A \to B)$ denotes the phase space for the 
reaction $A \to B$, $q$ is the 4-momentum of the vector boson 
and $\{k_i\}$ the momenta of the emitted gluons. On the right-hand 
side of \eqn{PS1} we have introduced the vector sum of the emitted 
gluon momenta $K = \sum_{i = 1}^n k_i$. One may interpret 
the product of phase spaces on the right-hand side of \eqn{PS1} 
in terms of the production of a pseudo-particle of invariant mass 
$K^2$, followed by its decay into $n$ massless gluons; one then
integrates over all possible invariant masses $K^2$. The convenience 
of this decomposition is that the two body phase space in \eqn{PS1}
is the same for any number of gluons, and depends upon the momenta 
of the latter solely through $K^2$. One may then evaluate the 
two-body phase space by parametrising momenta as
\beqa
  p & = & \frac{\sqrt{s}}{2} \left(1, 0, \ldots, 0, 1 \right) \, ;
  \nonumber \\
  \bar{p} & = & \frac{\sqrt{s}}{2} \left(1, 0, \ldots, 0, - 1 \right) \, ;  
  \label{parmom2p} \\
  q & = & \left( \sqrt{| \vec{q} |^2 + Q^2}, 0, \ldots, | \vec{q} |
  \sin \psi, | \vec{q} | \cos \psi \right) \, ; \nonumber \\
  K & = & \left( \sqrt{ | \vec{q} |^2 + K^2}, 0, \ldots, - | \vec{q} |
  \sin \psi, - | \vec{q} | \cos\psi \right) \nonumber \, .
\eeqa
Straightforward manipulations in $d = 4 - 2 \epsilon$ dimensions 
yield
\beq
  \text{dPS} \left(p + \bar{p} \rightarrow q + K \right) \, = \,
  \frac{1}{8 \pi \sqrt{s}} \, \frac{(4 \pi)^\epsilon}{\Gamma(1-
  \epsilon)} \left(\frac{\lambda^{1/2} (Q^2, K^2, s)}{2 \sqrt{s}}
  \right)^{1 - 2 \epsilon} \left( \sin \psi \right)^{1- 2 \epsilon}
  \, d \psi \, ,
\label{PS2a1}
\eeq
where $\lambda(x, y, z) = x^2 + y^2 + z^2 - 2 x z - 2 x y - 2 y z$ 
is the K\"{a}llen function. As in the case of one real gluon emission 
(corresponding to $K^2 = 0$), one may define the Mandelstam
invariants
\beqa
  t & = & - 2 K \cdot p \, = \, - \, \sqrt{s} \left(\sqrt{| \vec{q} |^2
  + Q^2} - | \vec{q} | \cos \psi \right) \, ; 
  \nonumber \\
  u & = & - 2 K \cdot \bar{p} \, = \, - \, \sqrt{s} \left(\sqrt{| 
  \vec{q} |^2 + Q^2} + | \vec{q} | \cos \psi \right) \, .
\label{utdefs2}
\eeqa
One may then show that
\beq
  \sin^2 \psi \, = \, \frac{4 \left( t u - Q^2 s \right)}{\lambda
  (Q^2, K^2, s)} \, ,
\label{sin2}
\eeq
so that the two-body phase space may be written as
\beq
  \text{dPS} \left(p + \bar{p} \rightarrow q + K \right) \, = \, 
  \frac{1}{8 \pi} \, \frac{(4 \pi)^{\epsilon}}{\Gamma(1
  - \epsilon)} \, s^{- 1 + \epsilon} \, 
  \left(t u - Q^2 s \right)^{1/2 - \epsilon} d \psi \, .
\label{PS2a2}
\eeq

The $n$-body phase space in \eqn{PS1}, on the other hand, is 
complicated in general. If however one performs all integrations
for fixed total invariant mass $K^2$, the dependence of the result
on $K^2$ is dictated by dimensional analysis, and is of the form
\beq
  \int \text{dPS} \left( K \rightarrow k_1 + \ldots k_n \right) 
  \, = \, (K^2)^{n - 2 - (n - 1) \epsilon} 
  \int \widehat{\text{dPS}} \left(K \rightarrow k_1 + 
  \ldots k_n \right) \, ,
\label{PS1b}
\eeq
where $\widehat{\text{dPS}}$ on the right-hand side denotes a
dimensionless bounded measure of integration depending only on 
the angles between the gluons in a chosen frame (for example their 
center of mass frame, as in \sect{DYNNLO}). This measure is therefore independent of $K^2$, $Q^2$ and $z = Q^2/s$. Using the Jacobian
\beq
  d K^2 \, d \psi \, = \, \frac{1}{\sqrt{t u - Q^2 s}} \, dt \,du \, ,
\label{jacobian}
\eeq
the full phase space of \eqn{PS1} becomes
\beqa
  \int \text{dPS} \left(p + \bar{p} \rightarrow q + k_1 + 
  \ldots k_n \right) & = & \frac{1}{16 \pi^2} \, 
  \frac{(4 \pi)^{\epsilon}}{\Gamma(1 - \epsilon)} \,
  s^{ - 1 + \epsilon} \, \int d t \, d u \, (K^2)^{n - 2 -  (n - 1) 
  \epsilon}  \nonumber \\ 
  &&  \hspace{- 1cm} \times \, 
  \left( t u - Q^2 s \right)^{- \epsilon} \, 
  \int \widehat{\text{dPS}} \left( K \rightarrow k_1 + \ldots k_n 
  \right) \, .
\label{PS12}
\eeqa
As in \sect{DYNNLO}, it is convenient to introduce the variables 
$x$ and $y$ defined implicitly in terms of $u$ and $t$ as in 
\eqn{z}. The Jacobian is
\beq
  d u \, d t \, = \, \frac{y \, (1 - y)}{1 - y (1 - z)} \, (1 - z)^3 
  \, s^2 \, d x \, d y \, ,
\label{jacobian2}
\eeq
so that the full phase space takes the final form
\beqa
  \int \text{dPS} \left( p + \bar{p} \rightarrow q + k_1 + 
  \ldots k_N \right) & = & \frac{1}{16 \pi^2} 
  \frac{(4 \pi)^{\epsilon}}{\Gamma(1-\epsilon)} \,
  s^{n - 1- n \epsilon}(1 - z)^{2 n - 1- 2 n \epsilon}
  \label{PSfinal} \\
  && \hspace{-2cm} \times \int_0^1d y \int_0^1d x \,
  \big[ y (1 - y) \big]^{n - 1- n \epsilon} \, x^{n - 2 - (n - 1) \epsilon}
  \, (1 - x)^{- \epsilon} \nonumber \\ && \hspace{-2cm} \times \,
  \big[1 - y (1 - z) \big]^{- (n - 1)(1- \epsilon)}
  \int \widehat{\text{dPS}} \left(K \rightarrow k_1 + 
  \ldots k_n \right) \, . \nonumber
\eeqa
One sees that the dependence on the energy fraction $z$ has only
two sources: the overall prefactor of $(1 - z)^{2 n - 1 - 2 n \epsilon}$, 
which conspires with the matrix element to give the leading power
$(1 - z)^{-1}$ in the threshold expansion of the cross section, and
the factor $[1 - y (1 - z)]^{- (n - 1)(1 - \epsilon)}$, which is neglected
in the eikonal approximation. One observes that \eqn{PSfinal}, for
$n = 2$, reduces as expected to the three-particle phase space 
measure in \eqn{PS}.

It is interesting to note that, at NLO in the threshold expansion, the
$z$ dependence of the phase space measure in \eqn{PSfinal} can be 
expressed in factorized form, with the same factor assigned to each
emitted gluon. Indeed
\beqa
  \big[ 1 - y (1 - z) \big]^{ - (n -1)(1- \epsilon)} & = & 
  1 + (n - 1) \, y \, (1 - z) \left(1 - \epsilon \right) + 
  {\cal O}[(1 - z)^2] \, \nonumber \\
  & = & z^{(1 - \epsilon) \, y} \, \left( z^{- (1 - \epsilon) \, y }
  \right)^n \, .
\label{facexpand2}
\eeqa
This provides, to some extent, a generalization of the factorization
property of the multi-gluon phase space, which is known to degenerate
into a product of uncorrelated single-gluon phase spaces at leading 
order in the threshold expansion. At NLO in the threshold expansion,
the multi-gluon phase space measure is still factorizable, however the 
factors assigned to each emitted gluon retain a dependence on the
angular variable $y$ describing the overall orientation on the 
multi-gluon system. In other words, single-gluon phase spaces are 
not fully uncorrelated, but their correlation is simple, depending only 
on the global variables of the multi-gluon system and not on 
individual gluon momenta.

This discussion applies to the explicit example of Drell-Yan production. 
We expect that such arguments will apply more generally in other 
scattering processes, pending a suitable parametrisation of the 
partonic momenta.

\section{The double-real-emission contribution to the Drell-Yan
$K$ factor}
\label{tworeal}

\begin{figure}
\begin{center}
\scalebox{1.0}{\includegraphics{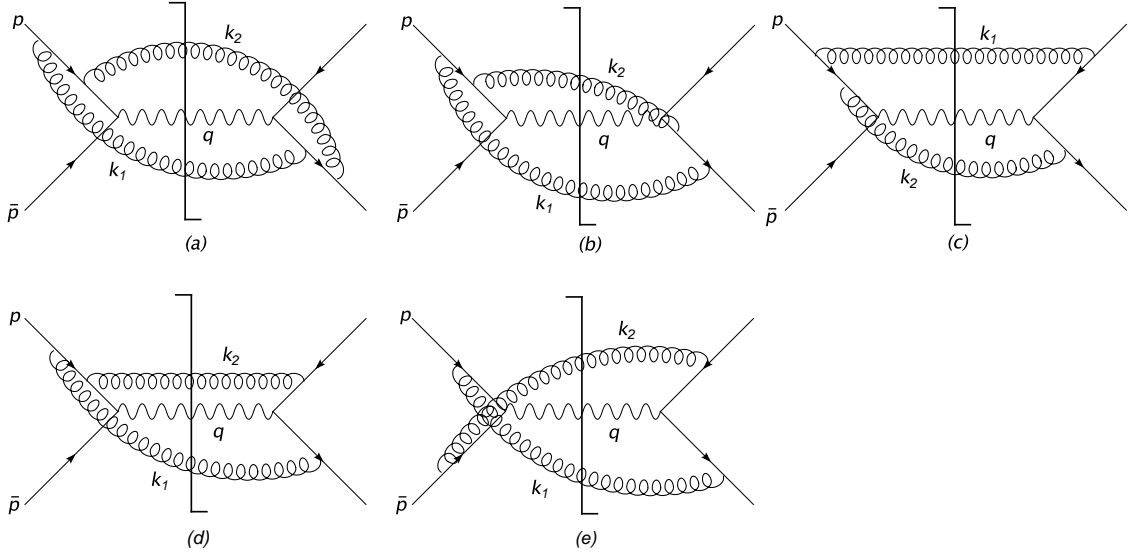}}
\caption{Diagrams for the double-real-emission contribution to 
the NNLO Drell-Yan $K$ factor discussed in the text. A cut is 
implied over the intermediate state in each case, and complex 
conjugates of the above diagrams (excluding (e), which is real) 
must also be included.}
\label{sigcdiags}
\end{center}
\end{figure}

In this appendix we briefly describe how to compute the terms 
proportional to $C_F^2$ of the Drell-Yan $K$-factor, for the 
$q \bar{q}$ initial state, by using ordinary Feynman diagrams 
and expanding them to NE order. The relevant diagrams are shown 
in fig.~\ref{sigcdiags}. The corresponding squared matrix elements
are easily computed, and must then be integrated with the phase 
space measure in \eqn{PS}.
As an example, diagram $(a)$ contributes a factor
\beq
  |{\cal M}|^2_{(a)} \, \propto \, 
  \frac{ \text{Tr} \left[ \pb \gamma^\alpha (\slsh{p} - 
  \slsh{k}_1 - \slsh{k}_2) \gamma^\nu (\slsh{p} - \slsh{k}_1)
  \gamma^\mu \slsh{p} \gamma_\alpha (- \pb + \slsh{k}_1 +
  \slsh{k}_2) \gamma_\mu (- \pb + \slsh{k}_2) \gamma_\nu
  \right]}{(p - k_1 - k_2)^2 \, (p - k_1)^2 \, (- \bar{p} + k_1
  + k_2)^2 \, (- \bar{p} + k_2)^2} \, .
  \label{ampa}
\eeq  
Note that the contributions from diagrams $(a) - (d)$ must be
counted twice in order to include Hermitian conjugate graphs, while
diagram $(e)$ is real.

To calculate the squared matrix element to NE order, one first 
relabels $k_i \rightarrow \xi k_i$, so that
\beq
  \bar{p} \cdot k_i \rightarrow \xi \, \bar{p} \cdot k_i, \quad p
  \cdot k_i \rightarrow \xi \, p \cdot k_i, \quad k_i
  \cdot k_j \rightarrow \xi^2 \, k_i\cdot k_j \, .
\label{recalemoms}
\eeq
One then expands each diagram to first subleading order in $\xi$, 
which corresponds to the NE approximation. Through repeated use 
of the identities
\beqa
  \frac{p \cdot k_1}{p \cdot k_2} & = & \frac{1}{p \cdot k_2}
  \frac{s - \tilde{t}}{2} - 1 \, , \nonumber \\
  \frac{\bar{p} \cdot k_1}{\bar{p} \cdot k_2} & = & 
  \frac{1}{\bar{p} \cdot k_2} \frac{s_{12} + \tilde{t} - 
  Q^2}{2} - 1 \, ,
\label{id2}
\eeqa
(with similar results for $k_1\leftrightarrow k_2$), each diagram 
can be written as a sum of terms containing no more than two factors 
of $p \cdot k_i$ and $\bar{p} \cdot k_i$. Then each term becomes 
an integral of the form of \eqn{int2}. The remaining phase space 
integrals can be carried out after expanding the integrand in powers 
of $1 - z$ and $\epsilon$, as described for the NE calculation in 
\sect{DYNNLO}. The final result for the full amplitude (keeping only 
logarithmic terms with rational coefficients as done in the text) is 
given by
\beqa
  K^{(2)}_\NE (z) & = & \left( \frac{\alpha_s}{4\pi} C_F \right)^2
  \left[ - \frac{32}{\epsilon^3} \, {\cal D}_0 (z) +
   \frac{128}{\epsilon^2} \,  {\cal D}_1 (z) - \frac{128}{\epsilon^2}
  \log(1 - z) \right. \nonumber \\
  && - \, \frac{256}{\epsilon} \, {\cal D}_2 (z) 
  + \frac{256}{\epsilon}
  \log^2 (1 - z) - \frac{320}{\epsilon} \log(1 - z) \nonumber \\ 
  && \left. + \, 
  \frac{1024}{3} \, {\cal D}_3 (z) - \frac{1024}{3} \log^3(1-z)
  + 640 \log^2(1 - z) \right] \, ,
\label{NEcomplete}
\eeqa
which is in complete agreement with the sum of \eqns{etot}{NEresult}.

\bibliographystyle{JHEP}
\bibliography{refshort.bib}
\end{document}